\documentclass[pdflatex,sn-mathphys-num]{sn-jnl}

\usepackage{amsfonts,amsmath,amssymb}
\usepackage{bm}
\usepackage{graphicx}
\usepackage{epstopdf}
\usepackage{caption}
\usepackage{subcaption}
\usepackage{algorithm}
\usepackage{algpseudocode}
\usepackage{setspace}
\usepackage{ulem}
\usepackage{multirow}
\usepackage{booktabs}
\usepackage{amsopn}
\usepackage{listings}
\usepackage{hyperref}
\usepackage{cleveref}
\usepackage{xcolor}

\ifpdf
  \DeclareGraphicsExtensions{.eps,.pdf,.png,.jpg}
\else
  \DeclareGraphicsExtensions{.eps}
\fi

\algnewcommand{\IIf}[1]{\State\algorithmicif\ #1\ \algorithmicthen}

\begin{document}
\title[Hyper-reduction methods for accelerating nonlinear FE simulations]{Hyper-reduction methods for accelerating nonlinear finite element simulations: open source implementation and reproducible benchmarks}

\author[1]{\fnm{Axel} \sur{Larsson}}\email{a.larsson@princeton.edu}

\author[2]{\fnm{Minji} \sur{Kim}}\email{mkim5@unc.edu}

\author[3]{\fnm{Chris} \sur{Vales}}\email{chris.vales@dartmouth.edu}

\author[1]{\fnm{Sigrid} \sur{Adriaenssens}}\email{sadriaen@princeton.edu}

\author[4]{\fnm{Dylan Matthew} \sur{Copeland}}\email{copeland11@llnl.gov}

\author[4]{\fnm{Youngsoo} \sur{Choi}}\email{choi15@llnl.gov}

\author*[4]{\fnm{Siu Wun} \sur{Cheung}}\email{cheung26@llnl.gov}

\affil[1]{\orgdiv{Department of Civil and Environmental Engineering}, \orgname{Princeton University}, \orgaddress{\city{Princeton}, \state{NJ 08540}, \country{USA}}}

\affil[2]{\orgdiv{Department of Statistics and Operations Research}, \orgname{The University of North Carolina at Chapel Hill}, \orgaddress{\city{Chapel Hill}, \state{NC 27599}, \country{USA}}}

\affil[3]{\orgdiv{Department of Mathematics}, \orgname{Dartmouth College}, \orgaddress{\city{Hanover}, \state{NH 03755}, \country{USA}}}

\affil[4]{\orgdiv{Center for Applied Scientific Computing}, \orgname{Lawrence Livermore National Laboratory}, \orgaddress{\city{Livermore}, \state{CA 94550}, \country{USA}}}

\abstract{Hyper-reduction methods have gained increasing attention
for their potential to accelerate reduced order models for nonlinear
systems, yet their comparative accuracy and computational efficiency
are not well understood.
Motivated by this gap, we evaluate a
range of hyper-reduction techniques for nonlinear finite element
models across benchmark problems of varying complexity,
assessing the inevitable tradeoff between accuracy and speedup.
More specifically, we consider interpolation methods based on
the gappy proper orthogonal decomposition
as well as the empirical quadrature procedure (EQP),
and apply them to the hyper-reduction of problems in
nonlinear diffusion, nonlinear elasticity and Lagrangian
hydrodynamics.
Our numerical results are generated using the open source
\texttt{libROM}, \texttt{Laghos} and \texttt{MFEM}
numerical libraries.
Our findings reveal that the comparative performance between
hyper-reduction methods depends on both the problem and the
choice of time integration method.
The EQP method generally achieves lower relative errors than
interpolation methods and is more efficient in terms of quadrature
point usage, resulting in a lower wall time for the nonlinear diffusion
and elasticity problems.
However, its online computational cost is observed to be relatively
high for Lagrangian hydrodynamics problems.
Conversely, interpolation methods exhibit greater variability,
especially with respect to the use of different time integration methods
in the Lagrangian hydrodynamics problems.
The presented results underscore the need for problem specific method
selection to balance accuracy and efficiency, while also offering
useful guidance for future comparisons and refinements
of hyper-reduction techniques.}


\maketitle

\newcommand{\bmat}[1]{\begin{bmatrix}#1\end{bmatrix}} 
\newcommand{\pmat}[1]{\begin{pmatrix}#1\end{pmatrix}} 
\newcommand{\vmat}[1]{\begin{vmatrix}#1\end{vmatrix}} 
\newcommand{\Vmat}[1]{\begin{Vmatrix}#1\end{Vmatrix}} 
\newcommand{\Bmat}[1]{\begin{Bmatrix}#1\end{Bmatrix}} 

\newcommand{\timeIndex}{n}
\newcommand{\basisIndex}{j}
\newcommand{\dummyBasisIndex}{j'}
\newcommand{\sampleIndex}{i}
\newcommand{\paramIndex}{m}
\newcommand{\windowIndex}{w}
\newcommand{\finalSymbol}{f}
\newcommand{\zoneSymbol}{z}
\newcommand{\dummyMat}{\boldsymbol{B}}
\newcommand{\dummyVec}{\boldsymbol{c}}
\newcommand{\rowIndex}{\basisIndex}
\newcommand{\dummyRowIndex}{\dummyBasisIndex}
\newcommand{\columnIndex}{j}
\newcommand{\identitySymbol}{I}
\newcommand{\zeroSymbol}{0}

\newcommand{\stateSymbol}{y}
\newcommand{\multiplierSymbol}{z}
\newcommand{\timeSymbol}{t}
\newcommand{\finalTime}{T}
\newcommand{\paramSymbol}{\mu}
\newcommand{\param}{\paramSymbol}
\newcommand{\dimensionSymbol}{d}
\newcommand{\stateDomainSymbol}{\Omega}
\newcommand{\gradientSymbol}{\nabla}
\newcommand{\paramDomainSymbol}{D}
\newcommand{\domain}[1]{\mathsf{#1}}
\newcommand{\paramDomain}{\domain{\paramDomainSymbol}}
\newcommand{\forceSymbol}{F}
\newcommand{\stateLinearOperatorSymbol}{A}
\newcommand{\bilinearOperatorSymbol}{B}
\newcommand{\multiplierLinearOperatorSymbol}{C}
\newcommand{\multiplierNonlinearOperatorSymbol}{g}
\newcommand{\constantForceSymbol}{s}
\newcommand{\stateNonlinearForceSymbol}{f}
\newcommand{\multiplierNonlinearForceSymbol}{g}

\newcommand{\vectorSpaceSymbol}{\mathcal{V}}
\newcommand{\HilbertSpaceSymbol}{\mathcal{H}}
\newcommand{\stateSpaceSymbol}{\mathcal{Y}}
\newcommand{\multiplierSpaceSymbol}{\mathcal{Z}}
\newcommand{\projectionSymbol}{\mathcal{P}}

\newcommand{\sizeFOMsymbol}{N}
\newcommand{\stateFOMSize}{\sizeFOMsymbol_{\stateSymbol}}
\newcommand{\multiplierFOMSize}{\sizeFOMsymbol_{\multiplierSymbol}}
\newcommand{\meshSize}{h}
\newcommand{\meshLevel}{m}
\newcommand{\feOrder}{k}
\newcommand{\feSpace}{\vectorSpaceSymbol^\meshSize}
\newcommand{\stateFeSpace}{\stateSpaceSymbol^\meshSize}
\newcommand{\multiplierFeSpace}{\multiplierSpaceSymbol^\meshSize}
\newcommand{\stateFeSymbol}{\stateSymbol_\meshSize}
\newcommand{\multiplierFeSymbol}{\multiplierSymbol_\meshSize}
\newcommand{\feBasisSet}{\mathcal{B}}
\newcommand{\stateFeBasisSymbol}{\varphi}
\newcommand{\multiplierFeBasisSymbol}{\lambda}

\newcommand{\stateFeBasisVec}{\boldsymbol{\stateFeBasisSymbol}}
\newcommand{\multiplierFeBasisVec}{\boldsymbol{\multiplierFeBasisSymbol}}

\newcommand{\stateFeVec}{L_{\feBasisSet_\stateSpaceSymbol}}
\newcommand{\multiplierFeVec}{L_{\feBasisSet_\multiplierSpaceSymbol}}

\renewcommand{\state}{\mathbf{\stateSymbol}}
\newcommand{\multiplier}{\mathbf{\multiplierSymbol}}
\newcommand{\massMatSymbol}{M}
\newcommand{\massMat}{\mathbf{\massMatSymbol}}
\newcommand{\stateConstantVec}{\mathbf{\constantForceSymbol}}
\newcommand{\stateLinearMat}{\mathbf{\stateLinearOperatorSymbol}}

\newcommand{\stateNonlinearVec}{\mathbf{\stateNonlinearForceSymbol}}
\newcommand{\multiplierNonlinearVec}{\mathbf{\multiplierNonlinearOperatorSymbol}}
\newcommand{\bilinearMat}
{\mathbf{\bilinearOperatorSymbol}}
\newcommand{\multiplierLinearMat}{\mathbf{\multiplierLinearOperatorSymbol}}

\newcommand{\forceSys}{\mathsf{\forceSymbol}}
\newcommand{\jacobianSymbol}{J}
\newcommand{\residualSymbol}{r}	
\newcommand{\residualVec}{\mathsf{\residualSymbol}}
\newcommand{\choleskyMat}{\mathbf{\choleskyMatSymbol}}
\newcommand{\HH}{H^1}
\newcommand{\LL}{L^2}

\newcommand{\sizeROMsymbol}{r}
\newcommand{\stateROMSize}{\sizeROMsymbol_{\stateSymbol}}
\newcommand{\multiplierROMSize}{\sizeROMsymbol_{\multiplierSymbol}}
\newcommand{\offsetSymbol}{\text{offset}}

\newcommand{\stateSubspaceSymbol}{\stateSpaceSymbol^{\sizeROMsymbol}}
\newcommand{\multiplierSubspaceSymbol}{\multiplierSpaceSymbol^{\sizeROMsymbol}}

\newcommand{\stateROMBasisSymbol}{\psi}
\newcommand{\stateROMBasisMat}{\boldsymbol{\Psi}}
\newcommand{\stateROMBasisVec}{\boldsymbol{\stateROMBasisSymbol}}
\newcommand{\multiplierROMBasisSymbol}{\theta}
\newcommand{\multiplierROMBasisMat}{\boldsymbol{\Theta}}
\newcommand{\multiplierROMBasisVec}{\boldsymbol{\multiplierROMBasisSymbol}}

\newcommand{\ROMSolutionSymbol}[1]{\hat{#1}}
\newcommand{\ROMApproxSymbol}[1]{\tilde{#1}}

\newcommand{\stateSymbolApprox}{\ROMApproxSymbol{\stateSymbol}}
\newcommand{\stateOSSymbol}{\stateSymbol_{\offsetSymbol}}
\newcommand{\multiplierSymbolApprox}{\ROMApproxSymbol{\multiplierSymbol}}
\newcommand{\multiplierOSSymbol}{\multiplierSymbol_{\offsetSymbol}}

\newcommand{\stateApprox}{\ROMApproxSymbol{\state}}
\newcommand{\multiplierApprox}{\ROMApproxSymbol{\multiplier}}
\newcommand{\stateOS}{\state_{\offsetSymbol}}
\newcommand{\multiplierOS}{\multiplier_{\offsetSymbol}}
\newcommand{\ROMmassMat}{\ROMSolutionSymbol{\massMat}}
\newcommand{\ROMforceSys}{\ROMSolutionSymbol{\forceSys}}
\newcommand{\ROMstateConstantVec}{\ROMSolutionSymbol{\stateConstantVec}}
\newcommand{\ROMstateLinearMat}{\ROMSolutionSymbol{\stateLinearMat}}
\newcommand{\ROMmultiplierLinearMat}{\ROMSolutionSymbol{\multiplierLinearMat}}
\newcommand{\ROMbilinearMat}{\ROMSolutionSymbol{\bilinearMat}}
\newcommand{\identityMat}[1]{\mathbf{\identitySymbol}_{#1}}
\newcommand{\zeroVec}[1]{\mathbf{\zeroSymbol}_{#1}}

\newcommand{\sizeROMForceSample}{n}
\newcommand{\sampleSymbol}{z}
\newcommand{\projectionMatSymbol}{\mathbf{\Pi}}

\newcommand{\sizeROMStateForceSymbol}{\sizeROMsymbol_{\stateNonlinearForceSymbol}}


\newcommand{\sizeROMSampleSymbol}{\sizeROMForceSample_{\stateNonlinearForceSymbol}}


\newcommand{\ROMstateNonlinearVec}{\ROMSolutionSymbol{\stateNonlinearVec}}

\newcommand{\ROMForceBasisSymbol}{\xi}
\newcommand{\ROMForceBasisMat}{\boldsymbol{\Xi}}
\newcommand{\ROMForceBasisVec}{\boldsymbol{\ROMForceBasisSymbol}}
\newcommand{\ROMstateNonlinearVecApprox}{\ROMApproxSymbol{\stateNonlinearVec}}

\newcommand{\ROMSamplingSetSymbol}{\mathcal{Z}}
\newcommand{\ROMSamplingSetNeighSymbol}{\ROMSamplingSetSymbol^\prime} 
\newcommand{\ROMSamplingMatSymbol}{\mathbf{Z}}
\newcommand{\gappyerr}{\varepsilon}

\newcommand{\ROMStateNonlinearForceCoeff}{\ROMSolutionSymbol{\mathbf{\stateNonlinearForceSymbol}}}
\newcommand{\ROMMultiplierNonlinearForceCoeff}{\ROMSolutionSymbol{\mathbf{\multiplierNonlinearForceSymbol}}}

\newcommand{\ROMStateForceBasisMat}{\ROMForceBasisMat_{\stateNonlinearForceSymbol}}
\newcommand{\ROMStateForceBasisVec}{\ROMForceBasisVec^{(\stateNonlinearForceSymbol)}}
\newcommand{\ROMStateSamplingSetSymbol}{\ROMSamplingSetSymbol_{\stateNonlinearForceSymbol}}
\newcommand{\ROMStateSamplingMatSymbol}{\ROMSamplingMatSymbol_{\stateNonlinearForceSymbol}}

\newcommand{\continuousError}{\boldsymbol{\continuousErrorSymbol}}
\newcommand{\discreteError}{\boldsymbol{\discreteErrorSymbol}}
\newcommand{\relError}{\relErrorSymbol}
\newcommand{\difference}{\boldsymbol{\differenceSymbol}}
\newcommand{\initialError}{\boldsymbol{\initialErrorSymbol}}
\newcommand{\residual}{\boldsymbol{\residualSymbol}}

\newcommand{\velocitySymbol}{v}
\newcommand{\positionSymbol}{x}
\newcommand{\pressureSymbol}{p}
\newcommand{\dummyVelocitySymbol}{v'}
\newcommand{\dummyPositionSymbol}{x'}
\newcommand{\dummyPressureSymbol}{p'}

\newcommand{\pressureFE}{\feSpace_P}
\newcommand{\sizePressureFE}{\sizeFOMsymbol_P}

\newcommand{\densitySymbol}{\rho}
\newcommand{\energySymbol}{e}
\newcommand{\stressSymbol}{\sigma}
\newcommand{\artificialStressSymbol}{\stressSymbol_a}
\newcommand{\adiabaticIndexSymbol}{\gamma}
\newcommand{\normalSymbol}{n}
\newcommand{\neumannSymbol}{h}
\newcommand{\initialSymbol}[1]{\tilde{#1}}
\newcommand{\initialDomain}{\initialSymbol{\stateDomainSymbol}}
\newcommand{\initialPosition}{\initialSymbol{\positionSymbol}}

\newcommand{\snapshotSymb}{E}
\newcommand{\snapshotSymbS}{S}

\newcommand{\oneSymbol}{1}
\newcommand{\lagrangianSymbol}{X}
\newcommand{\kinematicSymbol}{V}
\newcommand{\thermodynamicSymbol}{E}

\newcommand{\mapto}{\rightarrow}
\newcommand{\sizeSnapshot}{N_{\snapshotSymb}}
\newcommand{\sizeSnapshotS}{N_{\snapshotSymbS}}
\newcommand{\Span}[1]{\text{span}\{#1\}}
\newcommand{\RR}[1]{\mathbb{R}^{#1}}

\newcommand{\continuousErrorSymbol}{\varepsilon}
\newcommand{\discreteErrorSymbol}{\delta}
\newcommand{\relErrorSymbol}{\mathcal{\epsilon}}
\newcommand{\differenceSymbol}{\theta}
\newcommand{\initialErrorSymbol}{\zeta}
\newcommand{\norm}[1]{\left\lVert #1 \right\rVert}
\newcommand{\fullNorm}[1]{{\left\vert\kern-0.25ex\left\vert\kern-0.25ex\left\vert #1 
    \right\vert\kern-0.25ex\right\vert\kern-0.25ex\right\vert}}
\newcommand{\rkcoeffSymbol}{Y}
\newcommand{\rkcoeffk}[1]{\mathsf{\rkcoeffSymbol}_{#1}}
\newcommand{\rkcoeffI}{\rkcoeffk{1}}
\newcommand{\rkcoeffII}{\rkcoeffk{2}}
\newcommand{\rkcoeffIII}{\rkcoeffk{3}}
\newcommand{\rkcoeffIV}{\rkcoeffk{4}}
\newcommand{\rkErrorCoeffSymbol}{a}
\newcommand{\CFLconst}{\alpha}

\newcommand{\forceOne}{\forceSys^{\kinematicSymbol}}
\newcommand{\forceTv}{\forceSys^{\thermodynamicSymbol}}
\newcommand{\avgforceTv}{\bar{\forceSys}^{\thermodynamicSymbol}}
\newcommand{\forceOnek}[1]{\forceOne_{#1}}
\newcommand{\forceTvk}[1]{\forceTv_{#1}}
\newcommand{\avgforceTvk}[1]{\avgforceTv_{#1}}
\newcommand{\forceOneApprox}{\ROMApproxSymbol{\forceSys}^{\kinematicSymbol}}
\newcommand{\forceTvApprox}{\ROMApproxSymbol{\forceSys}^{\thermodynamicSymbol}}
\newcommand{\avgforceTvApprox}{\bar{\ROMApproxSymbol{\forceSys}}^{\thermodynamicSymbol}}
\newcommand{\forceOneApproxk}[1]{\forceOneApprox_{#1}}
\newcommand{\forceTvApproxk}[1]{\forceTvApprox_{#1}}
\newcommand{\avgforceTvApproxk}[1]{\avgforceTvApprox_{#1}}
\newcommand{\forceSysApprox}{\ROMApproxSymbol{\forceSys}}

\newcommand{\energySubspace}{stateSubspaceSymbol_{\energySymbol}}
\newcommand{\velocitySubspace}{stateSubspaceSymbol_{\velocitySymbol}}
\newcommand{\positionSubspace}{stateSubspaceSymbol_{\positionSymbol}}
\newcommand{\forceOneSubspace}{stateSubspaceSymbol_{\kinematicSymbol}}
\newcommand{\forceTvSubspace}{stateSubspaceSymbol_{\thermodynamicSymbol}}

\newcommand{\energy}{\boldsymbol{\energySymbol}}
\newcommand{\velocity}{\boldsymbol{\velocitySymbol}}
\newcommand{\position}{\boldsymbol{\positionSymbol}}

\newcommand{\velocityApprox}{\ROMApproxSymbol{\velocity}}
\newcommand{\energyApprox}{\ROMApproxSymbol{\energy}}
\newcommand{\positionApprox}{\ROMApproxSymbol{\position}}

\newcommand{\oneVec}{\boldsymbol{\oneSymbol}_{\thermodynamicSymbol}}
\newcommand{\kinematicFE}{\feSpace_{\kinematicSymbol}}
\newcommand{\thermodynamicFE}{\feSpace_{\thermodynamicSymbol}}
\newcommand{\kinematicMassMat}{\massMat_{\kinematicSymbol}}
\newcommand{\thermodynamicMassMat}{\massMat_{\thermodynamicSymbol}}
\newcommand{\kinematicCholeskyMat}{\choleskyMat_{\kinematicSymbol}}
\newcommand{\thermodynamicCholeskyMat}{\choleskyMat_{\thermodynamicSymbol}}
\newcommand{\jacobian}{\boldsymbol{\jacobianSymbol}}
\newcommand{\jacobianLocal}{\jacobian_{\zoneSymbol}}
\newcommand{\forceMat}{\boldsymbol{\forceSymbol}}
\newcommand{\forceMatLocal}{\forceMat_{\zoneSymbol}}
\newcommand{\sizeKinematicFE}{\sizeFOMsymbol_{\kinematicSymbol}}
\newcommand{\sizeThermodynamicFE}{\sizeFOMsymbol_{\thermodynamicSymbol}}
\newcommand{\sizeROMforceOne}{\stateROMSize_{\kinematicSymbol}}
\newcommand{\sizeROMforceTv}{\stateROMSize_{\thermodynamicSymbol}}
\newcommand{\samplingOp}[1]{\breve{#1}}
\newcommand{\sizeROMforceOneSample}{\stateROMSize_{\samplingOp{\kinematicSymbol}}}
\newcommand{\sizeROMforceTvSample}{\stateROMSize_{\samplingOp{\thermodynamicSymbol}}}
\newcommand{\sizeROMvelocity}{\stateROMSize_{\velocitySymbol}}
\newcommand{\sizeROMenergy}{\stateROMSize_{\energySymbol}}
\newcommand{\sizeROMposition}{\stateROMSize_{\positionSymbol}}
\newcommand{\ratioROMvelocity}{\zeta_{\velocitySymbol}}
\newcommand{\ratioROMenergy}{\zeta_{\energySymbol}}
\newcommand{\ratioROMposition}{\zeta_{\positionSymbol}}
\newcommand{\ratioROMforceOneSample}{\zeta_{\samplingOp{\kinematicSymbol}}}
\newcommand{\ratioROMforceTvSample}{\zeta_{\samplingOp{\thermodynamicSymbol}}}
\newcommand{\factorROMforceOneSample}{\lambda_{\samplingOp{\kinematicSymbol}}}
\newcommand{\factorROMforceTvSample}{\lambda_{\samplingOp{\thermodynamicSymbol}}}
\newcommand{\sizeWholeFE}{\sizeFOMsymbol}
\newcommand{\sizeWholeROM}{\stateROMSize}
\newcommand{\timek}[1]{\timeSymbol_{#1}}
\newcommand{\ntimestep}{N_{\timeSymbol}}
\newcommand{\timestep}{\Delta\timeSymbol}
\newcommand{\timestepk}[1]{\timestep_{#1}}
\newcommand{\timestepestimateSymbol}{\tau}
\newcommand{\timestepestimatek}[1]{\timestepestimateSymbol_{#1}}

\newcommand{\avgvelocityt}[1]{\bar{\velocity}_{#1}}
\newcommand{\statet}[1]{\state_{#1}}
\newcommand{\energyt}[1]{\energy_{#1}}
\newcommand{\velocityt}[1]{\velocity_{#1}}
\newcommand{\positiont}[1]{\position_{#1}}
\newcommand{\forceMatt}[1]{\forceMat^{#1}}
\newcommand{\forceSyst}[1]{\forceSys^{#1}}
\newcommand{\avgvelocityApproxt}[1]{\bar{\velocityApprox}_{#1}}
\newcommand{\stateApproxt}[1]{\stateApprox_{#1}}
\newcommand{\energyApproxt}[1]{\energyApprox_{#1}}
\newcommand{\velocityApproxt}[1]{\velocityApprox_{#1}}
\newcommand{\positionApproxt}[1]{\positionApprox_{#1}}
\newcommand{\ROMstatet}[1]{\ROMstate_{#1}}
\newcommand{\ROMenergyt}[1]{\ROMenergy_{#1}}
\newcommand{\ROMvelocityt}[1]{\ROMvelocity_{#1}}
\newcommand{\ROMpositiont}[1]{\ROMposition_{#1}}
\newcommand{\ROMforceOnet}[1]{\ROMforceOne_{#1}}
\newcommand{\ROMforceTvt}[1]{\ROMforceTv_{#1}}
\newcommand{\ROMavgvelocityt}[1]{\bar{\ROMvelocity}_{#1}}
\newcommand{\ROMavgforceTvt}[1]{\bar{\ROMSolutionSymbol{\thermodynamicSymbol}_{#1}}}
\newcommand{\ROMoneVec}{\ROMSolutionSymbol{\boldsymbol{\oneSymbol}}_{\thermodynamicSymbol}}

\newcommand{\nat}[1]{\mathbb{N}(#1)}
\newcommand{\velocityBasis}{\stateROMBasisMat_{\velocitySymbol}}
\newcommand{\energyBasis}{\stateROMBasisMat_{\energySymbol}}
\newcommand{\positionBasis}{\stateROMBasisMat_{\positionSymbol}}
\newcommand{\velocityBasisVec}{\stateROMBasisVec{\velocitySymbol}}
\newcommand{\energyBasisVec}{\stateROMBasisVec{\energySymbol}}
\newcommand{\positionBasisVec}{\stateROMBasisVec{\positionSymbol}}
\newcommand{\velocityBasisVeck}[1]{\velocityBasisVec^{#1}}
\newcommand{\energyBasisVeck}[1]{\energyBasisVec^{#1}}
\newcommand{\positionBasisVeck}[1]{\positionBasisVec^{#1}}
\newcommand{\forceSysEnergy}{\forceSys_{\energySymbol}}
\newcommand{\forceSysVelocity}{\forceSys_{\velocitySymbol}}
\newcommand{\forceSysPosition}{\forceSys_{\positionSymbol}}
\newcommand{\forceTvBasis}{\stateROMBasisMat_{\thermodynamicSymbol}}
\newcommand{\forceOneBasis}{\stateROMBasisMat_{\kinematicSymbol}}
\newcommand{\forceTvBasisVec}{\stateROMBasisVec^{(\thermodynamicSymbol)}}
\newcommand{\forceOneBasisVec}{\stateROMBasisVec^{(\kinematicSymbol)}}
\newcommand{\forceTvBasisVeck}[1]{\forceTvBasisVec_{#1}}
\newcommand{\forceOneBasisVeck}[1]{\forceOneBasisVec_{#1}}
\newcommand{\forceTvSamplingMat}{\ROMSamplingMatSymbol_{\thermodynamicSymbol}}
\newcommand{\forceOneSamplingMat}{\ROMSamplingMatSymbol_{\kinematicSymbol}}
\newcommand{\basisWindowk}[1]{\stateROMBasisMat_{#1}}
\newcommand{\ROMstateSymbol}{\ROMSolutionSymbol{\stateSymbol}}
\newcommand{\ROMstate}{\ROMSolutionSymbol{\state}}
\newcommand{\ROMmultiplier}{\ROMSolutionSymbol{\multiplier}}
\newcommand{\ROMenergy}{\ROMSolutionSymbol{\energy}}
\newcommand{\ROMvelocity}{\ROMSolutionSymbol{\velocity}}
\newcommand{\ROMposition}{\ROMSolutionSymbol{\position}}
\newcommand{\ROMforceMat}{\ROMSolutionSymbol{\forceMat}}
\newcommand{\ROMforceOne}{\ROMSolutionSymbol{\forceSys}^{\kinematicSymbol}}
\newcommand{\ROMforceTv}{\ROMSolutionSymbol{\forceSys}^{\thermodynamicSymbol}}
\newcommand{\ROMforceSysEnergy}{\ROMforceSys_{\energySymbol}}
\newcommand{\ROMforceSysVelocity}{\ROMforceSys_{\velocitySymbol}}
\newcommand{\ROMforceSysPosition}{\ROMforceSys_{\positionSymbol}}
\newcommand{\energyOS}{\energy_{\offsetSymbol}}
\newcommand{\velocityOS}{\velocity_{\offsetSymbol}}
\newcommand{\positionOS}{\position_{\offsetSymbol}}

\newcommand{\ntimestepROM}{\ROMApproxSymbol{N}_{\timeSymbol}}
\newcommand{\ROMReducedMassMat}{\hat{\massMat}}
\newcommand{\ROMKinematicMassMat}{\ROMReducedMassMat_{\kinematicSymbol}}
\newcommand{\ROMThermodynamicMassMat}{\ROMReducedMassMat_{\thermodynamicSymbol}}
\newcommand{\forceOneObliqProjMat}{\projectionMatSymbol_{\kinematicSymbol}}
\newcommand{\forceTvObliqProjMat}{\projectionMatSymbol_{\thermodynamicSymbol}}
\newcommand{\velocityIdentity}{\IdentitySymbol_\kinematicSymbol}
\newcommand{\energyIdentity}{\IdentitySymbol_\thermodynamicSymbol}
\newcommand{\positionIdentity}{\velocityIdentity}

\newcommand{\windowk}[1]{T_{#1}}
\newcommand{\nwindow}{N_w}
\newcommand{\energySubspaceWindow}[1]{stateSubspaceSymbol_{\energySymbol}^{#1}}
\newcommand{\velocitySubspaceWindow}[1]{stateSubspaceSymbol_{\velocitySymbol}^{#1}}
\newcommand{\positionSubspaceWindow}[1]{stateSubspaceSymbol_{\positionSymbol}^{#1}}
\newcommand{\forceOneSubspaceWindow}[1]{stateSubspaceSymbol_{\kinematicSymbol}^{#1}}
\newcommand{\forceTvSubspaceWindow}[1]{stateSubspaceSymbol_{\thermodynamicSymbol}^{#1}}
\newcommand{\sizeROMvelocityWindow}[1]{\stateROMSize_{\velocitySymbol}^{#1}}
\newcommand{\sizeROMenergyWindow}[1]{\stateROMSize_{\energySymbol}^{#1}}
\newcommand{\sizeROMpositionWindow}[1]{\stateROMSize_{\positionSymbol}^{#1}}
\newcommand{\sizeROMforceOneWindow}[1]{\stateROMSize_{\kinematicSymbol}^{#1}}
\newcommand{\sizeROMforceTvWindow}[1]{\stateROMSize_{\thermodynamicSymbol}^{#1}}
\newcommand{\sizeROMforceOneSampleWindow}[1]{\stateROMSize_{\samplingOp{\kinematicSymbol}}^{#1}}
\newcommand{\sizeROMforceTvSampleWindow}[1]{\stateROMSize_{\samplingOp{\thermodynamicSymbol}}^{#1}}
\newcommand{\energyOSWindow}[1]{\energy_{\offsetSymbol}^{#1}}
\newcommand{\velocityOSWindow}[1]{\velocity_{\offsetSymbol}^{#1}}
\newcommand{\positionOSWindow}[1]{\position_{\offsetSymbol}^{#1}}
\newcommand{\velocityBasisWindow}[1]{\stateROMBasisMat_{\velocitySymbol}^{#1}}
\newcommand{\energyBasisWindow}[1]{\stateROMBasisMat_{\energySymbol}^{#1}}
\newcommand{\positionBasisWindow}[1]{\stateROMBasisMat_{\positionSymbol}^{#1}}
\newcommand{\forceOneBasisWindow}[1]{\stateROMBasisMat_{\kinematicSymbol}^{#1}}
\newcommand{\forceTvBasisWindow}[1]{\stateROMBasisMat_{\thermodynamicSymbol}^{#1}}
\newcommand{\ROMenergyWindow}[1]{\ROMenergy^{#1}}
\newcommand{\ROMvelocityWindow}[1]{\ROMvelocity^{#1}}
\newcommand{\ROMpositionWindow}[1]{\ROMposition^{#1}}
\newcommand{\ROMforceOneWindow}[1]{\ROMforceOne^{#1}}
\newcommand{\ROMforceTvWindow}[1]{\ROMforceTv^{#1}}
\newcommand{\ROMKinematicMassMatWindow}[1]{\ROMKinematicMassMat^{#1}}
\newcommand{\ROMThermodynamicMassMatWindow}[1]{\ROMThermodynamicMassMat^{#1}}
\newcommand{\forceOneObliqProjMatWindow}[1]{\forceOneObliqProjMat^{#1}}
\newcommand{\forceTvObliqProjMatWindow}[1]{\forceOneObliqProjMat^{#1}}
\newcommand{\forceTvSamplingMatWindow}[1]{\forceTvSamplingMat^{#1}}
\newcommand{\forceOneSamplingMatWindow}[1]{\forceOneSamplingMat^{#1}}
\newcommand{\ntimestepWindow}{N_{\text{sample}}}
\newcommand{\timeIndexWindow}[1]{\timeIndex_{#1}}
\newcommand{\paramMetric}{d}
\newcommand{\interpolationWeightSymbol}{q}
\newcommand{\interpolationPowerSymbol}{r}

\newcommand{\continuousVelocityError}{\continuousError_\velocitySymbol}
\newcommand{\continuousEnergyError}{\continuousError_\energySymbol}
\newcommand{\continuousPositionError}{\continuousError_\positionSymbol}
\newcommand{\discreteVelocityErrort}[1]{\discreteError_{\velocitySymbol,#1}}
\newcommand{\discreteEnergyErrort}[1]{\discreteError_{\energySymbol,#1}}
\newcommand{\discretePositionErrort}[1]{\discreteError_{\positionSymbol,#1}}
\newcommand{\relerrorVelocityt}[1]{\relError_{\velocitySymbol,#1}}
\newcommand{\relerrorEnergyt}[1]{\relError_{\energySymbol,#1}}
\newcommand{\relerrorPositiont}[1]{\relError_{\positionSymbol,#1}}
\newcommand{\velocityDifference}{\difference_\velocitySymbol}
\newcommand{\energyDifference}{\difference_\energySymbol}
\newcommand{\positionDifference}{\difference_\positionSymbol}
\newcommand{\velocityInitialError}{\initialError_\velocitySymbol}
\newcommand{\energyInitialError}{\initialError_\energySymbol}
\newcommand{\positionInitialError}{\initialError_\positionSymbol}
\newcommand{\stateResidual}[1]{\residual_\stateSymbol^{(#1)}}
\newcommand{\velocityResidual}[1]{\residual_\velocitySymbol^{(#1)}}
\newcommand{\energyResidual}[1]{\residual_\energySymbol^{(#1)}}
\newcommand{\positionResidual}[1]{\residual_\positionSymbol^{(#1)}}
\newcommand{\forceOneResidual}[1]{\residual_{\kinematicSymbol}^{(#1)}}
\newcommand{\forceTvResidual}[1]{\residual_{\thermodynamicSymbol}^{(#1)}}
\newcommand{\velocityInducedNorm}[1]{\norm{#1}_{\kinematicMassMat}}
\newcommand{\energyInducedNorm}[1]{\norm{#1}_{\thermodynamicMassMat}}
\newcommand{\oneNorm}[1]{\norm{#1}_1}
\newcommand{\euclideanNorm}[1]{\norm{#1}_2}
\newcommand{\snapshots}{\boldsymbol \snapshotSymb}
\newcommand{\rkStage}{r}
\newcommand{\rkErrorCoeff}{\mathbf{\rkErrorCoeffSymbol}}
\newcommand{\rkVelocityErrorCoeff}[1]{\rkErrorCoeff^{(#1)}_{\velocitySymbol,\timeIndex}}
\newcommand{\rkEnergyErrorCoeff}[1]{\rkErrorCoeff^{(#1)}_{\energySymbol,\timeIndex}}
\newcommand{\rkPositionErrorCoeff}[1]{\rkErrorCoeff^{(#1)}_{\positionSymbol,\timeIndex}}
\newcommand{\rkDummyVelocityErrorCoeff}[1]{\rkErrorCoeff^{(#1)}_{\velocitySymbol,\dummyTimeIndex}}
\newcommand{\rkDummyEnergyErrorCoeff}[1]{\rkErrorCoeff^{(#1)}_{\energySymbol,\dummyTimeIndex}}
\newcommand{\rkDummyPositionErrorCoeff}[1]{\rkErrorCoeff^{(#1)}_{\positionSymbol,\dummyTimeIndex}}
\newcommand{\jacobianState}{\jacobian_{\stateSymbol}}

\newcommand{\leftSingularMat}{\boldsymbol{U}}
\newcommand{\leftSingularVec}{\boldsymbol{u}}
\newcommand{\leftSingularVeck}[1]{\leftSingularVec_{#1}}
\newcommand{\rightSingularMat}{\boldsymbol{V}}
\newcommand{\singularValueMat}{\boldsymbol{\Sigma}}
\newcommand{\singularValue}{\sigma}
\newcommand{\singularValueThreshold}{\epsilon_{\singularValue}}
\newcommand{\nparam}{\sizeROMForceSample_\paramSymbol}
\newcommand{\solArg}[1]{\sol_{#1}}

\newcommand{\conditionnumber}{\kappa}
\newcommand{\unitvecSymbol}{\mathfrak{e}}
\newcommand{\unitveck}[1]{\unitvecSymbol_{#1}}
\newcommand{\tuningparam}{\eta}

\newcommand{\multiIndexSymbol}{I}
\newcommand{\multiIndex}{\boldsymbol{\multiIndexSymbol}}

\newcommand{\argmin}{\operatornamewithlimits{arg\,min}}
\newcommand{\argmax}{\operatornamewithlimits{arg\,max}}

\theoremstyle{thmstyleone}
\newtheorem{definition}{Definition}[section]

\section{Introduction}\label{sec:intro}
Partial differential equations (PDEs) are employed in numerous
scientific fields to model the behavior of spatiotemporal systems.
Despite their vast utility, only a few PDEs of practical interest have known
analytical solutions under general conditions.
As such, in most applications approximate PDE solutions are sought
by various discretization methods, such as the finite differences method
or the finite element method (FEM). 
In this study we consider the spatial discretization of time-dependent PDEs
with algebraic constraints, resulting in systems of semi-explicit
differential-algebraic equations (DAEs)
\begin{equation}\label{eq:combinedFOM}
\begin{split}
    \massMat\frac{d\state}{d\timeSymbol}+
        \stateLinearMat\state+\bilinearMat\multiplier 
        &=\stateNonlinearVec(\state,\multiplier,\timeSymbol,\param)\\
    \bilinearMat^\top\state+\multiplierLinearMat\multiplier
        &=\multiplierNonlinearVec(\state,\multiplier,\timeSymbol,\param)\\
    \state(0) &=\state_0
\end{split}
\end{equation}
where
$\timeSymbol\in\left[0,\finalTime\right]$ denotes time
{and $\param \in \paramDomain$ denotes physical parameters driving the system}. The unknowns in the system include 
the differential variable 
$\state(\param):\left[0,\finalTime\right]\to\mathbb{R}^{\stateFOMSize}$,
with a prescribed initial condition
$\state_0(\param) \in \mathbb{R}^{\stateFOMSize}$, and
the algebraic variable
$\multiplier(\param) :\left[0,\finalTime\right]\to\mathbb{R}^{\multiplierFOMSize}$. 
Moreover, $\massMat\in\mathbb{R}^{\stateFOMSize\times\stateFOMSize}$
is the system mass matrix,
$\stateLinearMat\in\mathbb{R}^{\stateFOMSize\times\stateFOMSize}$,
$\multiplierLinearMat\in\mathbb{R}^{\multiplierFOMSize\times
\multiplierFOMSize}$
are linear operator matrices, and
$\bilinearMat\in\mathbb{R}^{\stateFOMSize\times\multiplierFOMSize}$
is a constraint matrix.
{These matrices are assumed to be independent of physical parameters $\param$ throughout the paper.}
Finally,
$\stateNonlinearVec:\mathbb{R}^{\stateFOMSize}\times
\mathbb{R}^{\multiplierFOMSize}\times[0,\finalTime] \times \paramDomain \to
\mathbb{R}^{\stateFOMSize}$
and
$\multiplierNonlinearVec:\mathbb{R}^{\stateFOMSize}\times
\mathbb{R}^{\multiplierFOMSize}\times[0,\finalTime] \times \paramDomain\to
\mathbb{R}^{\multiplierFOMSize}$
are nonlinear functions. 

Time integration schemes such as multistep and Runge-Kutta methods
can be used to solve the ODE part of the semi-explicit DAE system in
\eqref{eq:combinedFOM},
with $\multiplier$ solved algebraically. 
However, for many important physical systems the associated PDEs are
nonlinear and require a fine mesh to resolve the solution,
rendering the numerical modeling of such systems computationally expensive. 
This expense is compounded in multi-query applications such as
PDE-constrained optimization
\cite{tarantola2005inverse,biegler2003large,biegler2007real,
gunzburger2002perspectives}
where a large number of repeated PDE solves are necessary.
For instance, design optimization of inertial confinement fusion
experiments requires the determination of more than a dozen independent
parameters.
Depending on the fidelity of the physics model, each forward simulation
takes a few minutes to hundreds of CPU hours to complete
\cite{anirudh20232022}.
When employing a simple gradient based optimization algorithm,
finite difference approximation of the objective function gradient
requires as many radiation hydrodynamics simulations as the dimension
of the parameter search space, resulting in prohibitively high
computational cost for every parameter update.

For these reasons, it is of great interest to develop methodologies that
accelerate the solving of large-scale nonlinear PDEs.
One such methodology is reduced order modeling, where a
full order model (FOM) is approximated with an accurate but
computationally inexpensive reduced order model (ROM).
ROMs can be divided into two main classes:
data based and physics based.
Purely data based methods such as dynamic mode decomposition (DMD)
\cite{schmid2010dynamic,brunton2016discovering}
or neural network approximation
\cite{lagaris_artificial_1998,van_der_merwe_fast_2007}
have the benefit of not requiring prior knowledge of the PDE problem
specification.
However, they typically do not make accurate predictions in
extrapolation and are not robust to noisy input data
\cite{karniadakis2021physics}.
On the other hand, physics based ROM approaches incorporate prior
knowledge into the ROM formulation and typically fare better
in terms of generalization than purely data based approaches
\cite{karniadakis2021physics}.
Physics based ROM approaches include physics informed neural
networks \cite{raissi2019physics},
neural operators \cite{kovachki2023neural}
and projection based reduced order models (pROMs),
which are the class of ROMs studied in this paper.
At a high level, a pROM projects the PDE onto a low dimensional subspace
which approximates the PDE solution space.
The projected PDE is then solved in subspace coordinates,
leading to reduced computational cost.
For a more comprehensive description of pROMs we refer the reader
to the survey papers \cite{gugercin2004survey,benner2015survey}. 

Projection based ROMs have been shown to efficiently accelerate the
solving of physics problems such as
the Burgers equation
\cite{Veroy2003,choi2019space,choi2020sns,carlberg2018conservative},
the Navier-Stokes equations
\cite{xiao2014non,burkardt2006pod},
nonlinear diffusion problems
\cite{hoang2020domain,fritzen2018algorithmic},
convection-diffusion problems 
\cite{mojgani2017lagrangian},
the shallow water equations
\cite{zhao2014pod, cstefuanescu2013pod},
the Euler equations
\cite{Copeland2022,Cheung2023,Vales2025ceqp},
wave problems \cite{glas2020reduced,cheung2021explicit},
Boltzmann transport problems \cite{choi2021space}
and the Schr\"odinger equation \cite{cheng2016reduced}.
In addition, pROMs can enable substantial speed-up in design optimization
problems such as lattice structure design
\cite{mcbane2021component},
flutter avoidance wing shape optimization
\cite{choi2020gradient},
topology optimization of wind turbine blades
\cite{choi2019accelerating},
reservoir simulations
\cite{jiang2019implementation,yang2016fast,
wang2020generalized}
and rocket nozzle shape design \cite{amsallem2015design}.

In the standard pROM formulation, the effects of the nonlinear terms
$\stateNonlinearVec$ and $\multiplierNonlinearVec$
in the ROM subspaces are evaluated on the FOM mesh. 
This causes the pROM computational cost to scale with the FOM size,
significantly reducing their efficiency. 
Hyper-reduction methods address this issue by approximating the
nonlinear terms through evaluation at a reduced set of sample points,
thereby achieving substantial computational speedup while maintaining
a low approximation error.
As indicated in \cite{grimberg2021mesh},
hyper-reduction methods can be broadly classified into two categories.
The first category consists of interpolation or
``approximate-then-project'' methods,
which interpolate the nonlinear terms at a subset of selected nodes
before projecting them onto the reduced-order basis (ROB). 
This class of methods originates in the gappy proper orthogonal
decomposition (POD) approach \cite{everson1995},
which reconstructs high dimensional data using
empirical basis functions. 
These methods rely on interpolation based approximations,
where different sampling methods
\cite{Barrault2004,chaturantabut2010nonlinear,carlberg2013gnat,
drmac2016new,drmac2018discrete,lauzon2024s}
are introduced to select interpolation points for optimizing the
tradeoff between accuracy and speedup.
The second category comprises quadrature or ``project-then-approximate''
methods, which directly approximate the reduced nonlinear terms by
constructing sparse quadrature rules.
These methods can be viewed as data based quadrature techniques,
where both the sample points and associated weights are optimized
using training data.
Examples include the seminal work by An et al.\
\cite{An2008},
the energy-conserving sampling and weighting method
\cite{farhat2015structure},
the empirical cubature method \cite{hernandez2017dimensional}
and the empirical quadrature procedure (EQP)
\cite{Patera2017,Yano2019}.

Given that hyper-reduction methods are being employed in applications
across various domains, a comprehensive comparative analysis of
their accuracy and efficiency is of significant interest.
Existing works have provided valuable insights into specific
hyper-reduction techniques, with recent efforts focusing on a single
application at a time, such as nonlinear heat conduction
\cite{fritzen2018algorithmic,bhattacharyya2025hyper},
magneto-mechanics
\cite{brands2019reduced,delagnes2024comparison}
and the Navier-Stokes equations
\cite{romor2025explicable}.

\begin{figure}[!htpb]
    \centering
    \includegraphics[width=0.99\linewidth]{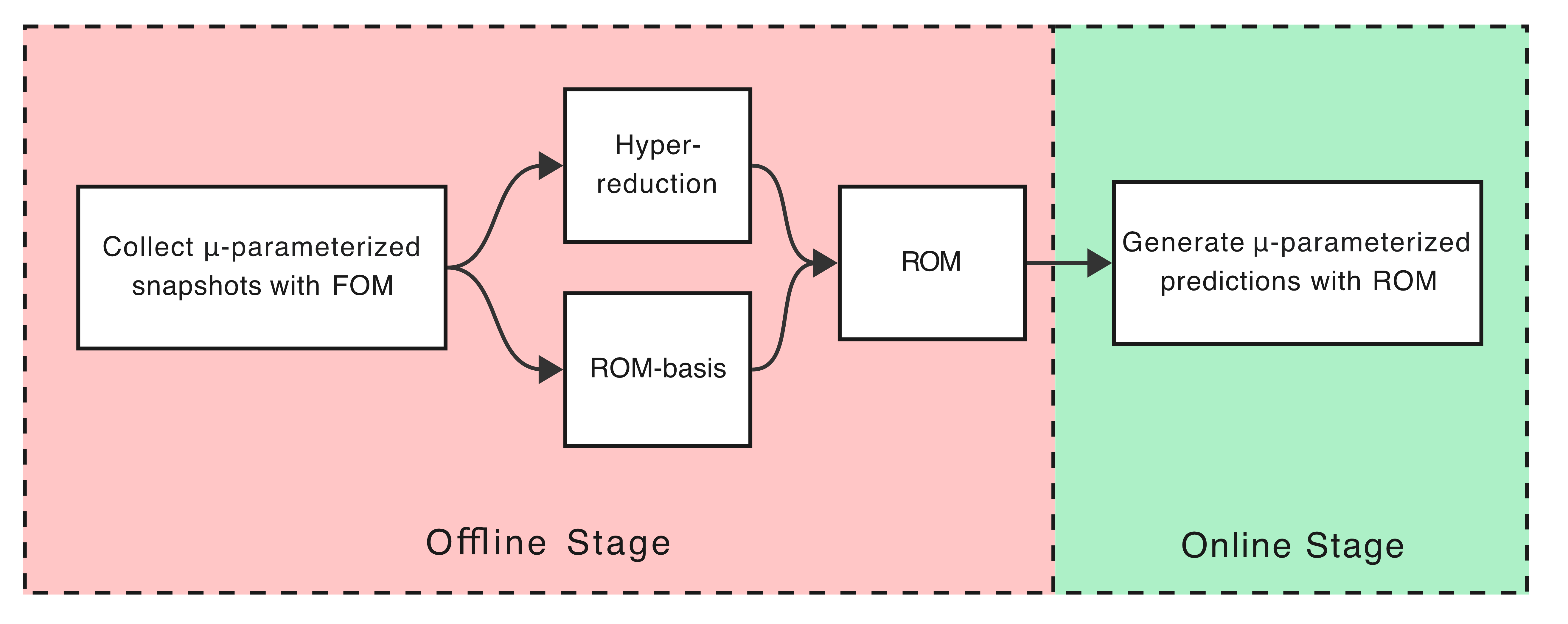}
    \caption{ROM workflow stages}
    \label{fig:romstages}
\end{figure}

In the present study we undertake a detailed empirical comparison
of several hyper-reduction methods by applying them to a diverse
selection of problem cases.
{
More specifically, from the class of interpolation methods,
we consider the Discrete Empirical Interpolation Method (DEIM)
\cite{chaturantabut2010nonlinear,carlberg2013gnat},
its subsequent development Q-DEIM \cite{drmac2016new},
and S-OPT \cite{lauzon2024s}.
These methods were selected to capture distinct interpolation-based sampling strategies: 
DEIM serves as a widely adopted baseline,
Q-DEIM improves the numerical conditioning of the interpolation operator through rank-revealing QR factorization, 
and S-OPT employs an optimization-based sampling procedure inspired by sensor placement and uncertainty quantification. 
For the class of quadrature methods, we consider EQP \cite{Yano2019}, which constructs sparse quadrature rules through non-negative least-squares optimization and has emerged as a widely used approach for hyper-reduction of nonlinear finite element simulations.
}
{Throughout this study, a two-stage workflow is employed to formulate and evaluate hyper-reduced projection-based ROMs. The workflow is split into an offline and an online stage, illustrated in Figure \ref{fig:romstages}. 
First, the offline stage collects snapshots of FOM problem simulations, parameterized by $\param$. From the snapshots, the linear ROM basis is formed by POD. The FOM snapshots are additionally utilized in initializing the chosen hyper-reduction method. In the second online stage, the ROM is used to make accelerated predictions, again parameterized by $\param$.}
We apply the hyper-reduction methods in this study to three distinct and challenging nonlinear benchmark problems discretized by finite element methods: nonlinear diffusion, nonlinear elasticity, and Lagrangian hydrodynamics.
The main features of our work are summarized below.
\begin{enumerate}
    \item \textit{Comparison}. We present a broad comparison of hyper-reduction
        methods, including both interpolation and quadrature methods.
    \item \textit{Application}. We evaluate each considered hyper-reduction
        method on three application domains with distinct challenges,
        revealing tradeoffs between accuracy and computational speedup.
    \item \textit{Open source implementation}. We employ an open source
        numerical implementation of all hyper-reduction methods and
        problem cases considered in this work, using the
        \texttt{libROM} \cite{librom}, \texttt{Laghos} \cite{Dobrev2012}
        and \texttt{MFEM} \cite{mfem-web} numerical libraries
        and providing the commands that can be used for their
        reproduction.
\end{enumerate}

The remainder of the paper is structured as follows.
Section \ref{sec:fe} introduces a unified variational framework
to describe the example problems and discusses their finite element
discretization.
Section \ref{sec:prom} serves as an introduction to the
pROM approach used for the linear part of the ROM formulation.
Section \ref{sec:hyp} presents the different hyper-reduction
methods compared in this work, which are used to approximate
the evaluation of nonlinear terms.
Section \ref{sec:ex} describes the example problems and presents our
numerical results,
followed by a discussion in Section \ref{sec:disc}.
Finally, Appendix \ref{sec:appendix} provides details on how to
reproduce our numerical results.

\section{Finite element discretization}\label{sec:fe}
Considering \eqref{eq:combinedFOM} in a variational setting, 
we seek $\stateSymbol(\param) \in L^2(0,\finalTime; \stateSpaceSymbol)$
and $\multiplierSymbol(\param) \in L^2(0,\finalTime; \multiplierSpaceSymbol)$
such that $d\stateSymbol(\param)/d\timeSymbol\in
L^2(0,\finalTime; \stateSpaceSymbol')$,
where $\HilbertSpaceSymbol$ and $\multiplierSpaceSymbol$
are Hilbert spaces and
$\stateSpaceSymbol\subset\HilbertSpaceSymbol\subset\stateSpaceSymbol'$
forms a Gelfand triple (rigged Hilbert space).
In addition, we ask that the following equations be satisfied for any
$\stateFeBasisSymbol\in\stateSpaceSymbol$
and $\multiplierFeBasisSymbol\in\multiplierSpaceSymbol$
\begin{equation}\label{eq:PDE}
\begin{split}
    \frac{d}{d \timeSymbol}(\stateSymbol,
        \stateFeBasisSymbol)_{\HilbertSpaceSymbol}
        +a(\stateSymbol, \stateFeBasisSymbol)
        +b(\stateFeBasisSymbol, \multiplierSymbol)&
        =\langle f(\stateSymbol, \multiplierSymbol, \timeSymbol, \param),
        \stateFeBasisSymbol\rangle_{\stateSpaceSymbol'\times
        \stateSpaceSymbol}\\
        b(\stateSymbol, \multiplierFeBasisSymbol)
        +c(\multiplierSymbol, \multiplierFeBasisSymbol)&
        =\langle g(\stateSymbol, \multiplierSymbol, \timeSymbol, \param),
        \multiplierFeBasisSymbol\rangle_{\multiplierSpaceSymbol'\times
        \multiplierSpaceSymbol}
\end{split}
\end{equation}
where
$\langle\cdot,\cdot\rangle_{\stateSpaceSymbol'\times\stateSpaceSymbol}$
denotes the duality pairing between
$\stateSpaceSymbol'$ and $\stateSpaceSymbol$
and $(\cdot,\cdot)_\HilbertSpaceSymbol$
the scalar product on $\HilbertSpaceSymbol$.
Additionally,
$a:\stateSpaceSymbol\times\stateSpaceSymbol\to\mathbb{R}$,
$b:\stateSpaceSymbol\times\multiplierSpaceSymbol\to\mathbb{R}$
and $c:\multiplierSpaceSymbol\times\multiplierSpaceSymbol\to\mathbb{R}$
denote bilinear forms, while
$\stateNonlinearForceSymbol:\stateSpaceSymbol\times\multiplierSpaceSymbol
\times[0,\finalTime]\times\paramDomain\to\stateSpaceSymbol'$
and $\multiplierNonlinearForceSymbol:\stateSpaceSymbol\times
\multiplierSpaceSymbol\times[0,\finalTime]\times\paramDomain\to\multiplierSpaceSymbol'$
are general nonlinear functions.

The spatial discretization of \eqref{eq:PDE} is carried out using the finite
element method, the result of which is the system of differential-algebraic
equations presented in \eqref{eq:combinedFOM}.
To that end, let
$\feBasisSet_\stateSpaceSymbol=\{\stateFeBasisSymbol_{\basisIndex}\}
_{\basisIndex=1}^{\stateFOMSize}$
be a basis for a
$\stateSpaceSymbol$-conforming finite element space
$\stateFeSpace$,
with $\stateFeVec:\stateFeSpace\to\mathbb{R}^{\stateFOMSize}$
the coordinate representation of $\stateFeSpace$
with respect to $\feBasisSet_\stateSpaceSymbol$.
We denote the finite element approximation
$\stateFeSymbol(\param) \in \stateFeSpace$ of $\stateSymbol(\param)$
by
\begin{equation*}
    \stateFeSymbol(\timeSymbol; \param)=\sum_{\basisIndex=1}^{\stateFOMSize}
        \state_\basisIndex(\timeSymbol; \param)\stateFeBasisSymbol_{\basisIndex}.
\end{equation*}
Similarly, for the algebraic variables let
$\feBasisSet_\multiplierSpaceSymbol=
\{\multiplierFeBasisSymbol_{\basisIndex}\}
_{\basisIndex=1}^{\multiplierFOMSize}$
be a basis for a
$\multiplierSpaceSymbol$-conforming finite element space
$\multiplierFeSpace$, with
$\multiplierFeVec:\multiplierFeSpace\to\mathbb{R}^{\multiplierFOMSize}$
the coordinate representation of $\multiplierFeSpace$
with respect to $\feBasisSet_\multiplierSpaceSymbol$.
The finite element approximation
$\multiplierFeSymbol(\param)\in\multiplierFeSpace$ of
$\multiplierSymbol(\param)$ is denoted by
\begin{equation*}
    \multiplierFeSymbol(\timeSymbol; \param)=\sum_{\basisIndex=1}^{\multiplierFOMSize}
        \multiplier_\basisIndex(\timeSymbol; \param)
        \multiplierFeBasisSymbol_{\basisIndex}.
\end{equation*}
Then the differential variable vector
$\state=\stateFeVec(\stateFeSymbol)$
and the algebraic variable vector
$\multiplier=\multiplierFeVec(\multiplierFeSymbol)$
satisfy \eqref{eq:combinedFOM},
wherein the matrix representations are defined entrywise by
\begin{equation}\label{eq:algebraic-FOM1}
\begin{aligned}
    \massMat_{\basisIndex,\dummyBasisIndex}&=
        (\stateFeBasisSymbol_{\dummyBasisIndex},
        \stateFeBasisSymbol_{\basisIndex})_\HilbertSpaceSymbol\qquad
    &\stateLinearMat_{\basisIndex,\dummyBasisIndex}&=
        a( \stateFeBasisSymbol_{\dummyBasisIndex},
        \stateFeBasisSymbol_{\basisIndex})\\
    \bilinearMat_{\basisIndex,\dummyBasisIndex}&=
        b(\stateFeBasisSymbol_{\dummyBasisIndex},
        \multiplierFeBasisSymbol_{\basisIndex})\qquad
    &\multiplierLinearMat_{\basisIndex,\dummyBasisIndex}&=
        c(\multiplierFeBasisSymbol_{\dummyBasisIndex},
        \multiplierFeBasisSymbol_{\basisIndex})\\
\end{aligned}
\end{equation}
{
and can often be computed exactly, and 
the vector representations are computed as a semi-discretization of the nonlinear functions acting on the test functions, 
\begin{equation}\label{eq:algebraic-FOM2}
\begin{aligned}
    \stateNonlinearVec_{\basisIndex}(\state,\multiplier,\timeSymbol,\param)&\approx
        \left\langle\stateNonlinearForceSymbol\left(\stateFeVec^{-1}(\state),
        \multiplierFeVec^{-1}(\multiplier),\timeSymbol,\param\right),
        \stateFeBasisSymbol_{\basisIndex} \right\rangle_{\stateSpaceSymbol'
        \times \stateSpaceSymbol}\\
    \multiplierNonlinearVec_{\basisIndex}(\state,\multiplier,\timeSymbol,\param)&\approx
        \left\langle\multiplierNonlinearForceSymbol\left(\stateFeVec^{-1}
        (\state),\multiplierFeVec^{-1}(\multiplier),\timeSymbol,\param\right),
        \multiplierFeBasisSymbol_{\basisIndex}\right\rangle
        _{\multiplierSpaceSymbol' \times \multiplierSpaceSymbol}.
\end{aligned}
\end{equation}
As in many applications, we assume that the nonlinear residual
functionals admit local integral representations over subdomains of the spatial domain $\Omega$. More precisely, we assume the existence of nonlinear densities
$\eta_{\stateNonlinearForceSymbol}: \stateFeSpace\times\multiplierFeSpace\times[0,\finalTime]\times\paramDomain\times\Omega \to [\stateFeSpace]'$ and 
$\eta_{\multiplierNonlinearForceSymbol}: \stateFeSpace\times\multiplierFeSpace\times[0,\finalTime]\times\paramDomain\times\Omega \to [\multiplierFeSpace]'$
such that the nonlinear vectors can be assembled from their action on the
finite element test functions. Using numerical quadrature, these actions are
approximated by
\begin{equation}\label{eq:fe-quadrature}
\begin{split}
\stateNonlinearVec_{\basisIndex}(\state,\multiplier,\timeSymbol,\param)
&=
\sum_{k=1}^{K_{\stateNonlinearForceSymbol}}
\rho_k^{(f)}
\left\langle\eta_{\stateNonlinearForceSymbol}
\left(
\stateFeVec^{-1}(\state),
\multiplierFeVec^{-1}(\multiplier),
\timeSymbol,
\param,
x_k^{(f)}
\right), 
\stateFeBasisSymbol_{\basisIndex}\right\rangle_{[\stateFeSpace]' \times \stateFeSpace}
\\
\multiplierNonlinearVec_{\basisIndex}(\state,\multiplier,\timeSymbol,\param)
&=
\sum_{k=1}^{K_{\multiplierNonlinearForceSymbol}}
\rho_k^{(g)}
\left\langle\eta_{\multiplierNonlinearForceSymbol}
\left(
\stateFeVec^{-1}(\state),
\multiplierFeVec^{-1}(\multiplier),
\timeSymbol,
\param,
x_k^{(g)}
\right), 
\multiplierFeBasisSymbol_{\basisIndex}
\right\rangle_{[\multiplierFeSpace]' \times \multiplierFeSpace}.
\end{split}
\end{equation}
where $\rho_k^{(f)} > 0$ and $\rho_k^{(g)} > 0$ denote quadrature weights, and
$x_k^{(f)}$ and $x_k^{(g)}$ the associated quadrature points.
}

\section{Projection based reduction}\label{sec:prom}
As mentioned in Section \ref{sec:intro},
solving \eqref{eq:combinedFOM}
can be computationally expensive when the FOM dimension
$\stateFOMSize+\multiplierFOMSize$ is very large.
In such cases, the solution procedure can be accelerated by projecting
the PDE to the affine subspaces
$\stateOSSymbol+\stateSubspaceSymbol$ and
$\multiplierOSSymbol+\multiplierSubspaceSymbol$
for the differential and algebraic variables respectively,
where
\begin{equation}
\begin{aligned}
    \dim(\stateSubspaceSymbol)=\stateROMSize\ll\stateFOMSize
        =\dim(\stateFeSpace)\\
    \dim(\multiplierSubspaceSymbol)=\multiplierROMSize\ll\multiplierFOMSize
        =\dim(\multiplierFeSpace)
    \end{aligned}
\end{equation}
with
$\stateOSSymbol\in\stateFeSpace$ and
$\multiplierOSSymbol\in\multiplierFeSpace$
denoting the prescribed offsets.
The subspaces
$\stateSubspaceSymbol$ and
$\multiplierSubspaceSymbol$
are spanned by the sets of basis vectors 
$\{\stateROMBasisSymbol_{\basisIndex}\}_{\basisIndex=1}^{\stateROMSize}
\subset \stateFeSpace$ and
$\{\multiplierROMBasisSymbol_{\basisIndex}\}_{\basisIndex=1}
^{\multiplierROMSize} \subset \multiplierFeSpace$ respectively
\begin{equation}\label{eq:subspaces}
\begin{aligned}
    \stateSubspaceSymbol &=
        \Span{\stateROMBasisSymbol_{\basisIndex}}_{\basisIndex=1}
        ^{\stateROMSize}\subseteq\stateFeSpace\\
    \multiplierSubspaceSymbol &=
        \Span{\multiplierROMBasisSymbol_{\basisIndex}}_{\basisIndex=1}
        ^{\multiplierROMSize}\subseteq\multiplierFeSpace.
\end{aligned}
\end{equation}
The differential and algebraic state vector are then approximated
using the subspace basis vectors
\begin{equation}\label{eq:fe-solrepresentation}
\begin{aligned}
    \stateSymbol(\timeSymbol; \param)\approx\stateSymbolApprox(\timeSymbol; \param)
        &=\stateOSSymbol+\sum_{\basisIndex=1}^{\stateROMSize}
        \ROMstate_\basisIndex(\timeSymbol; \param)
        \stateROMBasisSymbol_{\basisIndex}\\
    \multiplierSymbol(\timeSymbol; \param)\approx\multiplierSymbolApprox(\timeSymbol; \param)
        &=\multiplierOSSymbol+\sum_{\basisIndex=1}^{\multiplierROMSize}
        \ROMmultiplier_\basisIndex(\timeSymbol; \param)
        \multiplierROMBasisSymbol_{\basisIndex},
\end{aligned}
\end{equation}
{
and it suffices to solve for the vectors of generalized coordinates of the differential and algebraic variables 
}
\begin{equation*}
    \ROMstate(\param)=(\hat{\stateSymbol}_1,\ldots,
        \hat{\stateSymbol}_{\stateROMSize})^\top
        :[0,\finalTime]\to\mathbb{R}^{r_y}\qquad
    \ROMmultiplier(\param)=(\hat{\multiplierSymbol}_1,\ldots,
        \hat{\multiplierSymbol}_{\multiplierROMSize})^\top
        :[0, \finalTime]\to\mathbb{R}^{\multiplierROMSize}.
\end{equation*}
{
To achieve the desired computational acceleration, ROMs typically leverage an offline-online splitting paradigm. 
In the offline phase (Section~\ref{subsec:offline}), which is performed once, computationally expensive tasks such as high-fidelity snapshots generation and basis construction are executed. In contrast, the online phase (Section~\ref{subsec:online}) utilizes these precomputed quantities to assemble and solve the low-dimensional system for any new parameter configuration or time step}{, at a significantly lower computational cost than the offline phase}. {The specific procedures and computational complexities of these two phases are detailed in the subsequent subsections.
}

{
\subsection{Offline phase}\label{subsec:offline}
The offline phase is a one-time procedure aimed at constructing the low-dimensional subspaces $\stateSubspaceSymbol$ and $\multiplierSubspaceSymbol$. To computationally manipulate the finite element functions introduced in \eqref{eq:subspaces}, we first map them to their corresponding algebraic coordinate vectors in the Euclidean space via the isomorphisms $\stateFeVec: \stateFeSpace \to \mathbb{R}^{\stateFOMSize}$ and $\multiplierFeVec: \multiplierFeSpace \to \mathbb{R}^{\multiplierFOMSize}$. Specifically, the prescribed offsets are mapped as
\begin{equation*}
    \stateOS=\stateFeVec(\stateOSSymbol)
        \in\mathbb{R}^{\stateFOMSize}\qquad
    \multiplierOS=\multiplierFeVec(\multiplierOSSymbol)
        \in\mathbb{R}^{\multiplierFOMSize}.
\end{equation*}
To find an optimal choice for 
$\{\stateROMBasisSymbol_{\basisIndex}\}_{\basisIndex=1}^{\stateROMSize}$ and
$\{\multiplierROMBasisSymbol_{\basisIndex}\}_{\basisIndex=1}
^{\multiplierROMSize}$, 
one can follow the Proper Orthogonal Decomposition (POD) procedure. 
High-fidelity FOM simulations are conducted over a selected training set of parameters. 
For notational simplicity, each different variable snapshot is identified by a single index $1 \leq s \leq N_s^{(\stateSymbol)}$, and represented by a triplet $(\stateSymbol_s, \timeSymbol_{\ell_s^{(\stateSymbol)}},\param_{m_s^{(\stateSymbol)}}) \in \stateFeSpace \times [0,T] \times \paramDomain$, where $\timeSymbol_{\ell_s^{(\stateSymbol)}}$ is the corresponding time instance and $\param_{m_s^{(\stateSymbol)}}$ is the corresponding parameter where $s$-th state snapshot is simulated. 
Similarly, each algebraic variable snapshot is represented by
a triplet $(\multiplierSymbol_s, \timeSymbol_{\ell_s^{(\multiplierSymbol)}},\param_{m_s^{(\multiplierSymbol)}}) \in \multiplierFeSpace \times [0,T] \times \paramDomain$, for $1 \leq s \leq N_s^{(\multiplierSymbol)}$
From these snapshots, we collect the vector representation of the differential and algebraic variables and decentralize:
\begin{align*}
    \mathbf{S}_{\stateSymbol} &= [ \stateFeVec(\stateSymbol_1) - \stateOS, \, \ldots, \, \stateFeVec(\stateSymbol_{N_s^{(\stateSymbol)}}) - \stateOS ] \in \mathbb{R}^{\stateFOMSize \times N_s^{(\stateSymbol)}}, \\
    \mathbf{S}_{\multiplierSymbol} &= [ \multiplierFeVec(\multiplierSymbol_1) - \multiplierOS, \, \ldots, \, \multiplierFeVec(\multiplierSymbol_{N_s^{(\multiplierSymbol)}}) - \multiplierOS ] \in \mathbb{R}^{\multiplierFOMSize \times N_s^{(\multiplierSymbol)}}.
\end{align*}
The reduced bases can be subsequently extracted by performing a thin singular value decomposition (SVD) on these snapshot matrices:
\begin{equation}\label{eq:snapshot_svd}
    \mathbf{S}_y = \mathbf{U}_y \boldsymbol{\Sigma}_y \mathbf{V}_y^\top \qquad \mathbf{S}_z = \mathbf{U}_z \boldsymbol{\Sigma}_z \mathbf{V}_z^\top,
\end{equation}
with the singular values arranged in descending order. 
The choice of the basis dimensions $\stateROMSize$ and $\multiplierROMSize$ governs a critical trade-off: larger dimensions improve the accuracy of the reduced-order model by retaining more physical information, whereas smaller dimensions minimize computational complexity to maximize speed-up. To balance this, 
the dimensions $\stateROMSize$ and $\multiplierROMSize$ are typically determined via an energy criterion, selecting the minimal number of singular vectors required to capture a user-defined percentage $\zeta_y \in (0,1)$ and $\zeta_z \in (0,1)$ of the total snapshot energy, defined by the ratio of the cumulative sum of the singular values, i.e., 
\begin{equation}\label{eq:energy_criterion}
    \stateROMSize = \min \left\{ r : \frac{\sum_{i=1}^{r} \sigma_{y,i}^2}{\sum_{i=1}^{N_s^{(\stateSymbol)}} \sigma_{y,i}^2} \ge \zeta_y \right\}, \qquad
    \multiplierROMSize = \min \left\{ r : \frac{\sum_{i=1}^{r} \sigma_{z,i}^2}{\sum_{i=1}^{N_s^{(\multiplierSymbol)}} \sigma_{z,i}^2} \ge \zeta_z \right\}.
\end{equation}
where $(\sigma_{y,i})_{i=1}^{\stateROMSize}$ and 
$(\sigma_{z,i})_{i=1}^{\multiplierROMSize}$ 
are the diagonal entries of $\boldsymbol{\Sigma}_y$ and 
$\boldsymbol{\Sigma}_z$ respectively. 
The basis matrices $\stateROMBasisMat  \in \mathbb{R}^{\stateFOMSize \times \stateROMSize}$ and $\multiplierROMBasisMat \in \mathbb{R}^{\multiplierFOMSize \times \multiplierROMSize}$ are then constructed from the first $\stateROMSize$ and $\multiplierROMSize$ left singular vectors of $\mathbf{S}_y$ and $\mathbf{S}_z$ respectively, i.e., 
\begin{equation}\label{eq:basis_matrix}
    \stateROMBasisMat = \mathbf{U}_y \begin{bmatrix}\mathbf{e}_1, \mathbf{e}_2, \cdots \mathbf{e}_{\stateROMSize}\end{bmatrix} \qquad
    \multiplierROMBasisMat = \mathbf{U}_z \begin{bmatrix}\mathbf{e}_1, \mathbf{e}_2, \cdots \mathbf{e}_{\multiplierROMSize}\end{bmatrix}
\end{equation}
where each column represents an individual finite element basis function 
\begin{equation*}
    \stateROMBasisSymbol_{\basisIndex}=
        \stateFeVec^{-1}(\stateROMBasisVec_{\basisIndex})
        \in\stateFeSpace\qquad
    \multiplierROMBasisSymbol_{\basisIndex}=
    \multiplierFeVec^{-1}(\multiplierROMBasisVec_{\basisIndex})
        \in\multiplierFeSpace
\end{equation*}
for $\basisIndex = 1, \ldots, \stateROMSize$ and $\basisIndex = 1, \ldots, \multiplierROMSize$.
}

\subsection{Online phase}\label{subsec:online}
The reduced system of DAEs is derived by performing the Galerkin projection
of \eqref{eq:combinedFOM} onto the introduced subspaces.
With the basis matrices \eqref{eq:basis_matrix} in place, the approximate variables
\eqref{eq:fe-solrepresentation}
can be written in vector notation
\begin{equation}\label{eq:solrepresentation}
\begin{aligned}
  \state(\timeSymbol)\approx\stateApprox(\timeSymbol)
        &=\stateOS+\stateROMBasisMat\ROMstate(\timeSymbol)\\
  \multiplier(\timeSymbol)\approx\multiplierApprox(\timeSymbol)
        &=\multiplierOS+\multiplierROMBasisMat\ROMmultiplier(\timeSymbol).
\end{aligned}
\end{equation}
Invoking the approximation \eqref{eq:solrepresentation}
and applying Galerkin projection directly to
\eqref{eq:combinedFOM},
we obtain the reduced order model of dimension
$\stateROMSize+\multiplierROMSize$
\begin{equation}\label{eq:combinedROM}
\begin{aligned}
    \ROMmassMat\frac{d\ROMstate}{d\timeSymbol}+\ROMstateLinearMat
        \ROMstate+\ROMbilinearMat\ROMmultiplier=
        \overline{\stateNonlinearVec}(\stateOS+\stateROMBasisMat\ROMstate,
        \multiplierOS+\multiplierROMBasisMat\ROMmultiplier,\timeSymbol,\param)\\
    \ROMbilinearMat^\top\ROMstate+\ROMmultiplierLinearMat\ROMmultiplier=
        \overline{\multiplierNonlinearVec}(\stateOS+\stateROMBasisMat
        \ROMstate,\multiplierOS+\multiplierROMBasisMat\ROMmultiplier, 
        \timeSymbol,\param)
\end{aligned}
\end{equation}
with
\begin{equation}\label{eq:algebraic-ROM}
\begin{aligned}
    \ROMmassMat &=\stateROMBasisMat^\top\massMat\stateROMBasisMat
        \in\mathbb{R}^{\stateROMSize\times\stateROMSize}\qquad
    &\ROMstateLinearMat &=\stateROMBasisMat^\top\stateLinearMat
        \stateROMBasisMat\in\mathbb{R}^{\stateROMSize\times\stateROMSize}\\
    \ROMbilinearMat &=\stateROMBasisMat^\top\bilinearMat
        \multiplierROMBasisMat\in\mathbb{R}^{\stateROMSize\times
        \multiplierROMSize}\qquad
    &\ROMmultiplierLinearMat &=\multiplierROMBasisMat^\top
        \multiplierLinearMat\multiplierROMBasisMat\in
        \mathbb{R}^{\multiplierROMSize\times\multiplierROMSize}
\end{aligned}
\end{equation}
and
\begin{equation}\label{eq:algebraic-ROM-force}
    \overline{\stateNonlinearVec}=\stateROMBasisMat^\top\stateNonlinearVec
        \in\mathbb{R}^{\stateROMSize}\qquad
    \overline{\multiplierNonlinearVec}=\multiplierROMBasisMat^\top
        \multiplierNonlinearVec\in\mathbb{R}^{\multiplierROMSize}.
\end{equation}
Note that the nonlinear terms
$\overline{\stateNonlinearVec}:\mathbb{R}^{\stateFOMSize}\times
\mathbb{R}^{\multiplierFOMSize}\times[0,\finalTime]\times\paramDomain\to
\mathbb{R}^{\stateROMSize}$
and
$\overline{\multiplierNonlinearVec}:\mathbb{R}^{\stateFOMSize}\times
\mathbb{R}^{\multiplierFOMSize}\times[0,\finalTime]\times\paramDomain\to
\mathbb{R}^{\multiplierROMSize}$
are the projections of the image of the
vector valued functions
$\stateNonlinearVec$ and $\multiplierNonlinearVec$
onto the column spaces of
$\stateROMBasisMat$ and
$\multiplierROMBasisMat$ respectively.
The projected functions are defined entrywise as
{
\begin{equation}\label{eq:nonlinear-funcs}
\begin{split}
    \overline{\stateNonlinearVec}_{\basisIndex}(\stateApprox,
        \multiplierApprox,\timeSymbol,\param)=\sum_{k=1}^{K_{\stateNonlinearForceSymbol}}
    \rho_k^{(f)}
    \left\langle\eta_{\stateNonlinearForceSymbol}
    \left(
    \stateFeVec^{-1}(\stateApprox),
    \multiplierFeVec^{-1}(\multiplierApprox),
    \timeSymbol,
    \param,
    x_k^{(f)}
    \right), 
    \stateROMBasisSymbol_{\basisIndex}\right\rangle_{[\stateFeSpace]' \times \stateFeSpace}\\
    \overline{\multiplierNonlinearVec}_{\basisIndex}(\stateApprox,
        \multiplierApprox,\timeSymbol,\param)=\sum_{k=1}^{K_{\multiplierNonlinearForceSymbol}}
    \rho_k^{(g)}
    \left\langle\eta_{\multiplierNonlinearForceSymbol}
    \left(
    \stateFeVec^{-1}(\stateApprox),
    \multiplierFeVec^{-1}(\multiplierApprox),
    \timeSymbol,
    \param,
    x_k^{(g)}
    \right), 
    \multiplierROMBasisSymbol_{\basisIndex}
    \right\rangle_{[\multiplierFeSpace]' \times \multiplierFeSpace},
\end{split}
\end{equation}
}
which involves 
{numerical quadrautre} of the shown integrand functions over
the spatial domain.

The nonlinear terms
$\overline{\stateNonlinearVec}$ and
$\overline{\multiplierNonlinearVec}$
defined in
\eqref{eq:nonlinear-funcs}
{
need to be evaluated at every state update in a time marching scheme, 
and involve lifting back to the finite element functions, which scales
with the FOM sizes $\stateFOMSize$ and $\multiplierFOMSize$, before taking the nonlinear actions $\stateNonlinearForceSymbol$ and $\multiplierNonlinearForceSymbol$.}
This dominates the ROM computational cost and prevents it
from achieving significant speedup.
To overcome this issue, the evaluation of
$\overline{\stateNonlinearVec}$ and
$\overline{\multiplierNonlinearVec}$
is approximated by sampling
$\stateNonlinearVec$ and
$\multiplierNonlinearVec$
at a small number of evaluation points,
giving rise to various hyper-reduction methods.

In the next section we focus on hyper-reduction methods that are
based on interpolation or quadrature,
presenting their formulation and computational implementation.

\section{Hyper-reduction}\label{sec:hyp}
As explained in the previous section, the projection based reduction
of a nonlinear problems leads to nonlinear terms that retain their
dependence on the FOM size parameters
$\stateFOMSize\gg\stateROMSize$ and
$\multiplierFOMSize\gg\multiplierROMSize$.
To remove this dependence, various hyper-reduction methods have been developed. 
{The central idea of hyper-reduction is to replace the evaluation of nonlinear terms
$\overline{\stateNonlinearVec}$ and
$\overline{\multiplierNonlinearVec}$
defined in \eqref{eq:nonlinear-funcs} 
by approximations $\ROMStateNonlinearForceCoeff:\mathbb{R}^{\stateROMSize}\times
\mathbb{R}^{\multiplierROMSize}\times[0,\finalTime]\times\paramDomain\to
\mathbb{R}^{\stateROMSize}$ and
$\ROMMultiplierNonlinearForceCoeff:\mathbb{R}^{\stateROMSize}\times
\mathbb{R}^{\multiplierROMSize}\times[0,\finalTime]\times\paramDomain\to
\mathbb{R}^{\multiplierROMSize}$ which can be evaluated 
from the ROM coefficients 
$(\ROMstate, \ROMmultiplier) \in 
\mathbb{R}^{\stateROMSize}\times \mathbb{R}^{\multiplierROMSize}$
without the need of full lifting to
$(\stateApprox, \multiplierApprox) \in 
\mathbb{R}^{\stateFOMSize}\times \mathbb{R}^{\multiplierFOMSize}$
by \eqref{eq:solrepresentation} during the online phase.
Invoking these approximations to \eqref{eq:combinedROM},
we obtain the hyper-reduced order model of dimension
$\stateROMSize+\multiplierROMSize$ 
\begin{equation}\label{eq:combinedHRROM}
\begin{aligned}
    \ROMmassMat\frac{d\ROMstate}{d\timeSymbol}+\ROMstateLinearMat
        \ROMstate+\ROMbilinearMat\ROMmultiplier=
        \ROMStateNonlinearForceCoeff(\ROMstate,
        \ROMmultiplier,\timeSymbol,\param)\\
    \ROMbilinearMat^\top\ROMstate+\ROMmultiplierLinearMat\ROMmultiplier=
        \ROMMultiplierNonlinearForceCoeff(\ROMstate,\ROMmultiplier, 
        \timeSymbol,\param).
\end{aligned}
\end{equation}
}

In what follows, we focus on the hyper-reduction $\ROMStateNonlinearForceCoeff$ of
$\overline{\stateNonlinearVec}$;
the hyper-reduction $\ROMMultiplierNonlinearForceCoeff$ of
$\overline{\multiplierNonlinearVec}$
can be performed analogously.
{
The hyper-reduction methods considered in this paper can be classified into 
interpolation-based (Section \ref{sec:interpolation}) and 
quadrature-based (Section \ref{sec:quadrature}). 
For each class of methods, we will describe in detail 
the acceleration mechanism in the online phase,  
the additional precomputed quantities required in the offline phase, and 
the loss-minimizing design of hyper-reduction schemes.}

\subsection{Interpolation methods}\label{sec:interpolation}
{The acceleration mechanism of interpolation methods is to evaluate appropriately sampled entries of the nonlinear term $\stateNonlinearVec$ and interpolate to the other entries.}
Based on gappy proper orthogonal decomposition  \cite{everson1995}, 
the interpolation methods uses a reduced basis approximation of
$\stateNonlinearVec$
\begin{equation}\label{eq:interpolation-approx}
    \stateNonlinearVec\approx
        \ROMForceBasisMat_{\stateNonlinearForceSymbol}\,{\breve{\stateNonlinearVec}}
\end{equation}
where
\begin{equation*}
    \ROMForceBasisMat_{\stateNonlinearForceSymbol}=\bmat{\ROMForceBasisVec^{(\stateNonlinearForceSymbol)}_{1}&\cdots
        &\ROMForceBasisVec^{(\stateNonlinearForceSymbol)}_{\sizeROMStateForceSymbol}}
        \in\mathbb{R}^{\stateFOMSize\times\sizeROMStateForceSymbol}\qquad
        \sizeROMStateForceSymbol\ll\stateFOMSize
\end{equation*}
is a basis matrix for approximating the nonlinear term $\stateNonlinearVec$, 
{which can be obtained in a similar manner to \eqref{eq:basis_matrix} 
from additional snapshots of $\stateNonlinearVec$ in the offline phase, and
$\breve{\stateNonlinearVec}\in\mathbb{R}^{\sizeROMStateForceSymbol}$}
denotes the coordinates of $\stateNonlinearVec$
with respect to $\ROMForceBasisMat_{\stateNonlinearForceSymbol}$.
The approximation \eqref{eq:interpolation-approx}
is performed by interpolation of $\stateNonlinearVec$
at $\sizeROMSampleSymbol$ sampled indices, with
$\sizeROMStateForceSymbol\leq\sizeROMSampleSymbol\ll\stateFOMSize$.
More precisely, let
$\ROMSamplingSetSymbol_{\stateNonlinearForceSymbol}=
\{\sampleIndex_1,\dots,\sampleIndex_{\sizeROMSampleSymbol} \}
\subset\{1,\ldots,\stateFOMSize\}$
be a set of sampled indices and
$\ROMSamplingMatSymbol_{\stateNonlinearForceSymbol}\in
\mathbb{R}^{\stateFOMSize\times\sizeROMSampleSymbol}$
the associated sampling matrix whose entries are given by
$[\ROMSamplingMatSymbol_{\stateNonlinearForceSymbol}]_{\basisIndex,k}=
\delta_{\basisIndex,\sampleIndex_k}$.
Using the above, we formulate a least squares problem for the coordinates
\begin{equation*}
    {\breve{\stateNonlinearVec}}=
        \argmin_{\dummyVec\in\mathbb{R}^{\sizeROMStateForceSymbol}}
        \|\ROMSamplingMatSymbol_{\stateNonlinearForceSymbol}^{\top}(\stateNonlinearVec
        -
        \ROMForceBasisMat_{\stateNonlinearForceSymbol}\dummyVec)\|_2^2
\end{equation*}
whose solution is given by
\begin{equation*}
    {\breve{\stateNonlinearVec}}=(\ROMSamplingMatSymbol_{\stateNonlinearForceSymbol}^{\top}
    \ROMForceBasisMat_{\stateNonlinearForceSymbol})^\dagger
    \ROMSamplingMatSymbol_{\stateNonlinearForceSymbol}^{\top}\stateNonlinearVec, 
\end{equation*}
where superscript $\dagger$ denotes the Moore-Penrose pseudoinverse.
{Combining with \eqref{eq:algebraic-ROM-force},
this yields
\begin{equation}
    \overline{\stateNonlinearVec}(\stateApprox,
        \multiplierApprox,\timeSymbol,\param) \approx  
    \ROMStateNonlinearForceCoeff(\ROMstate,
        \ROMmultiplier,\timeSymbol,\param) = 
        \stateROMBasisMat^\top \ROMForceBasisMat_{\stateNonlinearForceSymbol}(\ROMSamplingMatSymbol_{\stateNonlinearForceSymbol}^{\top}
        \ROMForceBasisMat_{\stateNonlinearForceSymbol})^\dagger\ROMSamplingMatSymbol_{\stateNonlinearForceSymbol}^{\top}
        \stateNonlinearVec(\stateApprox,
        \multiplierApprox,\timeSymbol,\param), 
\end{equation}
where the matrix $\mathbf{T}_{\stateNonlinearForceSymbol} = \stateROMBasisMat^\top \ROMForceBasisMat_{\stateNonlinearForceSymbol}(\ROMSamplingMatSymbol_{\stateNonlinearForceSymbol}^{\top}
\ROMForceBasisMat_{\stateNonlinearForceSymbol})^\dagger \in \mathbb{R}^{\stateROMSize \times \sizeROMStateForceSymbol}$ is small, 
which can be precomputed and stored offline. 
In the online phase, it suffices to 
evaluate $\ROMSamplingMatSymbol_{\stateNonlinearForceSymbol}^{\top} \stateNonlinearVec \in \mathbb{R}^{\sizeROMSampleSymbol}$, i.e., the values of $\stateNonlinearVec$
at the sampling indices $\ROMSamplingSetSymbol_{\stateNonlinearForceSymbol}$, 
and multiply to the precomputed matrix 
$\mathbf{T}_{\stateNonlinearForceSymbol}$. 
In view of \eqref{eq:fe-quadrature}, the evaluation of $\ROMStateNonlinearForceCoeff$ only requires the values of  
the nonlinear density $\eta_{\stateNonlinearForceSymbol}$ 
in the support of the selected test functions $\{ \stateFeBasisSymbol_{\basisIndex} \}_{\basisIndex \in \ROMSamplingSetSymbol_{\stateNonlinearForceSymbol}}$, 
which is referred to as the sample mesh, and 
avoids the expensive full lifting. 
}

{
In addition to the basis matrices 
$\stateROMBasisMat$ and $\multiplierROMBasisMat$
described in Section~\ref{subsec:offline},
the offline procedure of interpolation-based hyper-reduction methods 
involves the construction of additional basis matrices 
$\ROMForceBasisMat_{\stateNonlinearForceSymbol}$ and 
$\ROMForceBasisMat_\multiplierNonlinearForceSymbol$, and 
the selection of sampling indices
$\ROMSamplingSetSymbol_{\stateNonlinearForceSymbol}$ and 
$\ROMSamplingSetSymbol_{\multiplierNonlinearForceSymbol}$, which
heavily affects the tradeoff between accuracy and speed-up.
Noting that 
the matrix $\projectionMatSymbol_{\stateNonlinearForceSymbol} = \ROMForceBasisMat_{\stateNonlinearForceSymbol}(\ROMSamplingMatSymbol_{\stateNonlinearForceSymbol}^{\top}
        \ROMForceBasisMat_{\stateNonlinearForceSymbol})^\dagger\ROMSamplingMatSymbol_{\stateNonlinearForceSymbol}^{\top}
\in\mathbb{R}^{\stateFOMSize\times\stateFOMSize}$
defines a rank-$\sizeROMStateForceSymbol$
oblique projection onto the column space of
$\ROMForceBasisMat_{\stateNonlinearForceSymbol}$, an}
error bound for interpolation methods as a function of
$\ROMSamplingMatSymbol_{\stateNonlinearForceSymbol}$
is given in \cite[Theorem 3.1]{lauzon2024s}
\begin{equation}\label{eq:proj-bound}
    \operatorname{err}(\ROMSamplingMatSymbol_{\stateNonlinearForceSymbol})=\left\|
        (\identityMat{\stateFOMSize}-\projectionMatSymbol_{\stateNonlinearForceSymbol})
        \stateNonlinearVec\right\|_2\leq\left\|
        (\ROMSamplingMatSymbol_{\stateNonlinearForceSymbol}^{\top}\ROMForceBasisMat_{\stateNonlinearForceSymbol})^\dagger\right\|_2
        \left\|(\identityMat{\stateFOMSize}-\ROMForceBasisMat_{\stateNonlinearForceSymbol}
        \ROMForceBasisMat_{\stateNonlinearForceSymbol}^\top)\stateNonlinearVec\right\|_2
\end{equation}
where the 2-norm
$\left\|(\ROMSamplingMatSymbol_{\stateNonlinearForceSymbol}^{\top}\ROMForceBasisMat_{\stateNonlinearForceSymbol})^\dagger\right\|_2$
is equal to the largest singular value of
$(\ROMSamplingMatSymbol_{\stateNonlinearForceSymbol}^{\top}\ROMForceBasisMat_{\stateNonlinearForceSymbol})^\dagger$,
or equivalently the reciprocal of the smallest strictly positive
singular value of
$\ROMSamplingMatSymbol_{\stateNonlinearForceSymbol}^{\top}\ROMForceBasisMat_{\stateNonlinearForceSymbol}$.
The same theorem provides the alternative expression
\begin{equation}\label{eq:proj-error-identity}
    \left\|(\identityMat{\stateFOMSize}-\projectionMatSymbol_{\stateNonlinearForceSymbol})
        \stateNonlinearVec\right\|_2^2=\left\|(\identityMat{\stateFOMSize}
        -\ROMForceBasisMat_{\stateNonlinearForceSymbol}\ROMForceBasisMat_{\stateNonlinearForceSymbol}^\top)
        \stateNonlinearVec\right\|_2^2+\left\|
        \epsilon(\stateNonlinearVec,\ROMSamplingSetSymbol_{\stateNonlinearForceSymbol})\right\|_2^2
\end{equation}
where
$\epsilon(\stateNonlinearVec,\ROMSamplingSetSymbol_{\stateNonlinearForceSymbol})
\in\mathbb{R}^{\sizeROMStateForceSymbol}$
is given by
\begin{equation}\label{eq:oblique-proj-error-identity}
    \epsilon(\stateNonlinearVec,\ROMSamplingSetSymbol_{\stateNonlinearForceSymbol})=\left(\left(
        \ROMSamplingMatSymbol_{\stateNonlinearForceSymbol}^{\top}\ROMForceBasisMat_{\stateNonlinearForceSymbol}\right)^{\top}
        \ROMSamplingMatSymbol_{\stateNonlinearForceSymbol}^{\top}\ROMForceBasisMat_{\stateNonlinearForceSymbol}\right)^{-1}
        \left(\ROMSamplingMatSymbol_{\stateNonlinearForceSymbol}^{\top}\ROMForceBasisMat_{\stateNonlinearForceSymbol}\right)^{\top}
        \ROMSamplingMatSymbol_{\stateNonlinearForceSymbol}^{\top}(\identityMat{\stateFOMSize}
        -\ROMForceBasisMat_{\stateNonlinearForceSymbol}\ROMForceBasisMat_{\stateNonlinearForceSymbol}^\top)\stateNonlinearVec.
\end{equation}
Different greedy sampling strategies for adaptively minimizing the error
$\operatorname{err}(\ROMSamplingMatSymbol_{\stateNonlinearForceSymbol})$
can be devised based on the expressions
\eqref{eq:proj-bound}-\eqref{eq:oblique-proj-error-identity},
three of which are considered in this work.
DEIM sampling methods (Section \ref{sec:DEIM})
aim to lower the error bound in \eqref{eq:proj-bound}
by minimizing the 2-norm of
$(\ROMSamplingMatSymbol_{\stateNonlinearForceSymbol}^{\top} \ROMForceBasisMat_{\stateNonlinearForceSymbol})^\dagger$.
Q-DEIM sampling methods (Section \ref{sec:QDEIM})
have the similar goal of minimizing the condition number of
$(\ROMSamplingMatSymbol_{\stateNonlinearForceSymbol}^{\top}\ROMForceBasisMat_{\stateNonlinearForceSymbol})^\dagger$
to minimize the error bound.
Finally, the S-OPT sampling algorithm (Section \ref{sec:SOPT})
aims first to increase the non-singularity of
$\left(\ROMSamplingMatSymbol_{\stateNonlinearForceSymbol}^{\top}\ROMForceBasisMat_{\stateNonlinearForceSymbol}\right)^{\top}
\ROMSamplingMatSymbol_{\stateNonlinearForceSymbol}^{\top}\ROMForceBasisMat_{\stateNonlinearForceSymbol}$
to increase numerical stability,
and second to increase the column orthogonality of
$\ROMSamplingMatSymbol_{\stateNonlinearForceSymbol}^{\top}\ROMForceBasisMat_{\stateNonlinearForceSymbol}$
in order to decrease the 2-norm of
$\epsilon(\stateNonlinearVec,\ROMSamplingSetSymbol_{\stateNonlinearForceSymbol})$.

\subsubsection{DEIM}\label{sec:DEIM}
The discrete empirical interpolation method (DEIM)
was introduced in \cite{chaturantabut2010nonlinear}
for greedily selecting
$\sizeROMSampleSymbol=\sizeROMStateForceSymbol$
sampling indices from the linearly independent rows
of $\ROMForceBasisMat_{\stateNonlinearForceSymbol}$
such that the 2-norm of
$(\ROMSamplingMatSymbol_{\stateNonlinearForceSymbol}^{\top}\ROMForceBasisMat_{\stateNonlinearForceSymbol})^\dagger$
is minimized.
A variant of this algorithm that allows for oversampling
($\sizeROMSampleSymbol \geq \sizeROMStateForceSymbol$)
was introduced in \cite{carlberg2011efficient},
a simplified version of which is presented in
Algorithm \ref{alg:deim}.

\begin{algorithm}[t]
\setstretch{1.35}
\caption{Oversampled DEIM sampling algorithm}
\label{alg:deim}
\textbf{Input:} Basis matrix
$\ROMForceBasisMat_{\stateNonlinearForceSymbol}\in
\mathbb{R}^{\sizeFOMsymbol\times\sizeROMStateForceSymbol}$
and desired number of sampled indices
$\sizeROMSampleSymbol > \sizeROMStateForceSymbol$\\
\textbf{Output:} Set of $\sizeROMSampleSymbol$
sampling indices
$\ROMSamplingSetSymbol_{\stateNonlinearForceSymbol}=\{\sampleIndex_1,\dots,
\sampleIndex_{\sizeROMSampleSymbol}\}$
\begin{algorithmic} [1]
 \State $\sampleIndex_1 = \texttt{argmax}_{\sampleIndex}
 \vert [\ROMForceBasisMat_{\stateNonlinearForceSymbol}]_{\sampleIndex,1} \vert$
 \State $\ROMSamplingSetSymbol_{\stateNonlinearForceSymbol} = \{ \sampleIndex_1 \}$
 \State $n_{\text{iter}} = \text{ceil}
(\frac{\sizeROMSampleSymbol-1}{\sizeROMStateForceSymbol-1})$
 \For{ $\basisIndex = 2: \sizeROMStateForceSymbol$}
   \State
   Construct $\ROMForceBasisMat_{\stateNonlinearForceSymbol}^{(\basisIndex-1)} = \ROMForceBasisMat_{\stateNonlinearForceSymbol}[\boldsymbol{e}_1, \cdots, \boldsymbol{e}_{\basisIndex-1}]$
   \For{ $\ell = (\basisIndex-2) n_{\text{iter}} + 1 : (\basisIndex-1)  n_{\text{iter}}$}
     \State Construct
     $\ROMSamplingMatSymbol_{\stateNonlinearForceSymbol} = [\boldsymbol{e}_{\sampleIndex_1}, \cdots, \boldsymbol{e}_{\sampleIndex_\ell}]$
     \State Compute
     $\bm{\gappyerr} = \ROMForceBasisMat_{\stateNonlinearForceSymbol}^{(\basisIndex-1)} (\ROMSamplingMatSymbol_{\stateNonlinearForceSymbol}^\top \ROMForceBasisMat_{\stateNonlinearForceSymbol}^{(\basisIndex-1)})^\dagger\ROMSamplingMatSymbol_{\stateNonlinearForceSymbol}^\top \ROMForceBasisVec^{(\stateNonlinearForceSymbol)}_{\basisIndex}$
     \State $\sampleIndex_{\ell+1} = \texttt{argmax}_{\sampleIndex} \left| [\ROMForceBasisMat_{\stateNonlinearForceSymbol}]_{\sampleIndex,\basisIndex} - \bm{\gappyerr}_\sampleIndex \right|$ 
     \State $\ROMSamplingSetSymbol_{\stateNonlinearForceSymbol} \leftarrow \ROMSamplingSetSymbol_{\stateNonlinearForceSymbol} \cup \{  \sampleIndex_{\ell+1} \}$
     \IIf {$|\ROMSamplingSetSymbol_{\stateNonlinearForceSymbol}|=
     \sizeROMSampleSymbol$} \Return
    \EndFor
    \EndFor
\end{algorithmic}
\end{algorithm}

\subsubsection{Q-DEIM}\label{sec:QDEIM}
The Q-DEIM algorithm was introduced in
\cite{drmac2016new}
as a variant of DEIM for selecting the sampling indices
in a way that produces a smaller estimate of the condition number
$\|(\ROMSamplingMatSymbol_{\stateNonlinearForceSymbol}^{\top}\ROMForceBasisMat_{\stateNonlinearForceSymbol})^\dagger||_2$
when $\sizeROMSampleSymbol=\sizeROMStateForceSymbol$.
In contrast to the greedy selection process employed in DEIM,
the Q-DEIM algorithm relies on a rank revealing QR factorization of
$(\ROMSamplingMatSymbol_{\stateNonlinearForceSymbol}^{\top} \ROMForceBasisMat_{\stateNonlinearForceSymbol})^{\top}$.
A further development of Q-DEIM termed GappyPOD+E was proposed in
\cite{peherstorfer2020stability},
which achieves low projection errors with few oversampling points
by maximizing the
lower bound of the smallest singular value
of $\ROMSamplingMatSymbol_{\stateNonlinearForceSymbol}^{\top}\ROMForceBasisMat_{\stateNonlinearForceSymbol}$.
This algorithm is presented in Algorithm \ref{alg:QDEIM}.

\begin{algorithm}[t]
\setstretch{1.35}
\caption{GappyPOD+E sampling algorithm}
\label{alg:QDEIM}
\textbf{Input:} Basis matrix $\ROMForceBasisMat_{\stateNonlinearForceSymbol} \in 
\mathbb{R}^{\sizeFOMsymbol \times \sizeROMStateForceSymbol}
$ and desired number of sampled indices 
$\sizeROMSampleSymbol$\\
\textbf{Output:} Set of  $\sizeROMSampleSymbol$ sampling indices
$\ROMSamplingSetSymbol_{\stateNonlinearForceSymbol}=\{\sampleIndex_1,\dots,
\sampleIndex_{\sizeROMSampleSymbol}\}$
\begin{algorithmic} [1]
 \State $\ROMForceBasisMat_{\stateNonlinearForceSymbol}^\top \ROMSamplingMatSymbol_{\stateNonlinearForceSymbol} = \mathbf{Q}\mathbf{R}$, 
 where $\mathbf{Q} \in 
\mathbb{R}^{\sizeROMStateForceSymbol \times \sizeROMStateForceSymbol}
$,
 $\mathbf{Q}^\top \mathbf{Q} = \identityMat{\sizeROMStateForceSymbol}$, 
 and $\mathbf{R} \in \mathbb{R}^{\sizeROMStateForceSymbol \times \sizeFOMsymbol}$
 is upper triangular.
 \State Take $\ROMSamplingSetSymbol_{\stateNonlinearForceSymbol} = \{ \sampleIndex_1, \ldots, \sampleIndex_{\sizeROMStateForceSymbol} \}$, with $\ROMSamplingMatSymbol_{\stateNonlinearForceSymbol} = [\boldsymbol{e}_{\sampleIndex_1}, \cdots, \boldsymbol{e}_{\sampleIndex_{\sizeROMStateForceSymbol}}]$
 \For{ $\ell = \sizeROMStateForceSymbol+1: \sizeROMSampleSymbol$} 
     \State 
     Construct 
     $\ROMSamplingMatSymbol_{\stateNonlinearForceSymbol} = [\boldsymbol{e}_{\sampleIndex_1}, \cdots, \boldsymbol{e}_{\sampleIndex_{\ell-1}}]$
     from $\ROMSamplingSetSymbol_{\stateNonlinearForceSymbol}$
     \State
     $\ROMSamplingMatSymbol_{\stateNonlinearForceSymbol}^\top \ROMForceBasisMat_{\stateNonlinearForceSymbol} = \mathbf{U} \mathbf{S} \mathbf{V}^\top$, where $\mathbf{U} \in \mathbb{R}^{j \times \sizeROMStateForceSymbol}$, $\mathbf{U}^\top \mathbf{U} = \identityMat{j}$, $\mathbf{V} \in \mathbb{R}^{\sizeROMStateForceSymbol \times \sizeROMStateForceSymbol}$, $\mathbf{V}^\top \mathbf{V} = \identityMat{\sizeROMStateForceSymbol}$, and $\mathbf{S} \in \mathbb{R}^{\sizeROMStateForceSymbol \times \sizeROMStateForceSymbol}$ is diagonal
     \State Compute $g = \sigma_{\sizeROMStateForceSymbol-1}^2 - \sigma_{\sizeROMStateForceSymbol}^2$ where $\mathbf{S} = \text{diag}(\sigma_1, \ldots, \sigma_{\sizeROMStateForceSymbol})$
     \State Compute $\mathbf{W} = \mathbf{V}^\top \ROMForceBasisMat_{\stateNonlinearForceSymbol}^\top \in \mathbb{R}^{\sizeROMStateForceSymbol \times \sizeFOMsymbol}$
     \State Compute $\mathbf{y} = (\mathbf{W} \odot \mathbf{W})^\top \mathbf{1}_{\sizeROMStateForceSymbol} \in \mathbb{R}^{\sizeFOMsymbol}$
     \State $\sampleIndex_{\ell} = \texttt{argmax}_{\sampleIndex \notin \ROMSamplingSetSymbol_{\stateNonlinearForceSymbol}} \, \left(
     g + \mathbf{y}_{\sampleIndex} - \sqrt{(g + \mathbf{y}_{\sampleIndex})^2 - 4g (\mathbf{W}_{\sizeROMStateForceSymbol, \sampleIndex})^2}\right)$ 
     \State $\ROMSamplingSetSymbol_{\stateNonlinearForceSymbol} \leftarrow \ROMSamplingSetSymbol_{\stateNonlinearForceSymbol} \cup \{  \sampleIndex_{\ell} \}$
    \EndFor
\end{algorithmic}
\end{algorithm}

\subsubsection{S-OPT}\label{sec:SOPT}
The majority of gappy  POD-based methods for selecting
$\ROMSamplingSetSymbol_{\stateNonlinearForceSymbol}$
are based on the principle of minimizing the
largest singular value of the sampled matrix
$(\ROMSamplingMatSymbol_{\stateNonlinearForceSymbol}^{\top}\ROMForceBasisMat_{\stateNonlinearForceSymbol})^\dagger$.
This is referred to as E-optimality
\cite{pukelsheim2006optimal}
in the optimal design community and suffers from the issue of numerical
instability \cite{lauzon2024s}.
On the other hand, the S-OPT algorithm introduced in
\cite{lauzon2024s}
pursues S-optimality, which results in better numerical stability than
sampling methods based on E-optimality.
The core principle of S-OPT is to simultaneously promote
the nonsingularity of the normal matrix
$\left(\ROMSamplingMatSymbol_{\stateNonlinearForceSymbol}^{\top}\ROMForceBasisMat_{\stateNonlinearForceSymbol}\right)^{\top}
\ROMSamplingMatSymbol_{\stateNonlinearForceSymbol}^{\top}\ROMForceBasisMat_{\stateNonlinearForceSymbol}$
and  the column orthogonality of the sampled matrix
$\ROMSamplingMatSymbol_{\stateNonlinearForceSymbol}^{\top}\ROMForceBasisMat_{\stateNonlinearForceSymbol}$
by maximizing measure $\mathcal{S}$
\cite{shin2016near}
\begin{equation}
    \mathcal{S}(\ROMSamplingMatSymbol_{\stateNonlinearForceSymbol}^{\top}\ROMForceBasisMat_{\stateNonlinearForceSymbol})=\left(
        \frac{\sqrt{\det\left(\left(\ROMSamplingMatSymbol_{\stateNonlinearForceSymbol}^{\top}
        \ROMForceBasisMat_{\stateNonlinearForceSymbol}\right)^{\top}\ROMSamplingMatSymbol_{\stateNonlinearForceSymbol}^{\top}
        \ROMForceBasisMat_{\stateNonlinearForceSymbol}\right)}}{\prod_{k=1}^{p}
        \|\ROMSamplingMatSymbol_{\stateNonlinearForceSymbol}^{\top}\ROMForceBasisMat_{\stateNonlinearForceSymbol}\mathbf{e}_k\|_2}
        \right)^{\frac{1}{p}}.
\end{equation}
The S-OPT algorithm as described in
\cite[Algorithm 1]{lauzon2024s} is presented in
Algorithm \ref{alg:s-opt}.
In the developed algorithm, the determinant computations are
performed using a fast method that exploits the rank-one updates
involved in each iteration \cite[Section 3.3.3]{shin2016near}.

\begin{algorithm}[t]
\setstretch{1.35}
\caption{S-OPT sampling algorithm}
\label{alg:s-opt}
\textbf{Input:} Basis matrix $\ROMForceBasisMat_{\stateNonlinearForceSymbol} \in \mathbb{R}^{\sizeFOMsymbol \times \sizeROMStateForceSymbol}$ and desired number of sampled indices $\sizeROMSampleSymbol$. \\
\textbf{Output:} A set of $\sizeROMSampleSymbol$ 
indices $\ROMSamplingSetSymbol_{\stateNonlinearForceSymbol} = \{\sampleIndex_1, \dots, 
\sampleIndex_{\sizeROMSampleSymbol} \}$.
\begin{algorithmic} [1]
\State $\sampleIndex_1 = \texttt{argmax}_{\sampleIndex} 
 \vert [\ROMForceBasisMat_{\stateNonlinearForceSymbol}]_{\sampleIndex,1} \vert$
 \State $\ROMSamplingSetSymbol_{\stateNonlinearForceSymbol} = \{ \sampleIndex_1 \}$
 \For{ $\ell = 2 : \sizeROMStateForceSymbol$} 
   \State Construct
   $\ROMForceBasisMat_{\stateNonlinearForceSymbol}^{(\ell-1)} = \ROMForceBasisMat_{\stateNonlinearForceSymbol}[\boldsymbol{e}_1, \cdots, \boldsymbol{e}_{\ell-1}]$
   \State Construct 
     $\ROMSamplingMatSymbol_{\stateNonlinearForceSymbol} = [\boldsymbol{e}_{\sampleIndex_1}, \cdots, \boldsymbol{e}_{\sampleIndex_{\ell-1}}]$
   \State Construct $\mathbf{A} = \ROMSamplingMatSymbol_{\stateNonlinearForceSymbol}^\top \ROMForceBasisMat_{\stateNonlinearForceSymbol}^{(\ell-1)}$ 
   \State $\sampleIndex_{\ell} = \texttt{argmax}_{\sampleIndex \notin \ROMSamplingSetSymbol_{\stateNonlinearForceSymbol}} 
   \dfrac{1 + \mathbf{r}^\top \mathbf{b}}
{\prod_{k=1}^{\basisIndex} (\|\mathbf{A} \mathbf{e}_k\|^2 + \mathbf{r}_k^2)}
\dfrac{\mathbf{c}^\top \mathbf{c} + \gamma^2 - \alpha}{\mathbf{c}^\top \mathbf{c} + \gamma^2}$, with
$$\mathbf{r}^\top = \mathbf{e}_{i}^\top \ROMForceBasisMat_{\stateNonlinearForceSymbol}^{(\ell-1)}, \quad 
\mathbf{b} = (\mathbf{A}^\top \mathbf{A})^{-1} \mathbf{r}, \quad 
\gamma = [\ROMForceBasisMat_{\stateNonlinearForceSymbol}]_{\sampleIndex, \ell}$$
   \State $\ROMSamplingSetSymbol_{\stateNonlinearForceSymbol} \leftarrow \ROMSamplingSetSymbol_{\stateNonlinearForceSymbol} \cup \{\sampleIndex_{\ell}\}$
 \EndFor
 \For{ $\ell = \sizeROMStateForceSymbol+1: \sizeROMSampleSymbol$}
 \State Construct 
     $\ROMSamplingMatSymbol_{\stateNonlinearForceSymbol} = [\boldsymbol{e}_{\sampleIndex_1}, \cdots, \boldsymbol{e}_{\sampleIndex_{\ell-1}}]$
 \State Construct $\mathbf{A} = \ROMSamplingMatSymbol_{\stateNonlinearForceSymbol}^\top \ROMForceBasisMat_{\stateNonlinearForceSymbol}$ 
 \State $\sampleIndex_{\ell} = \texttt{argmax}_{\sampleIndex \notin \ROMSamplingSetSymbol_{\stateNonlinearForceSymbol}} 
   \dfrac{1 + \mathbf{r}^\top (\mathbf{A}^\top \mathbf{A})^{-1} \mathbf{r}}
{\prod_{k=1}^{\sizeROMStateForceSymbol} (\|\mathbf{A} \mathbf{e}_k\|^2 + \mathbf{r}_k^2)}$, 
with $\mathbf{r}^\top = \mathbf{e}_{i}^\top \ROMForceBasisMat_{\stateNonlinearForceSymbol}$
   \State $\ROMSamplingSetSymbol_{\stateNonlinearForceSymbol} \leftarrow \ROMSamplingSetSymbol_{\stateNonlinearForceSymbol} \cup \{\sampleIndex_\ell\}$
 \EndFor
\end{algorithmic}
\end{algorithm}

\subsection{Quadrature methods}\label{sec:quadrature}
{The acceleration mechanism of quadrature-based methods is to 
approximate the integrations in the finite element assembly 
by an appropriately selected subset of quadrature integration points.}
As shown in \eqref{eq:nonlinear-funcs},
the nonlinear term
$\overline{\stateNonlinearVec}$ 
is defined as projection of the vector valued function
$\stateNonlinearVec$ onto the derived reduced subspace.
Recalling the duality pairings 
correspond to integration of the nonlinear integrands \eqref{eq:fe-quadrature}, 
the quadrature-based hyper-reduction methods seek to form a sparse and re-weighted approximation of \eqref{eq:nonlinear-funcs} on the reduced subspace  $\stateSubspaceSymbol$ as the test function space, i.e., 
{
\begin{equation}\label{eq:sparse-quadrature}
    \overline{\stateNonlinearVec}_{\basisIndex}(\stateApprox,\multiplierApprox,\timeSymbol,\param)\approx
    \ROMStateNonlinearForceCoeff_{\basisIndex}(\ROMstate,
        \ROMmultiplier,\timeSymbol,\param) = 
        \sum_{k=1}^{K_{\stateNonlinearForceSymbol}}
        \rho_k^{(f), \star}
        \left\langle\eta_{\stateNonlinearForceSymbol}
        \left(
        \stateFeVec^{-1}(\state),
        \multiplierFeVec^{-1}(\multiplier),
        \timeSymbol,
        \param,
        x_k^{(f)}
        \right), 
        \stateROMBasisSymbol_{\basisIndex}\right\rangle_{[\stateFeSpace]' \times \stateFeSpace}
\end{equation}
where the new weights are non-negative, i.e., 
$\rho_k^{({\stateNonlinearForceSymbol}),\star} \geq 0$,
and sparse, i.e., $\mathcal{K}^\star_{\stateNonlinearForceSymbol} = \{k\colon\rho_k^{({\stateNonlinearForceSymbol}),\star} \neq 0\}$ 
with $\vert \mathcal{K}^\star_{\stateNonlinearForceSymbol} \vert =K_{\stateNonlinearForceSymbol}^\star\ll K_{\stateNonlinearForceSymbol}$. 
In the online phase, 
the evaluation of $\ROMStateNonlinearForceCoeff$ only requires  
the ROM finite element approximation $\stateApprox$ and $\multiplierApprox$
where it takes to compute the nonlinear density $\eta_{\stateNonlinearForceSymbol}$ at selected quadrature points $\{x_k\}_{k \in \mathcal{K}^\star_{\stateNonlinearForceSymbol}}$, again called the sample mesh, and thus
avoids the expensive full lifting. 
}

{
In addition to the basis matrices 
$\stateROMBasisMat$ and $\multiplierROMBasisMat$
described in Section~\ref{subsec:offline},
the offline procedure of quadrature-based hyper-reduction methods 
involves finding two sets of sparse quadrature weights 
$\bm{\rho}^\star_{\stateNonlinearForceSymbol} = (\rho_k^{({\stateNonlinearForceSymbol}),\star})_{k=1}^{K_{\stateNonlinearForceSymbol}}$ and 
$\bm{\rho}^\star_{\multiplierNonlinearForceSymbol} = (\rho_k^{({\multiplierNonlinearForceSymbol}),\star})_{k=1}^{K_{\multiplierNonlinearForceSymbol}}$, which
heavily affects the tradeoff between accuracy and speed-up.
}
The computational expense of calculating
$\stateNonlinearVec$
scales with the number of sampled quadrature points, making it
desirable to minimize $K^\star_{\stateNonlinearForceSymbol}$.
At the same time, the set of nonzero quadrature weights
affects the quality of the approximation in
\eqref{eq:sparse-quadrature},
which in turn has a significant impact on the downstream ROM accuracy
(in analogy to the role of $\ROMSamplingSetSymbol_{\stateNonlinearForceSymbol}$
in interpolation methods).
It is therefore desirable to obtain a vector of weights
$\bm{\rho}^\star_{\stateNonlinearForceSymbol}$ that minimizes $K^\star_{\stateNonlinearForceSymbol}$
while keeping the approximation error below a chosen tolerance.
The task of computing $\bm{\rho}^\star_{\stateNonlinearForceSymbol}$ has been formulated
as a linear programming problem
\cite{Yano2019}
and as an equivalent nonnegative least squares (NNLS) problem
\cite{du2022,sleeman2022}.
In this work we follow the latter approach and determine
$\bm{\rho}^\star_{\stateNonlinearForceSymbol}$
by solving the NNLS problem
\begin{equation}\label{eq:nnls-system}
    \bm{\rho}^\star_{\stateNonlinearForceSymbol}=\argmin_{\mathbf{r}\geq0} \left\|\bm{G}_{\stateNonlinearForceSymbol} \left(\bm{\rho}_{\stateNonlinearForceSymbol}
        -\bm{r}\right) \right\|_2^2
\end{equation}
where $\bm{G}_{\stateNonlinearForceSymbol}\in\mathbb{R}^{n_c^{({\stateNonlinearForceSymbol})}\times K_{\stateNonlinearForceSymbol}}$
is an accuracy constraints matrix with $n_c^{(\stateNonlinearForceSymbol)}$ constraint rows
and $\bm{\rho}_{\stateNonlinearForceSymbol}$ is the original quadrature weights vector.
The entries of each row of $\bm{G}_{\stateNonlinearForceSymbol}$ correspond to the evaluation
of the integrand at all FOM quadrature points $x_k$
for a given combination of {snapshot quadruplets 
$(\stateSymbol_s, \multiplierSymbol_s,\timeSymbol_{\ell_s},\param_{m_s})$ introduced in Section~\ref{subsec:offline} and 
ROM test functions $\stateROMBasisSymbol_{\basisIndex}$
which gives rise to $n_c^{({\stateNonlinearForceSymbol})} = N_s^{(\stateSymbol)}\stateROMSize\ll K_\stateNonlinearForceSymbol$ constraints.
More specifically, we define $\bm{G}_{\stateNonlinearForceSymbol}$ entrywise by: 
for $1\leq s\leq N_s, 1\leq j\leq\stateROMSize, 1\leq k \leq K_{\stateNonlinearForceSymbol}$,
\begin{equation}
    [\bm{G}_{\stateNonlinearForceSymbol}]_{ik}=
        \left\langle\eta_{\stateNonlinearForceSymbol}
        \left(
        \stateSymbol_s, 
        \multiplierSymbol_s,
        \timeSymbol_{\ell_s},
        \param_{m_s},
        x_k^{(f)}
        \right), 
        \stateROMBasisSymbol_{\basisIndex}\right\rangle_{[\stateFeSpace]' \times \stateFeSpace}
        \quad
        i=j+(s-1)\stateROMSize. 
\end{equation}
}
After forming $\bm{G}_\stateNonlinearForceSymbol$, the NNLS problem
\eqref{eq:nnls-system}
is solved using the Lawson-Hanson algorithm
\cite{lawson1995solving}, {which is recited in Algorithm \ref{alg:LawsonHanson}}.
Even though the NNLS problem \eqref{eq:nnls-system}
does not attempt to explicitly minimize $K^\star_\stateNonlinearForceSymbol$,
we find that significantly sparse solutions
$\bm{\rho}^{\star}_\stateNonlinearForceSymbol$
are obtained in our numerical experiments
{using a solver tolerance $\epsilon = 10^{-6}$},
matching the behavior demonstrated in previous works
\cite{du2022,sleeman2022}. 

\begin{algorithm}[!htpb]
\setstretch{1.35}
\caption{Lawson-Hanson algorithm}
\label{alg:LawsonHanson}
\textbf{Input:} $\boldsymbol{G}_{\stateNonlinearForceSymbol} \in \mathbb{R}^{r_f \times K}$, $\boldsymbol{\rho}_{\stateNonlinearForceSymbol} \in \mathbb{R}^K$ and the solver tolerance $\varepsilon$\\
\textbf{Output:} $\bm{\rho}^\star_{\stateNonlinearForceSymbol} \in \mathbb{R}^K$
\begin{algorithmic} [1]
 \State $\boldsymbol{z} \leftarrow \emptyset$
 \State $\boldsymbol{v} \leftarrow \{1, \ldots, K\}$
 \State $\bm{\rho}^\star_{\stateNonlinearForceSymbol} \leftarrow \mathbf{0} \in \mathbb{R}^{K}$
 \State $\boldsymbol{u} \leftarrow \boldsymbol{G}^{\top}(\boldsymbol{G}_{\stateNonlinearForceSymbol} \boldsymbol{\rho} - \boldsymbol{G}_{\stateNonlinearForceSymbol} \bm{\rho}^\star_{\stateNonlinearForceSymbol})$
 \State $\boldsymbol{u}_R \leftarrow \boldsymbol{u}$
 \While{$|\boldsymbol{v}| > 0$ and $\max(\boldsymbol{u}_R) > \varepsilon$}
     \State $j \leftarrow \arg\max_i u_i \in \boldsymbol{u}_R$
     \State $\boldsymbol{z} \leftarrow \boldsymbol{z} \cup j$
     \State $\boldsymbol{v} \leftarrow \boldsymbol{v} \setminus j$
     \State $\boldsymbol{G}_{\stateNonlinearForceSymbol,z} \leftarrow \{\boldsymbol{G}_{\stateNonlinearForceSymbol}[:, i] \mid i \in \boldsymbol{z}\}$
     \State $\boldsymbol{s} \leftarrow \mathbf{0} \in \mathbb{R}^{K}$
     \State $\boldsymbol{s}_z \leftarrow \{s_i \mid i \in \boldsymbol{z}\}$
     \State $\boldsymbol{s}_R \leftarrow \{s_i \mid i \in \boldsymbol{v}\}$
     \State $\boldsymbol{s}_z \leftarrow \left((\boldsymbol{G}_{\stateNonlinearForceSymbol,z})^{\top} \boldsymbol{G}_{\stateNonlinearForceSymbol,z}\right)^{-1}(\boldsymbol{G}_{\stateNonlinearForceSymbol,z})^{\top} \boldsymbol{\rho}$
     \While{$\min(\boldsymbol{s}_z) \leq 0$}
         \State $\boldsymbol{s}_n \leftarrow \{s_i \leq 0\}$
         \State $\alpha \leftarrow \min\left(\frac{\bm{\rho}^\star_{\stateNonlinearForceSymbol}}{\bm{\rho}^\star_{\stateNonlinearForceSymbol} - \boldsymbol{s}_n}\right)$
         \State $\bm{\rho}^\star_{\stateNonlinearForceSymbol} \leftarrow \bm{\rho}^\star_{\stateNonlinearForceSymbol} + \alpha \cdot (\boldsymbol{s} - \bm{\rho}^\star_{\stateNonlinearForceSymbol})$
         \State $\boldsymbol{i}_z \leftarrow \{i \in \boldsymbol{z} : r_i \leq 0\}$
         \State $\boldsymbol{v} \leftarrow \boldsymbol{v} \cup \boldsymbol{i}_z$
         \State $\boldsymbol{z} \leftarrow \boldsymbol{z} \backslash \boldsymbol{i}_z$
         \State $\boldsymbol{s}_z \leftarrow \left((\boldsymbol{G}_{\stateNonlinearForceSymbol,z})^{\top} \boldsymbol{G}_{\stateNonlinearForceSymbol,z}\right)^{-1}(\boldsymbol{G}_{\stateNonlinearForceSymbol,z})^{\top} \boldsymbol{G}_{\stateNonlinearForceSymbol} \boldsymbol{\rho}$
         \State $\boldsymbol{s}_R \leftarrow 0$
     \EndWhile
     \State $\bm{\rho}^\star_{\stateNonlinearForceSymbol} \leftarrow \boldsymbol{s}$
     \State $\boldsymbol{u} \leftarrow \boldsymbol{G}_{\stateNonlinearForceSymbol}^{\top}(\boldsymbol{\rho} - \boldsymbol{G}_{\stateNonlinearForceSymbol} \bm{\rho}^\star_{\stateNonlinearForceSymbol})$
     \State $\boldsymbol{u}_R \leftarrow \{u_i \mid i \in \boldsymbol{v}\}$
 \EndWhile
\end{algorithmic}
\end{algorithm}
Before applying the algorithm, two conditioning operations are performed
on $\bm{G}_\stateNonlinearForceSymbol$, which we have found to increase the derived
quadrature rule's accuracy and decrease the number of iterations needed
for the algorithm's convergence.
First, each row of $\bm{G}_\stateNonlinearForceSymbol$ is rescaled by dividing the row by its
maximum absolute value.
In our numerical experiments we have found that within each row,
nonzero entries may differ considerably in magnitude.
Rescaling all entries to take values within
$[-1, 1]\subset\mathbb{R}$ was found to improve the convergence
of the Lawson-Hanson algorithm.
Second, we compute the LQ decomposition of matrix
$\bm{G}=\bm{L}\bm{Q}$
and use matrix $\bm{Q}$ as the constraints matrix.
It has been previously reported that using a constraints matrix
with orthonormal rows (constraints) consistently reduces the number
of iterations needed for the Lawson-Hanson algorithm to converge
\cite{Humphry2025},
and this indeed matches our own experience.
Additional details on the implementation of the EQP method
and the NNLS formulation can be found in
\cite{Vales2025ceqp}.
Finally, we note that generally the resulting sparse quadrature rule
can only be expected to yield accurate solutions in the neighborhood
of the differential variable snapshots $\stateFeSymbol(t_s)$
used for forming $\bm{G}_\stateNonlinearForceSymbol$.
If desired, additional snapshots can be incorporated into $\bm{G}_\stateNonlinearForceSymbol$
to make the derived quadrature rule more robust.
However, increasing the number of snapshots will increase the
offline time spent solving the NNLS problem and may also increase the
resulting number of nonzero entries of $\bm{\rho}_\stateNonlinearForceSymbol^\star$.

\subsection{Sample mesh}
\begin{figure}[t]
    \centering
    \begin{subfigure}{0.49\textwidth}
        \centering
        \includegraphics[width=\textwidth]{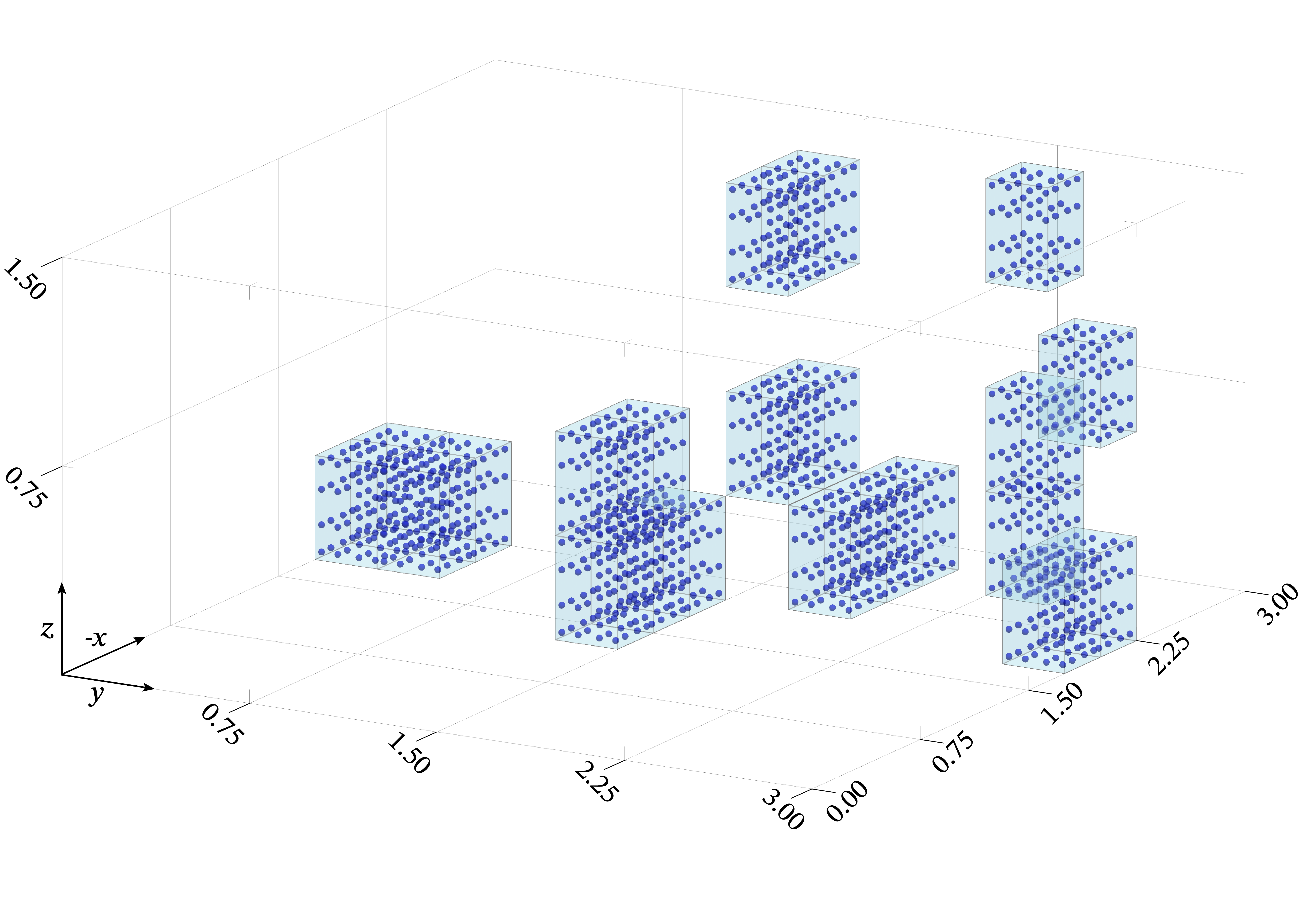}
    \end{subfigure}
    \begin{subfigure}{0.49\textwidth}
        \centering
        \includegraphics[width=\textwidth]{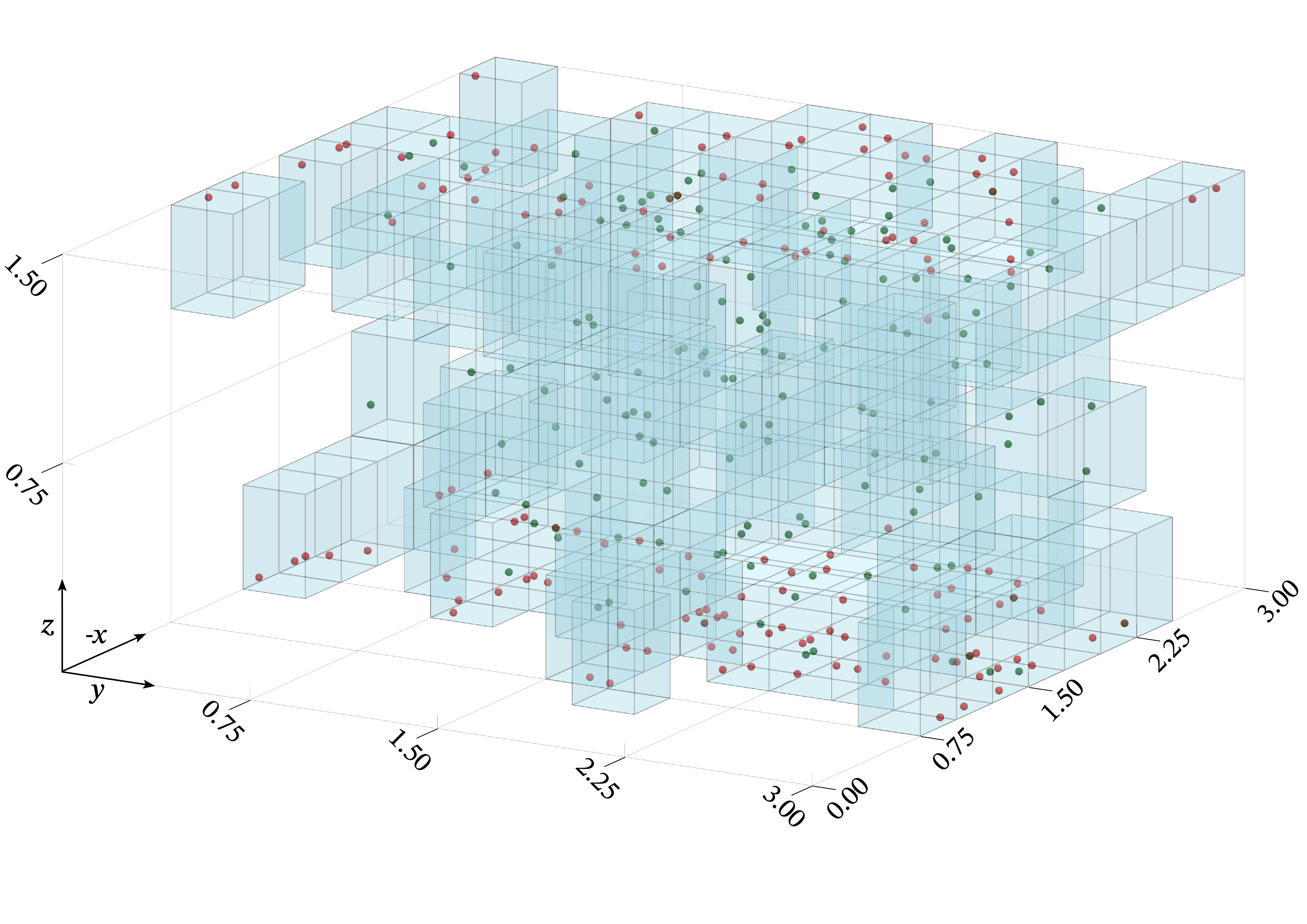}
    \end{subfigure}
    \caption{Sample mesh comparison for the triple point expansion problem.
    Left: S-OPT sample mesh with 22 sampled elements and $22\times64 = 1408$ sampled quadrature points (in dark blue).
    Right: EQP sample mesh with 178 sampled elements and 150 sampled quadrature points each for the energy field (in green) and the velocity field (in red).}
    \label{fig:smesh}
\end{figure}

{The hyper-reduction methods tested in this work require
the selection of the number of sampling points
($\sizeROMSampleSymbol$ for gappy POD methods and
$K^\star$ for EQP).
Ideally, this selection is done such that the nonlinear functions
$\stateNonlinearVec$ and $\multiplierNonlinearVec$
are accurately approximated while the computation time is minimized.
However, the relation between the number of sampling points and the
resulting computation time is not always straightforward.
Our implementation of the tested hyper-reduction methods relies on
the construction of a \textit{sample mesh}, which consists of the
finite elements containing the sampled quadrature or interpolation points.
Each finite element in our computational domain includes a prescribed
number of FEM basis functions (degrees of freedom) and associated
quadrature points.
The interpolation hyper-reduction methods sample degrees of freedom,
corresponding to the $\sizeROMSampleSymbol$ sampled indices
$\ROMSamplingSetSymbol_{\stateNonlinearForceSymbol}$.
Their implementation requires the evaluation of the sampled
degrees of freedom at all local quadrature points.
On the contrary, the EQP method samples quadrature points, determined
by the $K^\star$ non-zeros resulting from solving the NNLS problem
\eqref{eq:nnls-system}.
Its implementation requires the evaluation of all local degrees of
freedom at the subset of sampled quadrature points.
Because of this difference in approach between the two classes of
methods, the computational cost of the tested hyper-reduction methods
depends on element-wise factors that go beyond the number of sampled
points.
In many cases, the EQP method can also be implemented using pointwise
integration, effectively viewing the whole computational domain
as one ``global element'' to bypass the need for a sample
mesh \cite{du2022}.
This is possible when the nonlinear terms allow for a clean
offline-online splitting, where the geometric mapping and basis
function gradients are precomputed in a fixed reference configuration.
However, for the Lagrangian hydrodynamics problems
(Section \ref{sec:ex:laghos}),
the computational mesh is continuously deforming with time,
while the governing equations involve complex state-dependent terms
such as artificial viscosity that rely on local spatial stencils.
Specifically, robust shock-capturing via artificial viscosity necessitates
the calculation of velocity gradients and characteristic length scales
derived from the instantaneous configuration of the local mesh
\cite{Dobrev2012}.
Evaluating these element-based terms requires the online
reconstruction of local geometric quantities (e.g., Jacobians and
velocity gradients) and neighborhood connectivity at each time step.
Since these geometric quantities evolve dynamically and are
inextricably linked to the mesh topology, they cannot be fully
precomputed in a fixed reference configuration.
While alternative FOM discretizations (e.g., purely pointwise artificial
viscosity) might allow for a fully mesh-independent implementation,
they often lack the robustness required for complex shock-physics.
As a result, the standard formulations used in the present work require
access to the local mesh structure.
This complex relationship between the sample mesh and ROM cost
can lead to situations where using fewer sampled points does not
necessarily lead to lower computational cost.
Anticipating our later discussion, Figure \ref{fig:smesh}
illustrates such a scenario for the triple point expansion problem
(Section \ref{sec:ex:tp}), where the sampled elements are visualized
as volumes and the sampled points as dots.
In this particular example, although fewer points are sampled by EQP
than by S-OPT, the points sampled by EQP are scattered across many
high order elements.
This means that the sample mesh obtained by EQP contains significantly
more elements than the one obtained by S-OPT.
In such a case, the computational overhead associated with loading
the sampled elements and reconstructing their local geometry can
outweigh the savings gained from reducing the total number of
quadrature points.
As a result, the hyper-reduction speed-up becomes heavily dependent on
the spatial distribution of the sampled points and the spatial sparsity
of the sampled elements, rather than just the number of sampled
quadrature points.
These observations help highlight a key message: the employed FOM
discretization can be an important factor in determining the
speedup that can be delivered by a chosen hyper-reduction method.}

\section{Numerical results}\label{sec:ex}
The inherent tradeoff for any hyper-reduction method between
minimizing computational time and maximizing approximation accuracy
can be quantified with a Pareto front.
\begin{definition}[Pareto front]
Given a finite solution set
$\mathcal{D}=\{\mathbf{x}_1,\mathbf{x}_2,\ldots,\mathbf{x}_n\}$
and $m$ objective functions
$f_1,f_2,\ldots,f_m$,
the Pareto front $\mathcal{P}$ is defined as the set of solutions
$\mathbf{x}^*\in\mathcal{D}$ for which no
$\mathbf{x}\in\mathcal{D}$ satisfies both
\begin{equation*}
    f_i(\mathbf{x})\leq f_i(\mathbf{x}^*)\text{ for all }
        i\in\{1,\ldots,m\}\text{ and }
    f_j(\mathbf{x})<f_j(\mathbf{x}^*)\text{ for some }
        j\in\{1,\ldots,m\}.
\end{equation*}
\end{definition}
In this study, we consider the case $m=2$, with relative online phase
computation time and relative $L^2$ error between FOM and ROM solutions
as the two Pareto front objective functions.
It is expected that the Pareto front, and thus the optimal choice
of hyper-reduction method, will vary for different problems.

\begin{table}[t]
\setstretch{1.2}
\caption{Summary of simulation setups.}
\label{table:setup}
\centering
\begin{tabular}{c|c|c|c|c|c|c}
\multirow{2}{*}{Problem} &{\multirow{2}{*}{Mesh}}& {FE space}& {FOM size}& ODE  & \multirow{2}{*}{$T$}& \multirow{2}{*}{$N_t$}
\\
&& ($\stateFeSpace,\multiplierFeSpace$)&($\stateFOMSize, \multiplierFOMSize$) &solver&\\
\hline
Nonlinear& \multirow{2}{*}{{$32\times 32$}}& \multirow{2}{*}{$(\hat{Q}_0, RT_0)$}& \multirow{2}{*}{$(1024, 2112)$}& \multirow{2}{*}{BE} & \multirow{2}{*}{0.1}    &  \multirow{2}{*}{101} \\
diffusion &&&&&\\
\hline
{Nonlinear} & \multirow{2}{*}{{$32\times 4$}}& \multirow{2}{*}{$( [Q_2]^2 \times [Q_2]^2, -)$} & \multirow{2}{*}{$(2340, -)$} & \multirow{2}{*}{RK4} & \multirow{2}{*}{5.0}  & \multirow{2}{*}{500}\\ 
elasticity &&&&&\\
\hline
Sedov blast & {\multirow{2}{*}{$8\times 8\times 8$}}& \multirow{2}{*}{$([Q_2]^3 \times \hat{Q}_1 \times [Q_2]^3, -)$} &  \multirow{2}{*}{$(33574, -)$}& {RK4}& \multirow{2}{*}{0.2}  & 363  \\ 
\cline{5-5}\cline{7-7}
wave&&&& RK2Avg &&377 \\ 
\hline
Taylor-Green & {\multirow{2}{*}{$8\times 8\times 8$}}& \multirow{2}{*}{$([Q_2]^3 \times \hat{Q}_1 \times [Q_2]^3, -)$} & \multirow{2}{*}{$(33574, -)$}& {RK4}& \multirow{2}{*}{0.1}  & 266  \\ 
\cline{5-5}\cline{7-7}
vortex&&&& RK2Avg &&266 \\ 
\hline
Triple point & {\multirow{2}{*}{$28\times 12\times 4$}}& \multirow{2}{*}{$([Q_2]^3 \times \hat{Q}_1 \times [Q_2]^3, -)$}& \multirow{2}{*}{$(87702, -)$}& {RK4}& \multirow{2}{*}{0.5} & 95 \\ 
\cline{5-5}\cline{7-7}
expansion&&&& RK2Avg &&95  
\end{tabular}
\end{table}

\begin{figure}[t]
    \centering
    \begin{subfigure}{0.23\textwidth}
        \includegraphics[width=\linewidth,trim=150 20 20 0,clip]{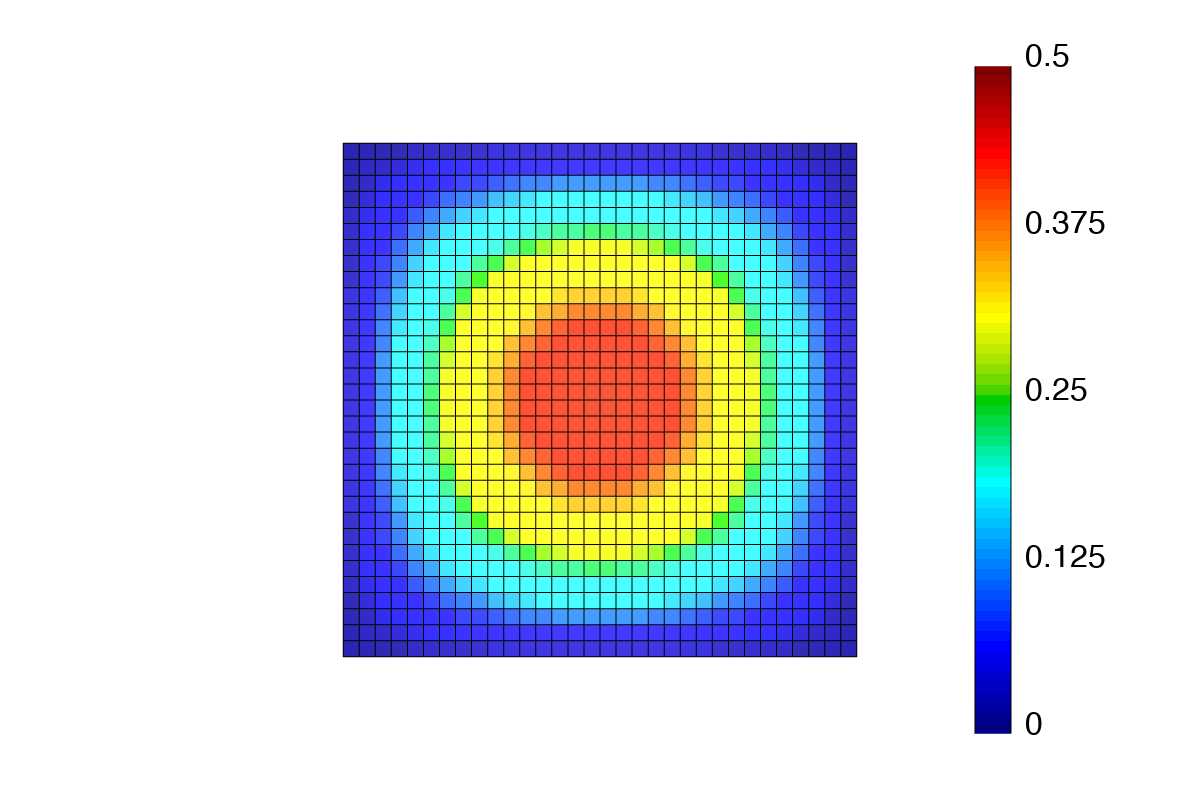}
    \end{subfigure}
    \begin{subfigure}{0.24\textwidth}
        \includegraphics[width=\linewidth,trim=150 30 20 0,clip]{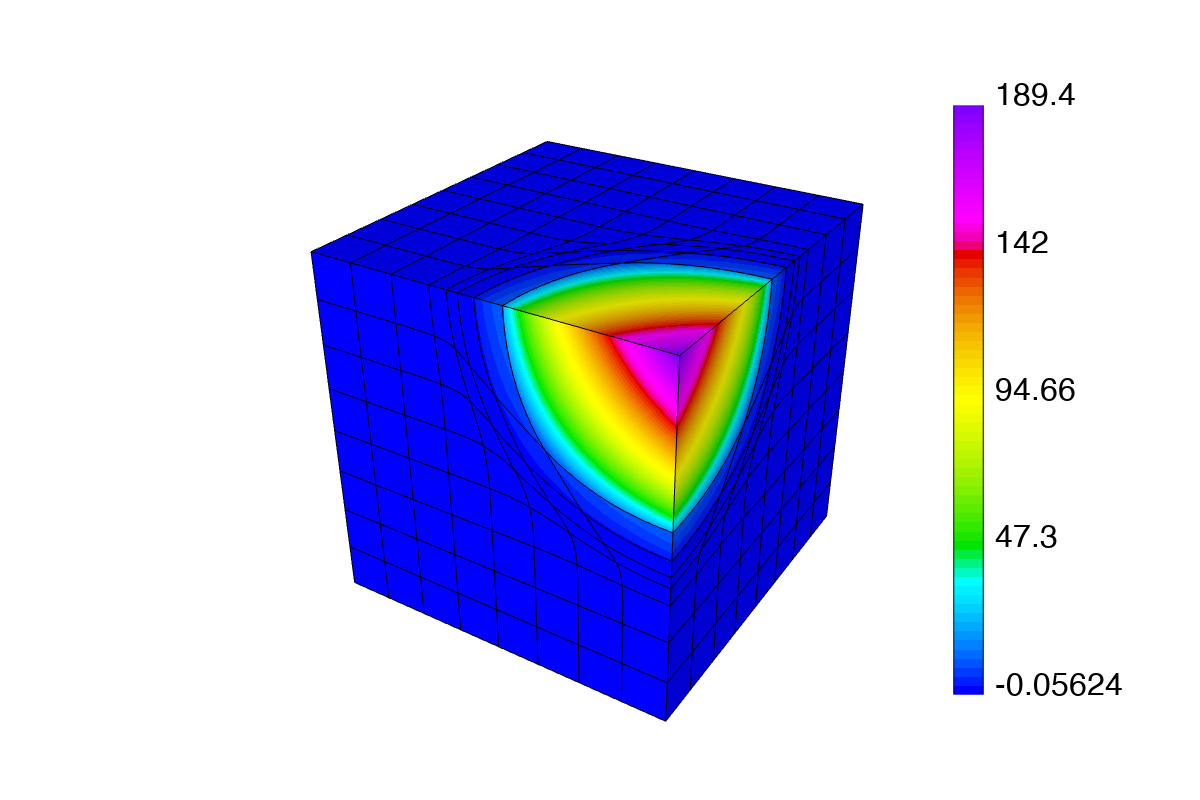}
    \end{subfigure}
    \begin{subfigure}{0.24\textwidth}
        \includegraphics[width=\linewidth,trim=150 30 20 0,clip]{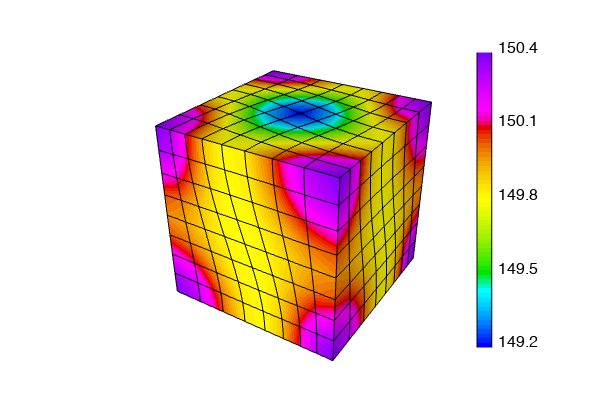}
    \end{subfigure}
    \begin{subfigure}{0.24\textwidth}
        \includegraphics[width=\linewidth,trim=150 20 20 0,clip]{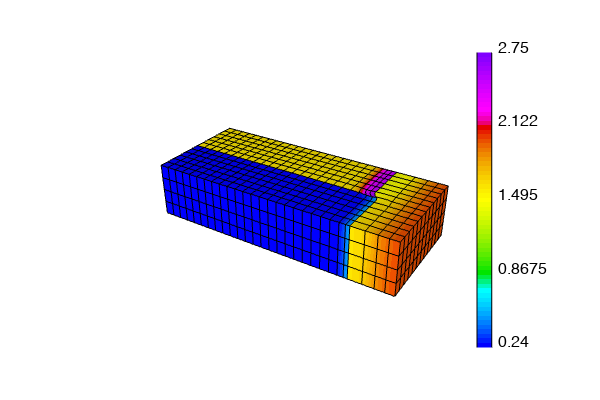}
    \end{subfigure}
    \\[4pt]
    \begin{subfigure}{0.24\textwidth}
        \includegraphics[width=\linewidth,trim=150 20 20 0,clip]{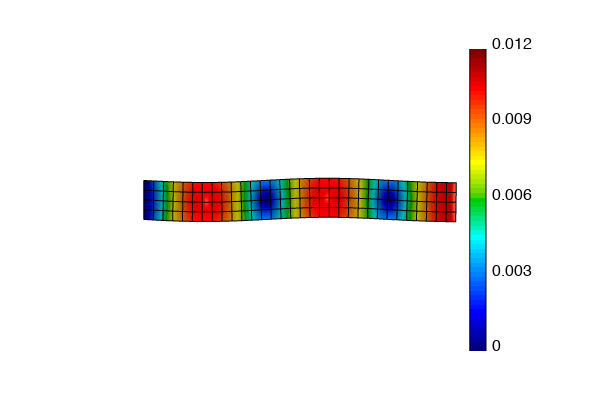}
    \end{subfigure}
    \begin{subfigure}{0.24\textwidth}
        \includegraphics[width=\linewidth,trim=150 20 20 0,clip]{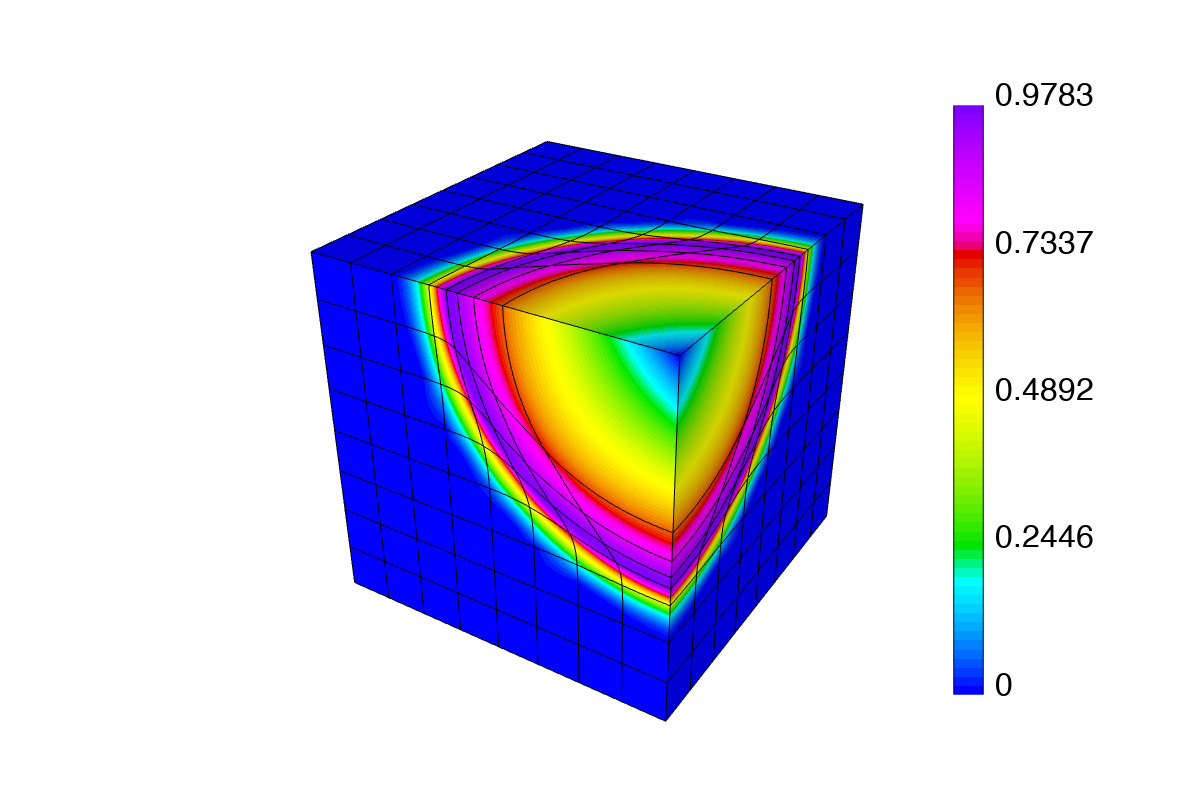}
    \end{subfigure}
    \begin{subfigure}{0.24\textwidth}
        \includegraphics[width=\linewidth,trim=150 20 20 0,clip]{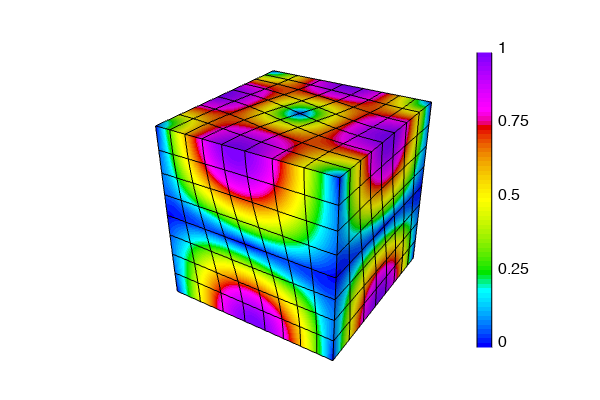}
    \end{subfigure}
    \begin{subfigure}{0.24\textwidth}
        \includegraphics[width=\linewidth,trim=150 20 20 0,clip]{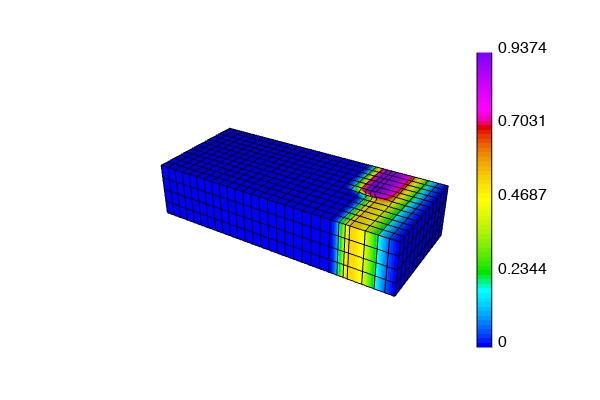}
    \end{subfigure}
    \\[2pt]
    \caption{Visualization of final-time fields from FOM simulations.
        The first column shows the nonlinear diffusion problem (top, {pressure}) and the nonlinear elasticity problem (bottom, velocity).
        The remaining columns correspond to Lagrangian benchmark problems, showing {internal energy per unit mass} (top) and velocity {magnitude} (bottom)
        for the Sedov blast wave (second column), the Taylor-Green vortex (third column), and the triple-point expansion problem (fourth column).}
    \label{fig:fom:solution}
\end{figure}

The hyper-reduction methods introduced in Section \ref{sec:hyp}
are applied to five problems:
nonlinear diffusion (Section \ref{sec:ex:mnd}),
nonlinear elasticity (Section \ref{sec:ex:nlel})
and three Lagrangian hydrodynamics problems:
Sedov blast wave (Section \ref{sec:ex:sedov}),
Taylor-Green vortex (Section \ref{sec:ex:tg})
and triple point expansion (Section \ref{sec:ex:tp}).
Detailed settings for these problems are summarized in
Table \ref{table:setup},
including mesh size, finite element spaces ($\stateFeSpace,\multiplierFeSpace$), 
FOM size ($\stateFOMSize, \multiplierFOMSize$),
the ODE solver, final time ($\finalTime$), 
and the number of time
steps ($N_t$) for FOM runs.
For each considered problem case, the physical quantities
are parametrized with $\param\in\paramDomain$.
Figure~\ref{fig:fom:solution} visualizes the final-time FOM solutions in
various benchmark problems.

In the results that follow we use the $L^2$
error between FOM and ROM solutions to quantify the error introduced by
the different hyper-reduction methods for each problem case.
In addition to the hyper-reduction error, there are two additional
sources of error included in the reported $L^2$ error values:
(1) the projection error introduced by restricting the dynamics to
a linear subspace of the original space;
(2) for predictive simulation runs, the sampling error introduced
when sampling each problem's parameter space.
The reason the $L^2$ FOM-to-ROM error can be employed to perform a fair
comparison between different hyper-reduction methods is that the
projection and sampling errors are constant among the different
hyper-reduction methods for each problem case.
This is because: (1) the same linear subspaces are used across
all simulations within each problem case;
(2) the same parameter values are sampled for training
and used for out-of-sample testing for each problem.
As such, even though the magnitude of the reported FOM-to-ROM errors
includes additional error sources, their relative \textit{difference}
depends only on the employed hyper-reduction methods, allowing
for fair comparisons within each problem case.

\subsection{Nonlinear diffusion}\label{sec:ex:mnd}
We consider the nonlinear diffusion equation in
$\stateDomainSymbol\subset\mathbb{R}^2$
\begin{equation}\label{eq:nonlinear-diffusion}
\begin{aligned}
    \frac{\partial\pressureSymbol}{\partial\timeSymbol}-
        \gradientSymbol\cdot v&=s& &\text{ in }\Omega\times(0, \finalTime)\\
    \gradientSymbol\pressureSymbol-\kappa(\pressureSymbol)^{-1}v&=0&
        &\text{ in }\Omega\times[0,\finalTime]\\
    v\cdot n&=0& &\text{ on }\partial\Omega\times[0,\finalTime]\\
    p&=p_0(\param)& &\text{ in }\Omega\times\{0\}
\end{aligned}
\end{equation}
where $\pressureSymbol$
denotes the pressure,
$\velocitySymbol$ denotes
the flux, 
$s$ denotes an external source function,
{
and $\pressureSymbol_0$ denotes the initial condition on the pressure. 
The system is parametrized through the 
initial condition $\pressureSymbol_0(\param)$.
}

To fit the nonlinear diffusion equation
\eqref{eq:nonlinear-diffusion}
into the general variational framework \eqref{eq:PDE}
introduced in Section \ref{sec:fe}
we identify the following components.
The dependent variables are given by
$\stateSymbol=\pressureSymbol$ and
$\multiplierSymbol=\velocitySymbol$,
with associated vector spaces
$\mathcal{Y}=L^2(\stateDomainSymbol)$,
$\mathcal{Z}=H_0(\text{div};\stateDomainSymbol)
=
\{\psi\in\HilbertSpaceSymbol(\text{div};\stateDomainSymbol)
\colon\psi\cdot n=0\}$ and
$\mathcal{H}=L^2(\stateDomainSymbol)$
with the standard $L^2$ inner product
$(\cdot, \cdot)_{\HilbertSpaceSymbol}$,
leading to the trivial Gelfand triple
$\mathcal{Y}=\mathcal{H}=\mathcal{Y}'$.
The bilinear forms
$a:\stateSpaceSymbol\times\stateSpaceSymbol\to\mathbb{R}$,
$b:\stateSpaceSymbol\times\multiplierSpaceSymbol\to\mathbb{R}$ and
$c:\multiplierSpaceSymbol\times\multiplierSpaceSymbol\to\mathbb{R}$
are defined as
\begin{equation*}
    a(\pressureSymbol,\pressureSymbol')=0\qquad
    b(\pressureSymbol,\velocitySymbol)=
        -\int_\Omega\pressureSymbol(\nabla\cdot\velocitySymbol)dx\qquad
    c(\velocitySymbol,\velocitySymbol')=0
\end{equation*}
for any $\pressureSymbol$, $\pressureSymbol'\in\stateSpaceSymbol$
and $\velocitySymbol$, $\velocitySymbol'\in\multiplierSpaceSymbol$.
Finally, the duality pairing involving the nonlinear functions
{
$\stateNonlinearForceSymbol:\stateSpaceSymbol\times
\multiplierSpaceSymbol\times [0,\finalTime] \times \paramDomain \to\stateSpaceSymbol'$
and
$\multiplierNonlinearForceSymbol:\stateSpaceSymbol\times
\multiplierSpaceSymbol\times [0,\finalTime]\times \paramDomain\to\multiplierSpaceSymbol'$}
are given by
{
\begin{equation*}
    \langle f(\pressureSymbol,\velocitySymbol,\timeSymbol, \param),\pressureSymbol'
        \rangle_{\stateSpaceSymbol'\times\stateSpaceSymbol}=
        \int_\Omega s(t)\pressureSymbol'dx\qquad
    \langle g(\pressureSymbol,\velocitySymbol,\timeSymbol, \param),
        \velocitySymbol'\rangle_{\multiplierSpaceSymbol'\times
        \multiplierSpaceSymbol}=\int_\Omega-\kappa(\pressureSymbol)^{-1}
        v\cdot v'dx
\end{equation*}
}
for any
$(\pressureSymbol,\velocitySymbol,\timeSymbol)\in\stateSpaceSymbol
\times\multiplierSpaceSymbol\times [0,\finalTime]$,
$\pressureSymbol'\in\stateSpaceSymbol$ and
$\velocitySymbol'\in\multiplierSpaceSymbol$.

\begin{figure}[t]
    \centering
    \includegraphics[width=.8\textwidth]{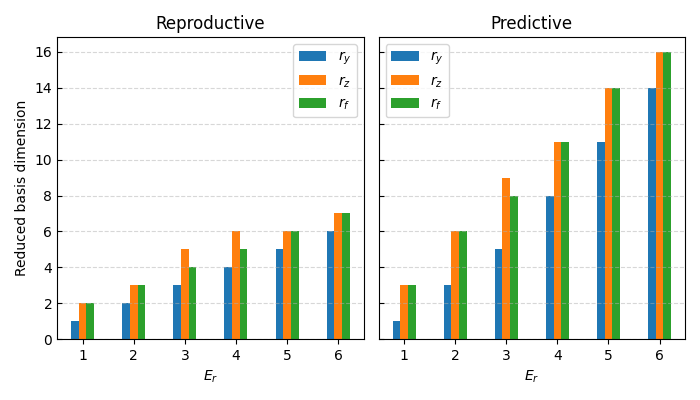}
    \caption{Reduced basis dimensions with respect to the residual
        energy fraction for the nonlinear diffusion simulations.}
    \label{fig:mnd:rdim}
\end{figure}

To generate our numerical results we consider a two dimensional
problem defined on the unit square
$\stateDomainSymbol=[0,1]^2$.
The initial condition for pressure is 
{
parametrized by $\param \in \paramDomain = [0, 0.5]$ through
\begin{equation*}
    p_0(x; \param)=\begin{cases}
        1&\text{if }\max_{i=1,2}\vert x_i-0.5\vert<\param\\
        0&\text{otherwise}
        \end{cases}
\end{equation*}
where $x = (x_1, x_2)\in\stateDomainSymbol$,
}
with conductivity $\kappa(p)=2.0+p$ and source $s = 0$.
We perform the spatial discretization using a mixed finite element
formulation, with the finite element spaces chosen such as to satisfy
the inf-sup condition for stable mixed formulations.
{
In particular, the variables $\stateSymbol = \pressureSymbol$ and $\multiplierSymbol = \velocitySymbol$
are approximated in
$\stateFeSpace\subset L^2(\stateDomainSymbol)$
and 
$\multiplierFeSpace \subset H_0(\text{div};\stateDomainSymbol)$.
We choose the lowest-order Raviart-Thomas elements $\stateFeSpace = \hat{Q}_0$ and $\multiplierFeSpace = \text{RT}_0$,
where $\hat{Q}_0$ denotes the space of piecewise constant scalar functions on the mesh, and $\mathrm{RT}_0$ is the lowest-order Raviart--Thomas space consisting of vector fields whose normal components are continuous across element interfaces.
Backward Euler (BE) method is used for temporal discretization with a fixed time step size $\Delta t = 0.001$ up to final time $\finalTime = 0.1$. 
}

{
Solutions at all the time steps with parameters $\param = 0.15, 0.25, 0.35$ are collected as snapshots
}
{for the predictive case and $\param = 0.3$ for the reproductive case} {in the offline stage. We set $\param = 0.3$ for the online stage in both the reproductive and the predictive case.} {The number of snapshots per training parameter is $N_s^{(y)} = 101$ and $ N_s^{(z)} = 100$.} 
We perform both reproductive and predictive simulation runs, with
a range of residual energy fractions summarized in
Figure \ref{fig:mnd:rdim}.
{In the figure, the basis dimensions satisfying different prescribed energy fractions 
$\zeta_\stateSymbol = \zeta_\multiplierSymbol = 10^{-E_r}$
are reported for each field, with the relation defined in \eqref{eq:energy_criterion}.}

\begin{figure}[t]
    \centering
    \includegraphics[width=.85\textwidth]{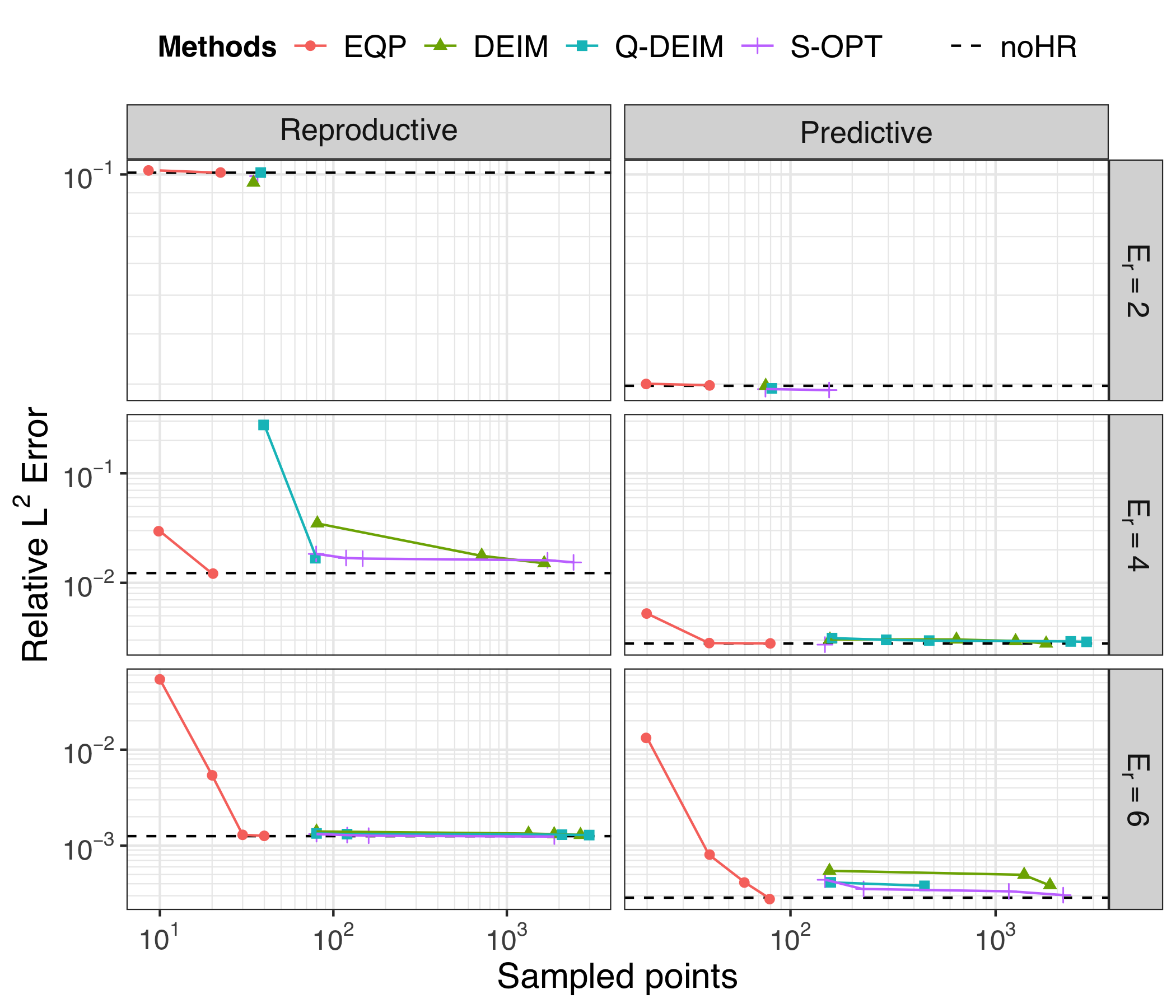}
    \caption{{ROM error with varying sampled points
    for the nonlinear diffusion problem.
    The dashed line indicates the error of the ROM without hyper-reduction (`noHR').}}
    \label{fig:mnd:nqp}
\end{figure}

Figure \ref{fig:mnd:nqp}
{illustrates how the ROM error varies with the number of sampled points for} different
hyper-reduction methods across a range of residual energy fractions.
In every figure panel, the Pareto front between $L^2$ error of the pressure field,
on the vertical axis, and number of sampled points on the horizontal axis,
is shown for the different hyper-reduction methods. 
We note that the Pareto front points are optimal with respect to
computational time and error, not necessarily with respect to
the number of sampled interpolation/quadrature points.
{Every figure panel also includes a dashed reference line indicating the error of the ROM without hyper-reduction for the corresponding residual energy fraction to gauge the accuracy cost of hyper-reduction.}

\begin{figure}[t]
    \centering
    \includegraphics[width=.85\textwidth]{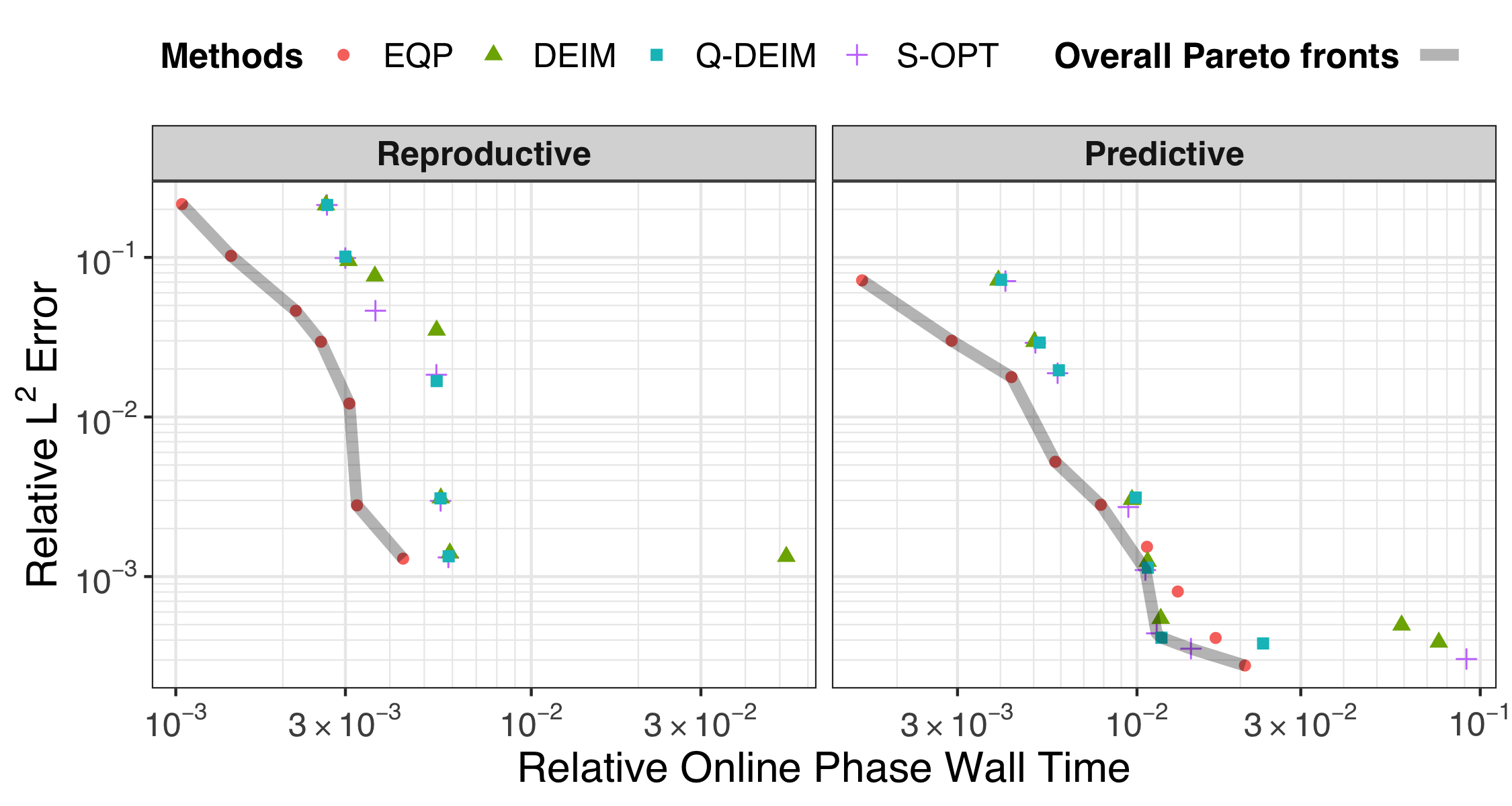}
    \caption{Pareto fronts across all hyper-reduction methods
        for the nonlinear diffusion problem, illustrating the tradeoff
        between relative $L^2$ error and online phase wall time.}
    \label{fig:mnd:pareto}
\end{figure}

{As the residual energy fraction (or equivalently, ROM basis dimensions) increases, the errors of both interpolation methods and EQP converge toward the error of the ROM without hyper-reduction.}
At $E_r=6$ all methods in the reproductive case reach a relative error
of about $10^{-3}$,
whereas for the predictive case the lowest relative error is
approximately $10^{-4}$.
In both reproductive and predictive runs the error levels of the
interpolation methods seems to correlate with $E_r$
and not with the number of sampled interpolation points.
On the contrary, the EQP error levels demonstrate a tendency to
decrease as the number of quadrature points increases,
especially when $E_r$ is large. 
The lowest error levels obtained by each method are approximately
of the same order of magnitude for this problem case.
However, the EQP method seems to generally require a smaller number
of quadrature points to reach a given error level.

The tradeoff between computational cost and relative error is made
explicit in Figure \ref{fig:mnd:pareto},
which depicts the Pareto fronts obtained for each hyper-reduction
method among all values of $E_r$. {In the figure, the relative $L^2$-error between ROM and FOM is plotted on the vertical axis and the relative wall time, i.e. the fraction between ROM wall time and FOM wall time for the online phase, is plotted on the horizontal axis.}
The overall Pareto front is drawn as a line connecting the most efficient
configurations across all methods.
In the reproductive case the overall Pareto front consists entirely of
EQP method simulations.
The lowest error level achieved is about $10^{-3}$,
at which point EQP records a relative online time of
approximately $4\times 10^{-3}$,
while the interpolation methods record about
$6\times 10^{-3}$.
In the predictive scenario, although EQP shows greater
efficiency for higher error levels, a tradeoff appears as the
error levels decrease. 
More specifically, at error levels below $10^{-3}$
or when the relative online time exceeds $10^{-2}$,
the interpolation methods emerge on the overall Pareto front.

To summarize, for the nonlinear diffusion problem, the Pareto optimal
method is EQP in the reproductive case as well as in the predictive
case for error levels above $10^{-3}$.
For lower predictive error levels, interpolation methods
also appear on the Pareto front,
with no method consistently outperforming the others.
The observed lower error levels in the predictive scenario compared
to the reproductive one could be explained by the larger number of
basis vectors retained for the same energy residual as shown in
Figure \ref{fig:mnd:rdim}.
Furthermore, the samples taken in the offline stage seem to
describe the system conditions tested in the online stage well.
If one were to sample $\mu$ further away from its online stage value,
then the obtained error would be expected to rise.

\subsection{Nonlinear elasticity}\label{sec:ex:nlel}
We consider the dynamics of a visco-hyperelastic solid of homogeneous neo-hookean material {undergoing finite strains, i.e. with both geometric and material nonlinearities}.
The state of the solid is described as a function of time
$\timeSymbol$ and reference {geometry}
$\initialPosition\in\initialDomain=\stateDomainSymbol(0)$
at initial time $\timeSymbol=0$.
The system evolves according to the equations
\begin{equation}\label{eq:nonlinear-elasticity}
\begin{aligned}
    \rho\dfrac{d\velocitySymbol}{d\timeSymbol}&=
        {\nabla\cdot\mathbf{P}\left(\positionSymbol\right)}+\eta\Delta\velocitySymbol & &\text{ in }
        \initialDomain\times(0,\finalTime)\\
    \frac{d\positionSymbol}{d\timeSymbol}&=\velocitySymbol & &\text{ in }
        \initialDomain\times(0,\finalTime)\\ 
    \velocitySymbol &=0 & &\text{ on }\Gamma_{\text{in}}
        \times[0,\finalTime]\\
    \normalSymbol\cdot\nabla\velocitySymbol &=0 & &\text{ on }
        (\partial\initialDomain\setminus\Gamma_{\text{in}})
        \times[0,\finalTime] \\
        \velocitySymbol &=\velocitySymbol_0(\param) & & \text{ in } \initialDomain \times \{0\} \\
        \positionSymbol &=0 & & \text{ in } \initialDomain \times \{0\} \\
\end{aligned}
\end{equation}
where $\positionSymbol$ and $\velocitySymbol$ denote the
solid's deformation and velocity fields,
while $\rho$ and $\eta$ are the constant reference density and
viscosity of the material in its initial configuration. 
In addition, $d/d\timeSymbol$ denotes the material derivative
and $\normalSymbol$ the normal vector on the boundary
$\partial\initialDomain$;
the spatial differential operators act with respect to the
reference coordinates $\initialPosition$.
{The divergence of the first Piola-Kirchhoff stress tensor can be written as a nonlinear function
$\nabla\cdot\mathbf{P}\left(\positionSymbol\right) =H(\positionSymbol)$, $H\colon H^1_{0,\Gamma_{\text{in}}}(\initialDomain)\to
H^{-1}_{\Gamma_{\text{in}}}(\initialDomain)$.}
{Finally, $\velocitySymbol_0$ denotes the initial condition on the velocity. 
The system is parametrized through the 
initial condition $\velocity_0(\param)$.}

To write the nonlinear elasticity equations
\eqref{eq:nonlinear-elasticity}
in the general variational form \eqref{eq:PDE}
we make the following identifications.
We denote the differential variable
$\stateSymbol=(\velocitySymbol,\positionSymbol)$, 
and define the space
$\mathcal{Y}=[H^1(\initialDomain)]^\dimensionSymbol\times
[H^1(\initialDomain)]^\dimensionSymbol$
and Hilbert space
$\HilbertSpaceSymbol=[L^2(\initialDomain)]^\dimensionSymbol
\times[L^2(\initialDomain)]^\dimensionSymbol$.
There is no algebraic variable $\multiplierSymbol$
in this problem case.
The inner product
$(\cdot,\cdot)_{\HilbertSpaceSymbol}\colon
\HilbertSpaceSymbol\times\HilbertSpaceSymbol\to\mathbb{R}$
and bilinear form
$a\colon\mathcal{Y}\times\mathcal{Y}\to\mathbb{R}$
are given by
\begin{equation*}
    ((\velocitySymbol,\positionSymbol),\left(\velocitySymbol',
        \positionSymbol'\right))_{\HilbertSpaceSymbol}
        =\int_{\initialDomain}\rho\velocitySymbol\cdot\velocitySymbol'
        +\positionSymbol\cdot\positionSymbol'd\initialPosition\qquad
    a((\velocitySymbol,\positionSymbol),
        (\velocitySymbol',\positionSymbol'))
        =\int_{\initialDomain}\eta\nabla\velocitySymbol:\nabla
        \velocitySymbol'-\velocitySymbol\cdot\positionSymbol'
        d\initialPosition.
\end{equation*}
Finally, the nonlinear function
{
$\stateNonlinearForceSymbol\colon\mathcal{Y}\times[0,\finalTime] \times \paramDomain
\to\mathcal{Y}'$
is defined by its action
\begin{equation*}
    \langle\stateNonlinearForceSymbol(\velocitySymbol,\positionSymbol,
        \timeSymbol, \param),(\velocitySymbol',\positionSymbol')
        \rangle_{\mathcal{Y}'\times\mathcal{Y}}
        =\int_{\initialDomain}-\mathbf{P}(\positionSymbol):\nabla
        \velocitySymbol'd\initialPosition
\end{equation*}
}
where $\stateSpaceSymbol'$ denotes the dual of
$\stateSpaceSymbol$ with respect to
$(\cdot,\cdot)_{\HilbertSpaceSymbol}$.

\begin{figure}[t]
    \centering
    \includegraphics[width=.8\textwidth]{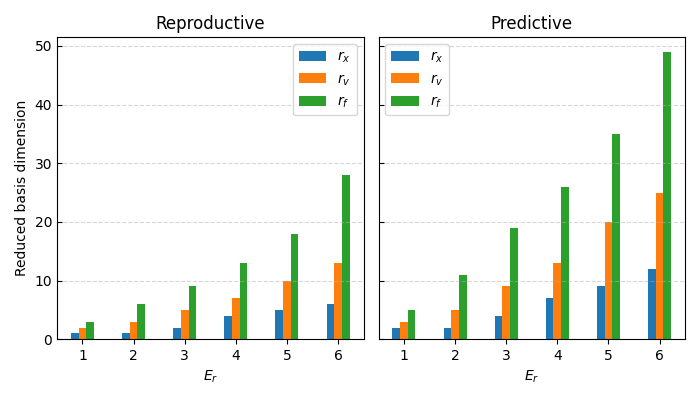}
    \caption{Reduced basis dimensions used for each residual energy
        fraction in the nonlinear elasticity simulations.}
    \label{fig:nlel:rdim}
\end{figure}

Our numerical results are generated for a two dimensional problem
defined on the rectangular domain
$\initialDomain=[0,8]\times[0,1]$.
The velocity initial condition is parametrized by 
{
$\param\in\paramDomain = [0.5, 2]$ through
\begin{equation*}
    \velocitySymbol_0(\initialPosition,0; \param)=\left(0, -\dfrac{\param}{80}\sin(\param \initialPosition_1)\right),
\end{equation*}
with $\initialPosition=(\initialPosition_1,\initialPosition_2)\in\initialDomain$.
}
The stress tensor $\mathbf{P}$
is derived from a strain energy density function $\Psi$ that involves
the shear modulus $\nu$ and bulk modulus $K$ of the material and depends
nonlinearly on the deformation gradient $\bm{\jacobianSymbol}$
and its determinant
$\jacobianSymbol=\vert\bm{\jacobianSymbol}\vert$, assuming finite strains,
\begin{equation*}
    \Psi(\bm{\jacobianSymbol})=\frac{\nu}{2}
        (\bar{I}_1 - 2)+\frac{K}{2}(J/g-1)^2
\end{equation*}
where $g$ is a reference volumetric scaling and
$\bar{I}_1 = \operatorname{tr}(\bm{\jacobianSymbol}^\top
\bm{\jacobianSymbol})/J$,
yielding the stress tensor
\begin{equation*}
\bm{P} = \frac{\nu}{ \jacobianSymbol}\cdot\bm{\jacobianSymbol} 
    +\left(\frac{KJ(\jacobianSymbol - g)}{g^2}
    -\frac{\nu\operatorname{tr}(\bm{\jacobianSymbol}^\top
    \bm{\jacobianSymbol})}{2\jacobianSymbol}\right)
    \bm{\jacobianSymbol}^{-\top}.
\end{equation*}

\begin{figure}[t]
    \centering
    \includegraphics[width=0.85\textwidth]{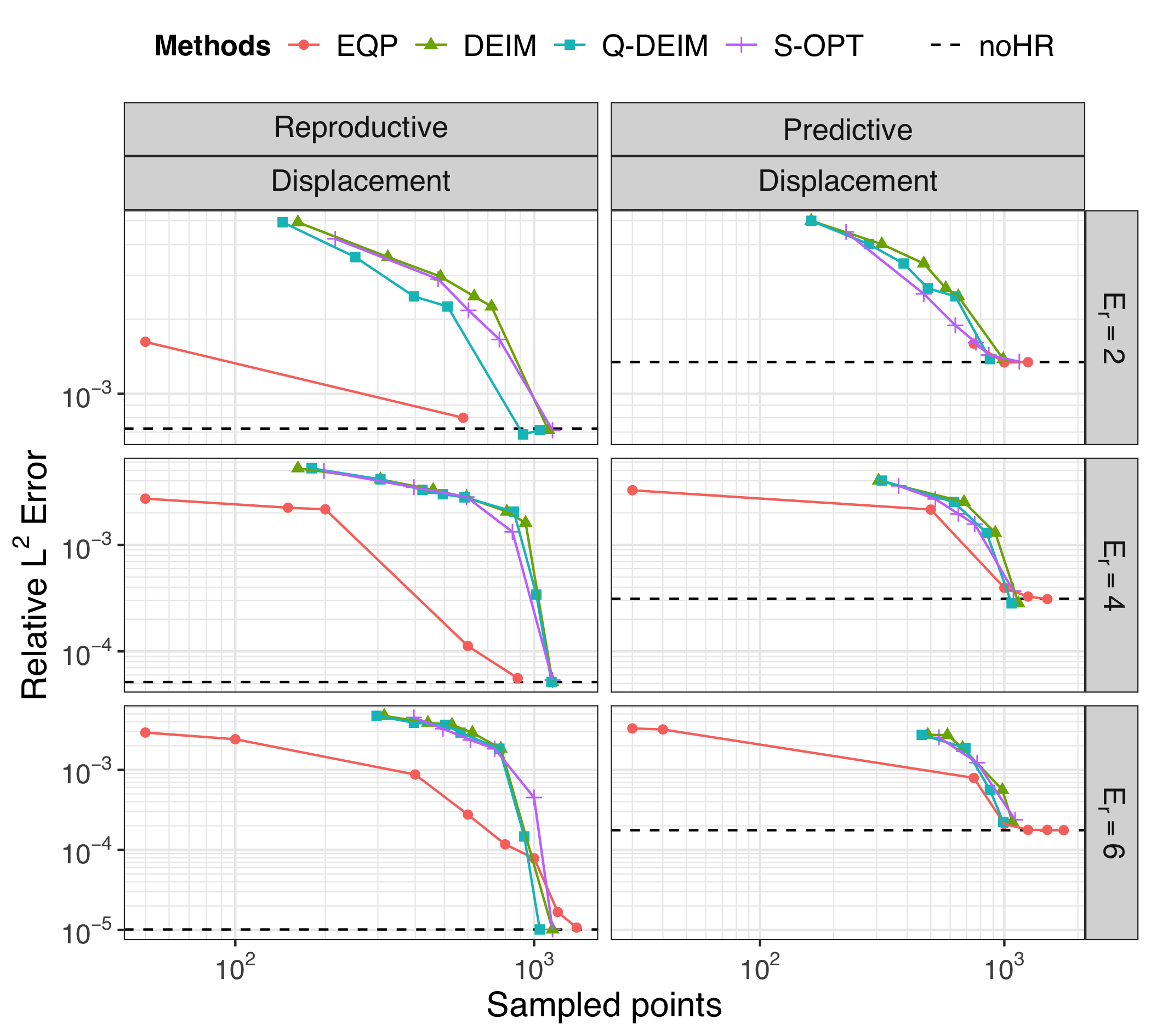}
    \caption{
    {ROM displacement error with varying sampled points
    for the nonlinear elasticity problem.}}
    \label{fig:nlel:nqp:pos}
\end{figure}

For the spatial discretization of the nonlinear elasticity equations
we use a single finite element space
$\kinematicFE\subset[H^1(\initialDomain)]^\dimensionSymbol$ of dimension $\sizeKinematicFE$
to approximate the {displacement} and velocity fields. {
Therefore, the differential variable $\stateSymbol = (\velocitySymbol, \positionSymbol)$ is approximated in
$\stateFeSpace = \kinematicFE \times \kinematicFE \subset \stateSpaceSymbol = [H^1(\initialDomain)]^\dimensionSymbol \times [H^1(\initialDomain)]^\dimensionSymbol$.
We choose $\kinematicFE = [Q_2]^\dimensionSymbol$, 
where $Q_2$ denotes the space of continuous piecewise tensor-product polynomials of degree at most $2$ in each coordinate direction, and $[Q_2]^\dimensionSymbol$ is its $\dimensionSymbol$-component vector-valued counterpart.
Therefore $\stateFeSpace = [Q_2]^\dimensionSymbol \times [Q_2]^\dimensionSymbol$.
Fourth-order explicit Runge-Kutta (RK4) method is used for temporal discretization with a fixed time step size $\Delta t = 0.01$ up to final time $\finalTime = 5.0$.} 

{
Solutions and intermediate Runge-Kutta stages at all the time steps with parameters $\param = 0.9, 1.1$ are collected as snapshots} {for the predictive case and $\param = 1.0$ for the reproductive case } {in the offline stage. In both cases we set $\param = 1.0$ for the online stage.} {The number of snapshots per training parameter is $N_s^{(y)} = 501$.}
{Rather than constructing a single POD basis for the product space as described in Section~\ref{subsec:offline}, we exploit its natural decomposition and compute independent reduced bases for each state component. 
}
We partition the differential variable $\stateSymbol$
into the velocity $\velocitySymbol$
and {displacement} $\positionSymbol$
variables of equal full dimension.
{The reduced bases for the two variables are constructed separately from samples of FOM displacements and velocities. The dimensions of the reduced variables are denoted $r_x$ for the reduced displacement dimension and $r_v$ for the reduced velocity dimension.}
{
More precisely, the finite element representations of the velocity and displacement snapshots are assembled into two snapshot matrices,
$\mathbf{S}_{\velocitySymbol}, 
\mathbf{S}_{\positionSymbol} \in \mathbb{R}^{\sizeKinematicFE \times N_s}$, 
and a thin singular value decomposition is performed separately on each matrix. The reduced basis dimensions are selected independently using the energy criterion~\eqref{eq:energy_criterion}, yielding the basis matrices
$\velocityBasis \in \mathbb{R}^{\sizeKinematicFE \times r_v},\positionBasis \in \mathbb{R}^{\sizeKinematicFE \times r_x}$. 
The overall differential variable ROM subspace is then defined as the direct product of the corresponding reduced subspaces. Equivalently, the resulting basis functions are supported on exactly one component of the product space, i.e., 
\begin{equation}
    \stateROMBasisMat = \begin{bmatrix}
        \velocityBasis & \mathbf{0} \\
        \mathbf{0} & \positionBasis
    \end{bmatrix} \in \mathbb{R}^{\stateFOMSize \times \stateROMSize}.
\end{equation}
}
The overall reduced state dimension
$\stateROMSize=r_x+r_v$
is then given by the sum of the two reduced dimensions.
The reduced basis dimensions obtained for each tested residual energy
fraction are shown in Figure \ref{fig:nlel:rdim}.

\begin{figure}[t]
    \includegraphics[width=0.85\textwidth]{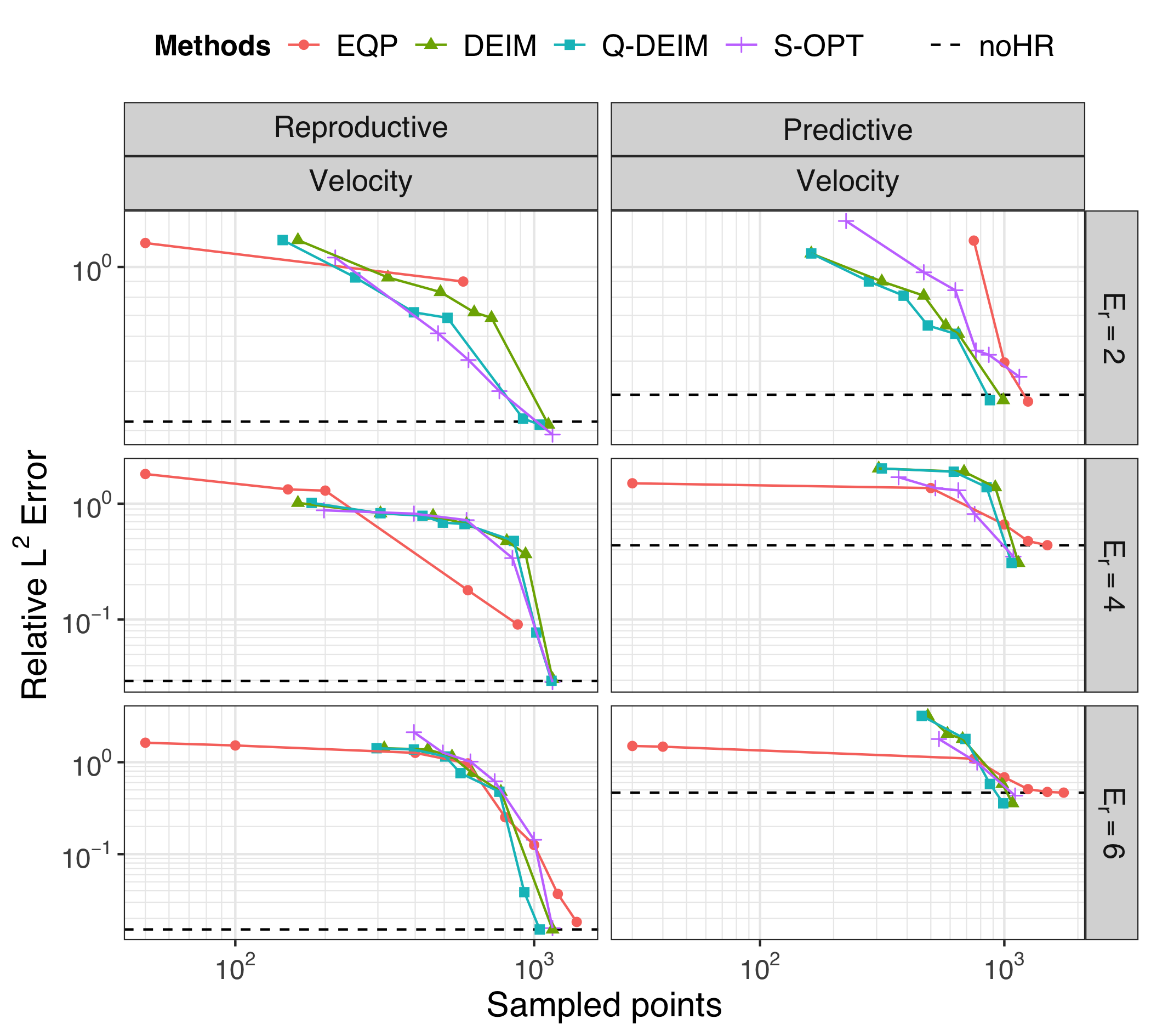}
    \caption{{ROM velocity error with varying sampled points
    for the nonlinear elasticity problem.}}
    \label{fig:nlel:nqp:vel}
\end{figure}

Figures \ref{fig:nlel:nqp:pos} and \ref{fig:nlel:nqp:vel}
present an analogous comparison to that in
Figure \ref{fig:mnd:nqp}, with the {displacement} and velocity error results
shown separately.
The plotted points represent Pareto-optimal configurations for each
hyper-reduction method, where 
{Pareto optimality is determined with respect to computational time and the combined} 
$L^2$-product norm of the error in the state
$(\velocitySymbol, \positionSymbol)$.
The results demonstrate that for all methods the relative errors
decrease as the number of interpolation/quadrature points
increases.
We note that the error levels for {displacement} are noticeably
smaller than those for velocity.
For {displacement} errors, in both the reproductive and predictive scenarios,
EQP generally attains a given error level with fewer quadrature points than the interpolation methods. 
{However, as the error approaches that of the ROM without hyper-reduction, the performance gap diminishes, and the interpolation methods may achieve comparable or even better efficiency.
The interpolation methods exhibit relatively similar performance among themselves, and, with sufficiently many sampled points, all method converge toward the error of ROM without hyper-reduction.
For the velocity field, interpolation methods reach the error level of ROM without hyper-reduction across all-scenarios, whereas EQP fails to do so in the reproductive scenario at low residual error fractions.
In all cases, interpolation methods outperform EQP near the error level of the ROM without hyper-reduction.
Overall, EQP tends to be more efficient at relatively high error levels, whereas interpolation methods become more efficient as higher accuracy is sought.}

\begin{figure}[t]
    \centering
\centering
    \includegraphics[width=0.85\textwidth]{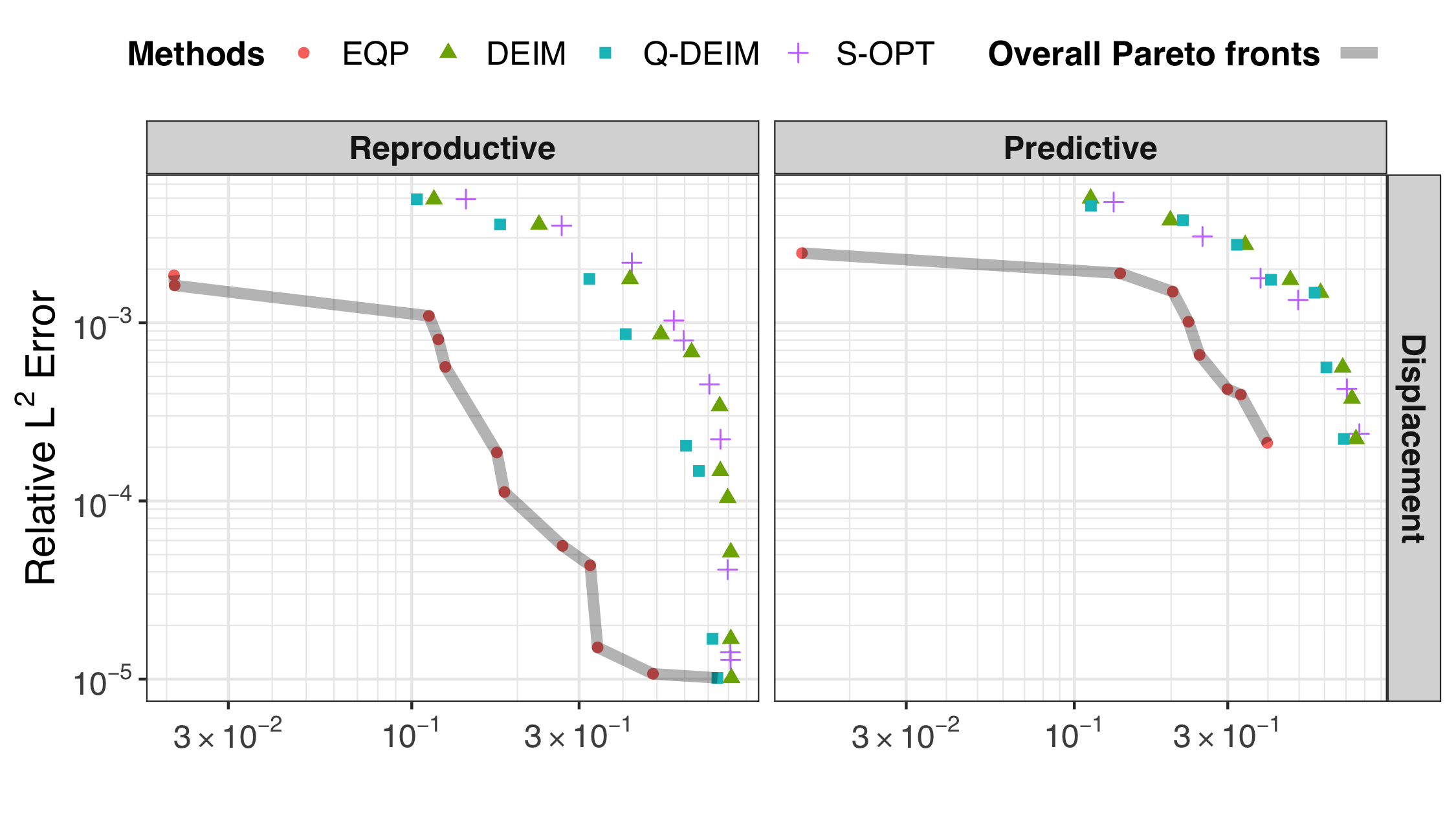}

    \vspace{0.5em}

    \includegraphics[width=0.85\textwidth]{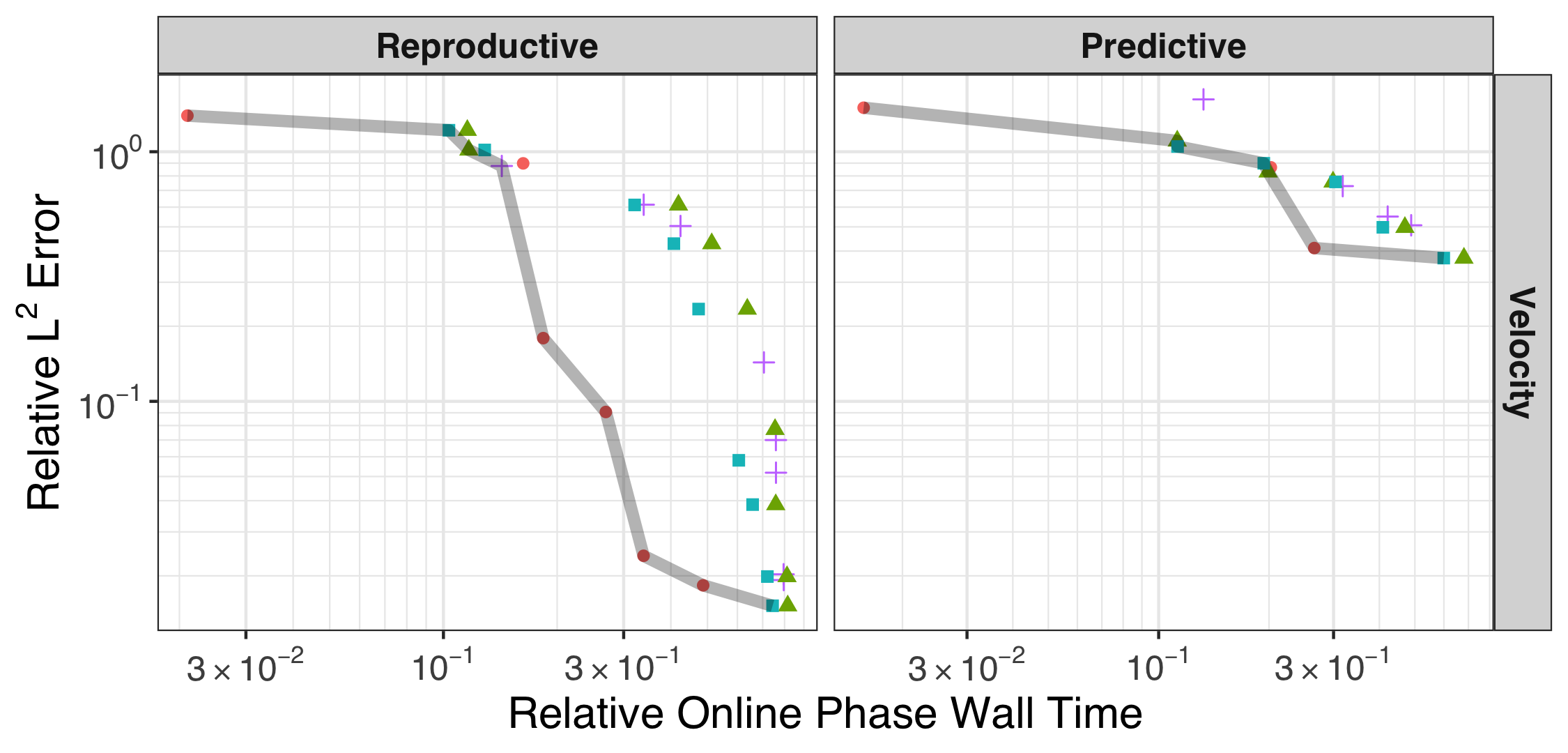}

    \caption{Pareto fronts for displacement (top) and velocity (bottom) across all hyper-reduction methods in the nonlinear elasticity problem.}
    \label{fig:nlel:pareto}

\end{figure}

Figure \ref{fig:nlel:pareto}
displays the Pareto fronts obtained for each hyper-reduction method with respect to the {displacement and velocity} state errors
and relative online wall time.
{
For the displacement field, EQP composes the overall Pareto front across most error ranges in both scenarios, with Q-DEIM becoming Pareto optimal at the lowest error levels. The lowest attained relative error is approximately $10^{-5}$ in the reproductive scenario and $2 \cdot 10^{-4}$ in the predictive scenario. For the velocity field, EQP similarly dominates the Pareto front in the reproductive scenario,} {except near $10^0$ error where interpolation methods are the highest performing, and at the lowest error levels where Q-DEIM and DEIM are Pareto optimal.
Q-DEIM achieves the lowest error slightly above $10^{-2}$ at a
relative online time of approximately $7\times 10^{-1}$.}
{In the predictive scenario for the velocity field, no single method consistently outperforms the others. The lowest attained error is approximately $4\times 10^{-1}$ at relative online time $6\times 10^{-1}$, again by Q-DEIM.
While no single method consistently dominates for the velocity field, EQP dominates the overall Pareto front for displacement errors across both scenarios, making it the more attractive choice when displacement accuracy is the primary concern.}

{
The Pareto fronts highlight the tradeoff between computational efficiency and accuracy. For the velocity field, the reproductive scenario allows accuracy to be improved progressively as computational efficiency is sacrificed, from relative errors above 10\% at relative online time below 30\% to about 1\% at relative online time above 50\%.
In contrast, the predictive scenario cannot reduce the relative error below approximately 40\%, even at the highest computational cost.
}
{On the other hand, in the case of structural dynamics the displacement field can be of greater interest, and in this case a small relative error is achieved with the EQP at a significantly higher speedup. For instance, a 0.1\% relative error can with the EQP method be achieved with approximately 90\% speedup in the reproductive case and 80\% speedup in the predictive case.}

\subsection{Lagrangian hydrodynamics}\label{sec:ex:laghos}
We consider the hydrodynamics of a compressible fluid governed
by the Euler equations in a Lagrangian reference frame.
Similarly to the nonlinear elasticity case,
the medium is described as a function of time
$\timeSymbol$ and reference coordinates of the material particles
$\initialPosition\in\initialDomain=\stateDomainSymbol(0)$
at initial time $\timeSymbol=0$.
In the Lagrangian description of the medium's motion, the physical
quantities are expressed as a function of its initial position
$\initialPosition$ and time $\timeSymbol$.
The system of the Euler equations of compressible fluid dynamics in a
Lagrangian reference frame is written as follows
\cite{Aris1989}
\begin{equation}\label{eq:euler}
\begin{aligned}
    \text{momentum conservation}:& &\densitySymbol
        \frac{d\velocitySymbol}{d\timeSymbol}&=
        \gradientSymbol\cdot\stressSymbol\\
    \text{mass conservation}:& &\dfrac{1}{\densitySymbol}
        \frac{d\densitySymbol}{d\timeSymbol}&=
        -\gradientSymbol\cdot\velocitySymbol\\
    \text{energy conservation}:& &\densitySymbol
        \frac{d\energySymbol}{d\timeSymbol}&=
        \stressSymbol:\gradientSymbol\velocitySymbol\\
    \text{equation of motion}:& &\frac{d\positionSymbol}{d\timeSymbol}
        &=\velocitySymbol
\end{aligned}
\end{equation}
where $d/d\timeSymbol$ denotes the material derivative,
$\densitySymbol$ the density of the fluid,
$\positionSymbol$ and $\velocitySymbol$ the position and velocity of
the particles, 
$\energySymbol$ the internal energy per unit mass and
$\stressSymbol$ the deformation stress tensor. 
In the equations, the spatial differential operators act with respect
to the current coordinates $\positionSymbol$. 
We consider an isotropic stress tensor
$\stressSymbol=-\pressureSymbol\identitySymbol+\artificialStressSymbol$
where $\pressureSymbol$ denotes the thermodynamic pressure and
$\artificialStressSymbol$ the artificial viscosity stress,
both of which depend on the differential variables nonlinearly. 
The thermodynamic pressure is described by the equation of
state and can be expressed as a function of the density and internal
energy.
In what follows, we focus on the case of polytropic ideal gas with
an adiabatic index $\adiabaticIndexSymbol>1$,
which yields the equation of state 
\begin{equation}\label{eq:EOS}
    \pressureSymbol=(\adiabaticIndexSymbol-1)\densitySymbol\energySymbol.
\end{equation}
The system is prescribed with an initial condition and a boundary condition
$\normalSymbol\cdot\velocitySymbol=\neumannSymbol$,
where $\normalSymbol$ is the outward normal unit vector on the
domain boundary.

The semidiscrete Lagrangian formulation involves a 
computational mesh that deforms based on the fluid velocity. 
Using the Reynolds transport theorem
\cite{Aris1989}, 
the mass conservation in \eqref{eq:euler}
is satisfied by 
\begin{equation}
    \densitySymbol(\initialPosition,\timeSymbol)=
        \frac{\densitySymbol(\initialPosition,0)}
        {\vert\jacobianSymbol(\initialPosition,\timeSymbol)\vert}
\end{equation}
where $\jacobianSymbol=\nabla_{\initialPosition}\positionSymbol$ 
denotes the Jacobian of the deformation gradient. 
Moreover, any integrals over the current domain
$\stateDomainSymbol(\timeSymbol)$
involve the transformation to the reference domain
$\initialDomain$
using the determinant of the transformation gradient
$\vert\jacobianSymbol\vert$.

To write the remaining equations in \eqref{eq:euler}
in the variational form \eqref{eq:PDE}
we define the following quantities.
We denote the differential variable by
$\stateSymbol=(\velocitySymbol,\energySymbol,\positionSymbol)$
and define the vector space
$\mathcal{Y}=[H^1(\initialDomain)]^\dimensionSymbol\times
L^2(\initialDomain)\times[H^1(\initialDomain)]^\dimensionSymbol$.
There is no algebraic variable $\multiplierSymbol$.
We define the Hilbert space
$\HilbertSpaceSymbol=[L^2(\initialDomain)]^\dimensionSymbol\times
L^2(\initialDomain)\times[L^2(\initialDomain)]^\dimensionSymbol$
with inner product
$(\cdot,\cdot)_{\HilbertSpaceSymbol}\colon
\HilbertSpaceSymbol\times\HilbertSpaceSymbol\to\mathbb{R}$
\begin{equation*}
    ((\velocitySymbol,\energySymbol,\positionSymbol),\left(\velocitySymbol',
        \energySymbol',\positionSymbol'\right))_{\HilbertSpaceSymbol}
        =\int_{\stateDomainSymbol(\timeSymbol)}\densitySymbol
        \velocitySymbol\velocitySymbol'+\densitySymbol\energySymbol
        \energySymbol'+\positionSymbol\positionSymbol'dx
\end{equation*}
and denoted by $\stateSpaceSymbol'$ its dual with respect to
$(\cdot,\cdot)_{\HilbertSpaceSymbol}$.
The bilinear form
$a\colon\stateSpaceSymbol\times\stateSpaceSymbol\to\mathbb{R}$
is given by
\begin{equation*}
    a((\velocitySymbol,\energySymbol,\positionSymbol),
        (\velocitySymbol',\energySymbol',\positionSymbol'))
        =-\int_{\stateDomainSymbol(\timeSymbol)}\velocitySymbol\,
        \positionSymbol'dx
\end{equation*}
while the action of the nonlinear function
$\stateNonlinearForceSymbol\colon\stateSpaceSymbol\times
[0,\finalTime]\to\stateSpaceSymbol'$
by
\begin{equation*}
    \langle\stateNonlinearForceSymbol
        (\velocitySymbol,\energySymbol,\positionSymbol, \timeSymbol), 
        (\velocitySymbol',\energySymbol',\positionSymbol')
        \rangle_{\stateSpaceSymbol'\times\stateSpaceSymbol}
        =\int_{\stateDomainSymbol(\timeSymbol)}
        -\stressSymbol(\velocitySymbol,\energySymbol,\positionSymbol):
        \nabla\velocitySymbol'+(\stressSymbol(\velocitySymbol,\energySymbol,
        \positionSymbol):\nabla\velocitySymbol)\energySymbol'dx.
\end{equation*}

Following \cite{Dobrev2012}
we adopt a spatial discretization for \eqref{eq:euler}
using a kinematic space
$\kinematicFE \subset [H^1(\initialDomain)]^\dimensionSymbol$ of dimension $\sizeKinematicFE$
for approximating the position and velocity functions,
and a thermodynamic space
$\thermodynamicFE \subset L^2(\initialDomain)$
of dimension $\sizeThermodynamicFE$ for approximating the energy.
The density can be eliminated, and the equation of mass conservation
can be decoupled from \eqref{eq:euler}.
{
Therefore, the differential variable $\stateSymbol = (\velocitySymbol, \energySymbol, \positionSymbol)$ is approximated in
$\stateFeSpace = \kinematicFE \times \thermodynamicFE \times \kinematicFE \subset \stateSpaceSymbol = [H^1(\initialDomain)]^\dimensionSymbol \times L^2(\initialDomain) \times [H^1(\initialDomain)]^\dimensionSymbol$.
We choose $\kinematicFE = [Q_2]^\dimensionSymbol$ and $\thermodynamicFE = \hat{Q}_1$, 
where $Q_2$ denotes the space of continuous piecewise tensor-product polynomials of degree at most $2$ in each coordinate direction, so that $[Q_2]^\dimensionSymbol$ is the corresponding $\dimensionSymbol$-component vector-valued space, and $\hat{Q}_1$ denotes the space of discontinuous piecewise tensor-product polynomials of degree at most $1$ in each coordinate direction.
Therefore, $\stateFeSpace = [Q_2]^\dimensionSymbol \times \hat{Q}_1 \times [Q_2]^\dimensionSymbol$.
Each problem case is time integrated using two methods: 
the averaged second-order explicit Runge Kutta (RK2Avg) method
for conserving the total energy \cite{Dobrev2012,sandu2021conservative} and 
the standard fourth-order explicit Runge Kutta (RK4) method. 
In either method, adaptive time step control is employed for numerical stability. 
}
We refer the reader to \cite{Dobrev2012}
for additional details.

We perform numerical simulations of three problem cases:
Sedov blast wave, Taylor-Green vortex and triple point expansion,
introduced in the following sections.
For each problem we focus on reproductive
simulations, where the constructed reduced models are tested using
initial conditions that were part of their training data.

{
Solutions and intermediate Runge-Kutta stages at all the time steps and parameters are collected as snapshots.
We make two modifications to the offline basis construction procedure outlined in Section~\ref{subsec:offline}.
First, we employ the time windowing technique,
where the simulation time interval $[0, \finalTime]$ is decomposed into a number $N_w$ of
time sub-intervals (windows)
\begin{equation}
    0 = \tau_0 < \tau_1 < \ldots < \tau_{N_w} = \finalTime.
\end{equation}
Instead of building a single ROM for the whole time interval $[0,\finalTime]$, multiple ROMs are constructed for piecewise approximating the local dynamics.
In practice, the number of time windows $N_w$ used is determined by prescribing the number of snapshots to be collected per window.
The snapshots are clustered into groups by the time they are simulation into the corresponding time window. 
Second, as in the previous example, we assemble finite element representations of the velocity, internal energy, and mesh position snapshots for the time window $[\tau_{i-1}, \tau_{i}]$ into three snapshot matrices,
$\mathbf{S}_{\velocitySymbol}^{(i)} \in \mathbb{R}^{\sizeKinematicFE \times N_{s}^{(\stateSymbol,i)}}, 
\mathbf{S}_{\energySymbol}^{(i)} \in \mathbb{R}^{\sizeThermodynamicFE \times N_{s}^{(\stateSymbol,i)}},
\mathbf{S}_{\positionSymbol}^{(i)} \in \mathbb{R}^{\sizeKinematicFE \times N_{s}^{(\stateSymbol,i)}}$, 
and create the basis matrices
$\velocityBasis^{(i)} \in \mathbb{R}^{\sizeKinematicFE \times r_{\velocitySymbol}^{(i)}},\energyBasis^{(i)} \in \mathbb{R}^{\sizeThermodynamicFE \times r_{\energySymbol}^{(i)}},\positionBasis^{(i)} \in \mathbb{R}^{\sizeKinematicFE \times r_{\positionSymbol}^{(i)}}$
which defines the product ROM basis matrix by 
\begin{equation}
    \stateROMBasisMat^{(i)} = \begin{bmatrix}
        \velocityBasis^{(i)} & \mathbf{0} & \mathbf{0} \\
        \mathbf{0} & \energyBasis^{(i)} & \mathbf{0} \\
        \mathbf{0} & \mathbf{0} & \positionBasis^{(i)}
    \end{bmatrix} \in \mathbb{R}^{\stateFOMSize \times \stateROMSize^{(i)}},
\end{equation}
where $\stateROMSize^{(i)} = r_{\velocitySymbol}^{(i)} + r_{\energySymbol}^{(i)} + r_{\positionSymbol}^{(i)}$. 
This separate basis construction allows flexible separate state update necessary in the RK2Avg scheme.
}
For each problem, the number of snapshots and residual energy fraction
$E_r$ are such as to ensure an adequate basis size for each window.
Finally, we also employ an adaptive timestepping strategy to dynamically
adjust the timesteps during each simulation
\cite{Dobrev2012,Copeland2022}.

\begin{table}[t]
\centering \setstretch{1.2}
\caption{Parameters used for the reduced simulations of Lagrangian
    hydrodynamics problems.}
\label{table:setup:laghos}
\begin{tabular}{c|c|c|c|c|c|c}
{Problem} & {$E_r$} & {ODE solver} & {$N_w$}  & {$N_s^{(y)}$}& {$r_v$}& {$r_e$}\\
\hline
\multirow{2}{*}{Sedov blast wave} & \multirow{2}{*}{6} & RK4 & 15 & 1451& 118 & 37 \\ 
& & RK2Avg & 15 & 754& 151 & 40 \\ 
\hline
\multirow{2}{*}{Taylor-Green vortex} & \multirow{2}{*}{6}  & RK4 & 2 & 1064& 8 & 22\\
& & RK2Avg & 2& 532 & 8 & 23\\
\hline
\multirow{2}{*}{Triple point expansion} & \multirow{2}{*}{6} & RK4 & 2 & 379 & 11 & 10 \\ 
& & RK2Avg & 2 & 190& 12 & 10 \\ 
\end{tabular}
\end{table}

The parameters used for the reduced simulations of each problem are
summarized in Table \ref{table:setup:laghos},
including the number of time windows ($N_w$) {and the total number of snapshots ($N_s^{(y)}$)
and reduced basis dimensions ($r_v$ and $r_e$) summed over all time windows, i.e., $r_v = \sum_{i=1}^{N_w} r_v^{(i)}$ and $r_v = \sum_{i=1}^{N_w} r_e^{(i)}$}
A detailed development of the reduction process using interpolation
hyper-reduction methods is given in \cite{Copeland2022}
and using quadrature methods in \cite{Vales2025ceqp}.

\subsubsection{Sedov blast wave}\label{sec:ex:sedov}
The Sedov blast wave problem is a three dimensional standard shock
hydrodynamic test, involving the release of energy from a localized
source and the formation of a propagating shock wave
\cite{sedov1993similarity}.
We consider a localized source of internal energy initially deposited
at the origin of a three dimensional cube,
in analogy to the two-dimensional test in \cite{Dobrev2012}.
The computational domain is the unit cube
$\initialDomain = [0,1]^3$
with wall boundary conditions on all surfaces,
$\velocitySymbol\cdot\normalSymbol=0$.
We prescribe initial velocity $\velocitySymbol=0$
and initial density $\densitySymbol=1$.
The initial energy is given by a delta function supported at the origin.
In our implementation, the delta function energy source is approximated
by setting the internal energy to zero in all degrees of freedom except
at the origin, with $\energySymbol(0,0,0)=0.25$.
The adiabatic index in the ideal gas equation of state is set to
$\adiabaticIndexSymbol=1.4$. 

\begin{figure}[t]
    \centering
    \includegraphics[width=0.85\textwidth]{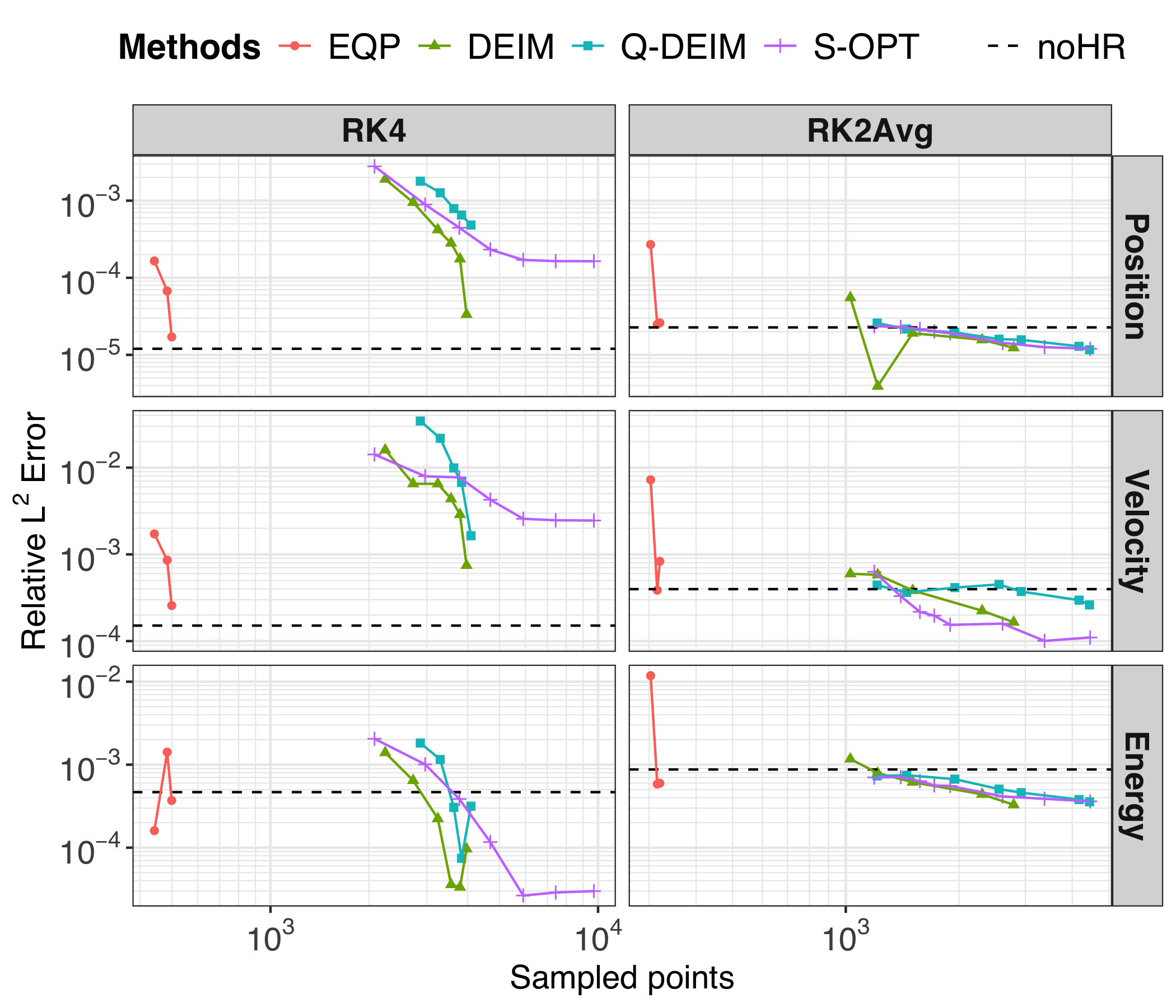}
    \caption{
    {ROM error with varying sampled points
    for the Sedov blast wave problem.}
    }
    \label{fig:sb:nqp}
\end{figure}

Figure \ref{fig:sb:nqp}
illustrates {how the ROM error varies}
with the number of sampled interpolation/quadrature points, averaged over the number of time windows, across
different hyper-reduction methods.
Our comparisons focus on the achieved ROM error
between the two ODE solvers (RK4 and RK2Avg) under a fixed residual
energy fraction $E_r$ as given in
Table \ref{table:setup:laghos}.
To obtain the presented results we increase the number of sampled
interpolation points for interpolation methods and the number of
nonzero sparse quadrature weights for EQP within each time window.
The figure compares the errors for each of the three state
variables: position, velocity and energy.
The position field generally exhibits the smallest error,
while the ordering of the
energy and velocity errors depends on the case.
For both solvers, the EQP method is the most efficient in terms of the number of
quadrature points, achieving a given error level with approximately one order of magnitude fewer quadrature points.
For the RK4 solver, EQP attains lower errors than the interpolation methods
for the position and velocity fields; however, for the energy field, EQP does not
achieve the lowest error levels, with the best interpolation results being
nearly one order of magnitude smaller.
Among the interpolation methods, DEIM generally demonstrates the most efficient
performance in terms of the number of sampled interpolation points.
For the RK2Avg solver, the interpolation methods achieve lower
errors than EQP across all fields considered.
Among them, S-OPT shows a slight advantage for the velocity field at
lower error levels, whereas the position and energy fields exhibit comparable performance.

Comparing the two solvers, EQP shows little difference in error between RK4 and
RK2Avg, with RK4 yielding slightly smaller errors for the velocity and energy
fields.
In contrast, the interpolation methods exhibit more pronounced solver
dependence.
For the velocity and position fields, RK2Avg enables an improvement
of approximately one order of magnitude in the attainable error,
reaching error levels of approximately $10^{-5}$ for the position field
and $10^{-4}$ for the velocity field.
For the energy field, however, RK2Avg results in errors that are roughly one order of
magnitude larger than those obtained with RK4, with minimum errors around
$3\times10^{-4}$.
Overall, while EQP is highly efficient in terms of the number of sampled points,
the achievable error levels across methods vary with the solver and
state component.

\begin{figure}[t]
    \centering
    \includegraphics[width=0.85\textwidth]{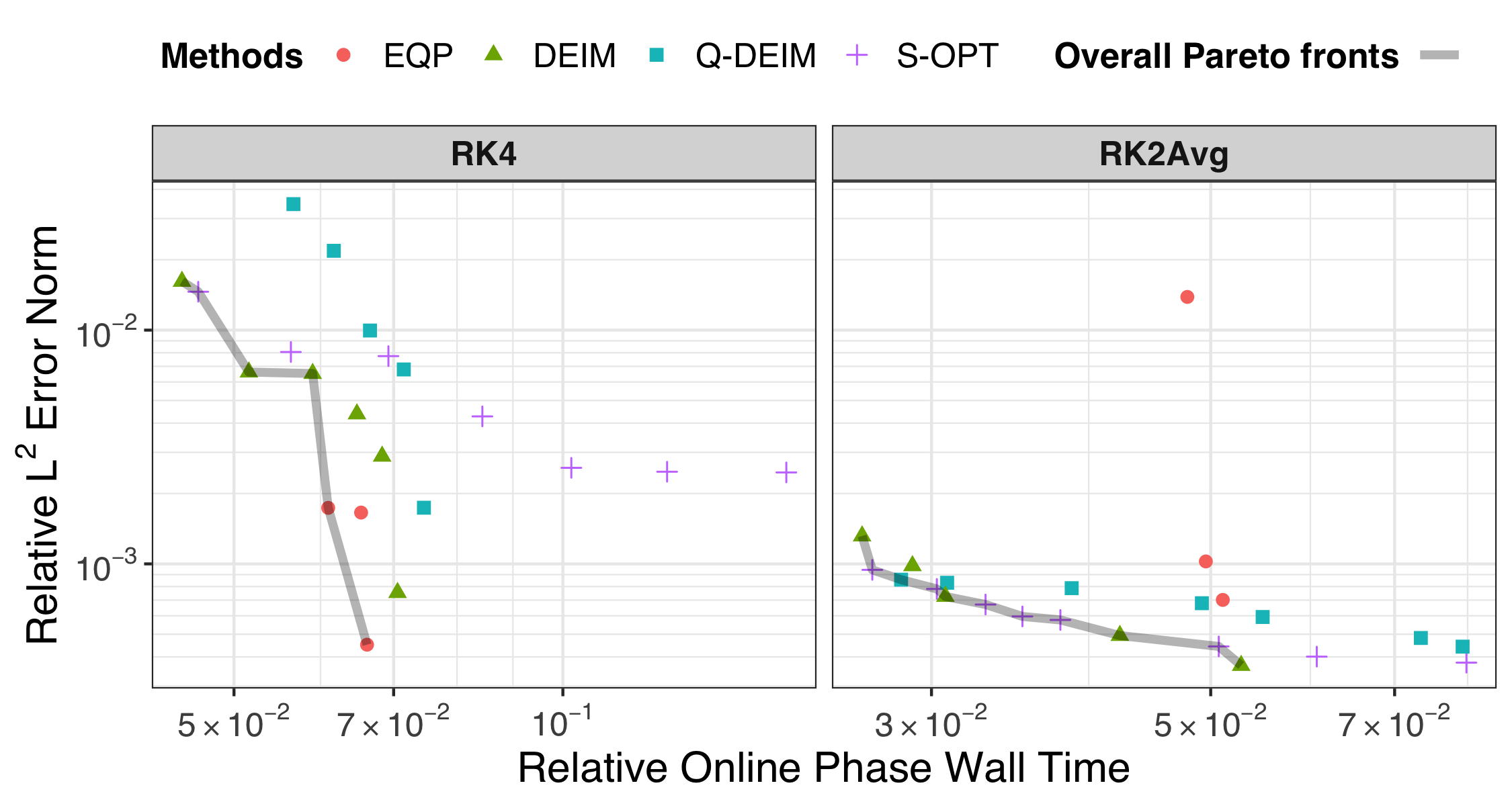} 
    \caption{Pareto fronts across hyper-reduction methods
        for the Sedov blast wave problem.}
    \label{fig:sb:pareto}
\end{figure}

Figure \ref{fig:sb:pareto}
reveals the tradeoff between approximation accuracy and time,
showing the Pareto fronts for $L^2$-product norm of the error in the state $(\velocitySymbol, \energySymbol, \positionSymbol)$, where all differential variables are aggregated into a single error metric rather than shown separately.
For the RK4 solver, the EQP method dominates the overall Pareto front
at error levels below $2\times 10^{-3}$, with a relative online time
of approximately $6 \times 10^{-2}$.
Among the interpolation methods, DEIM achieves the best performance and occupies the upper segment of the overall Pareto front,
reaching its lowest error level near $8\times 10^{-4}$
at a relative online time of about $7\times 10^{-2}$.
In the case of the RK2Avg solver, the interpolation methods constitute the Pareto-optimal set, with no single method dominating the others.
The lowest velocity error is obtained by DEIM, achieving an error level of $4\times 10^{-4}$ at a relative online time of about $5\times 10^{-2}$.
Although the interpolation methods require considerably more sampled
interpolation points than EQP, their relative online times are comparable
or even smaller. This could be due to computational overhead stemming from the EQP sample mesh implementation, as discussed in the introduction of Section \ref{sec:ex}.
We also note the solver dependent performance of each hyper-reduction method.
For EQP, the error and relative online time are similar across solvers.
For S-OPT, both solvers exhibit gradual tradeoffs: RK4 spans error levels between $10^{-2}$ and $10^{-3}$ over relative online times from $4\times10^{-2}$ to $2\times10^{-1}$, whereas RK2Avg reaches lower error levels between $10^{-3}$ and $4\times 10^{-4}$ at smaller relative online times from $2\times10^{-2}$ to $8\times10^{-2}$.
For DEIM and Q-DEIM, both methods exhibit a relatively steep tradeoff under RK4, with the error dropping from $10^{-2}$ to $10^{-3}$ over a narrower relative online time range of approximately $4\times10^{-2}$ to $7\times10^{-2}$.
Under RK2Avg, however, DEIM and Q-DEIM follow a more gradual tradeoff, similar to that observed for S-OPT.
Overall, RK2Avg paired with interpolation methods yields the most favorable performance across the results considered.

\subsubsection{Taylor-Green vortex}\label{sec:ex:tg}
The Taylor-Green vortex problem is a three dimensional standard
benchmark test for the incompressible inviscid Navier-Stokes equations.
In this work we manufacture a smooth solution by extending
the steady state Taylor-Green vortex solution to the
compressible Euler equations,
in analogy to the two-dimensional test in \cite{Dobrev2012}.
Our computation domain is the unit cube
$\initialDomain=[0, 1]^3$
with wall boundary conditions on all surfaces,
$\velocitySymbol\cdot\normalSymbol=0$.
Denoting
$\initialPosition=(\initialPosition_1,\initialPosition_2,\initialPosition_3)
\in\initialDomain$,
the initial velocity is given by
\begin{equation*}
\velocitySymbol(\initialPosition,0)=\left(\sin(\pi\initialPosition_1)
    \cos(\pi\initialPosition_2)\cos(\pi\initialPosition_3),
    -\cos(\pi\initialPosition_1)\sin(\pi\initialPosition_2)
    \cos(\pi\initialPosition_3)\right)
\end{equation*}
and the initial thermodynamic pressure by
\begin{equation*}
\pressureSymbol(\initialPosition,0)=100+
    \frac{(\cos(2\pi\initialPosition_1+\cos(2\pi\initialPosition_2))
    (\cos(2\pi\initialPosition_3)+2)-2}{16}
\end{equation*}
with initial density $\densitySymbol=1$.
Based on the above, the initial energy is determined by
the equation of state \eqref{eq:EOS}
with adiabatic index set to
$\adiabaticIndexSymbol=5/3$.

\begin{figure}[t]
    \centering
    \includegraphics[width=0.85\textwidth]{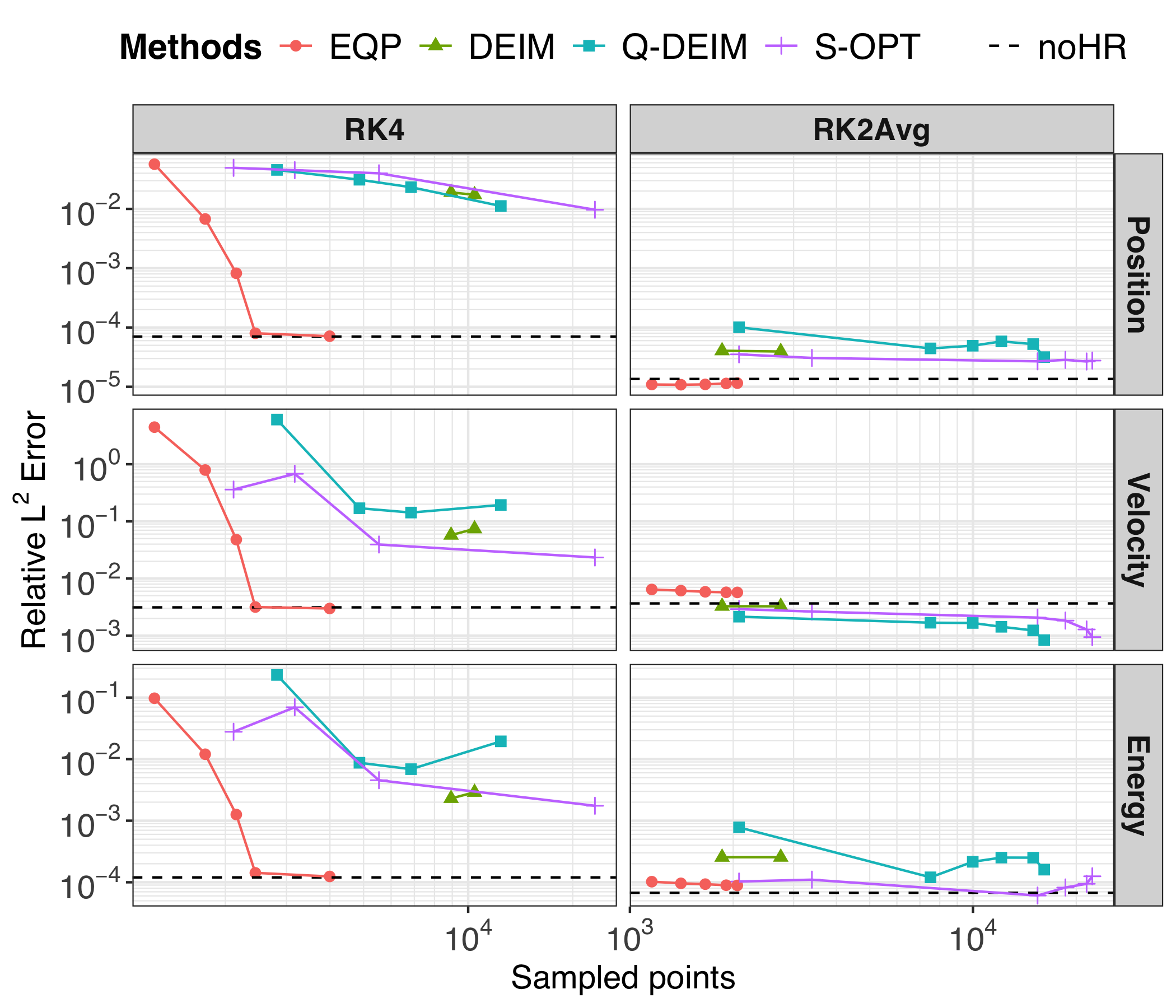}
    \caption{{ROM error with varying sampled points
    for the Taylor-Green vortex problem.}}
    \label{fig:tg:nqp}
\end{figure}

Figure \ref{fig:tg:nqp}
{presents the ROM errors across hyper-reduction methods
with respect to the average number of
interpolation/quadrature points.}
With the RK4 solver, the EQP method requires fewer quadrature points
than the interpolation methods to reach a given error level across all
variables.
More specifically, EQP achieves roughly an order of magnitude lower
velocity error than the interpolation methods while using fewer
quadrature points.
Among the interpolation methods,
S-OPT tends to exhibit the most efficient performance in terms of the
number of quadrature points required for a given velocity error.
With the RK2Avg solver, however, a different trend is observed. {We note that the EQP method is converging to the no-hyper-reduction ROM performance level for all sampled values of $K^*$.
Although EQP requires fewer quadrature points to reach near no-hyper-reduction ROM performance},
its velocity error levels remain higher than those of the
interpolation methods.
For approximately $2\times 10^{3}$ quadrature points, the interpolation
methods achieve lower velocity errors than EQP.
The minimum velocity error obtained with the interpolation methods is
approximately one order of magnitude smaller than that of EQP,
albeit with a larger number of sampled interpolation points.
Overall, EQP demonstrates greater efficiency with the RK4 solver,
whereas with RK2Avg the interpolation methods yield higher accuracy
in velocity.

\begin{figure}[t]
    \centering
    \includegraphics[width=0.85\textwidth]{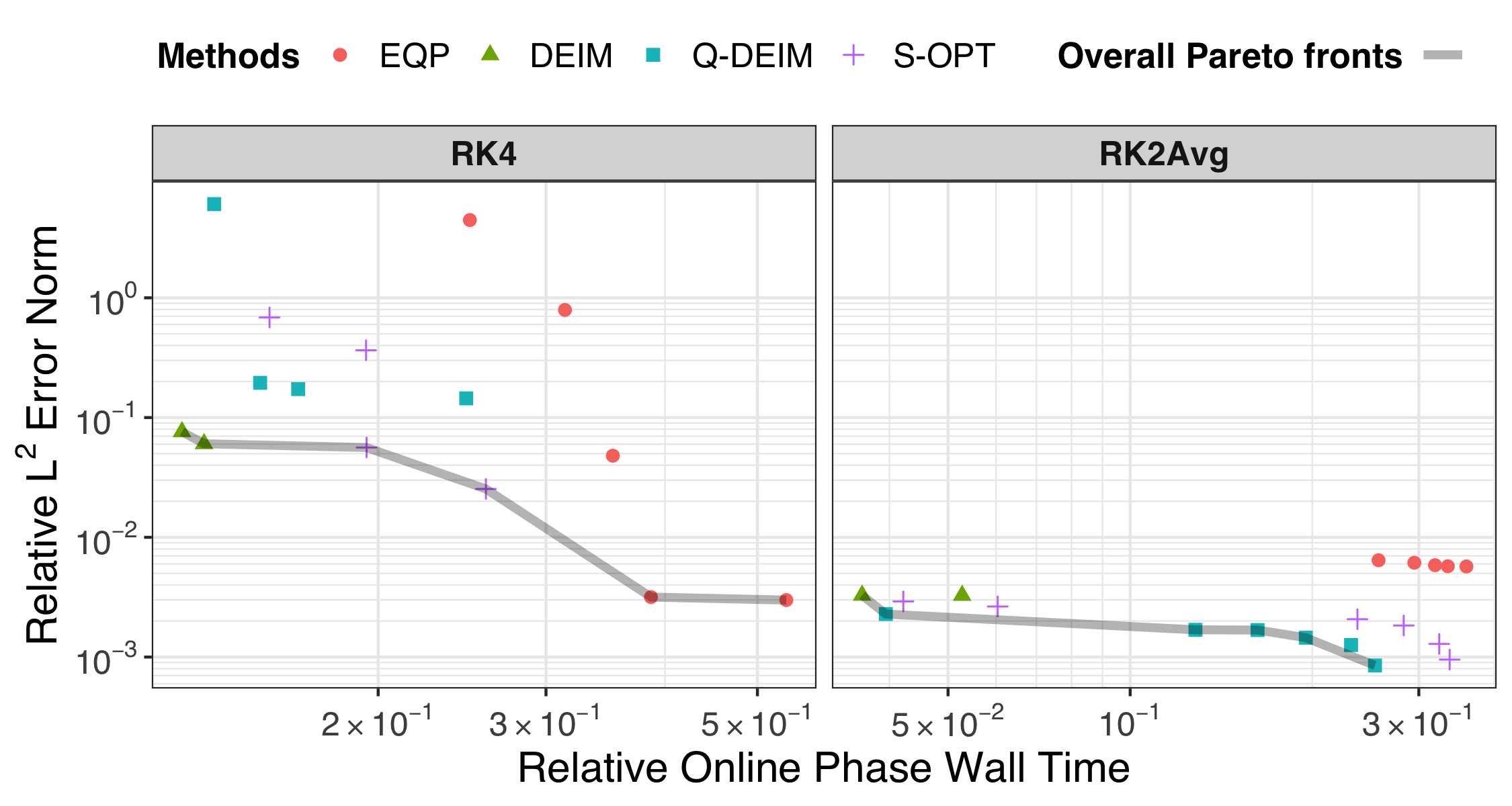} 
    \caption{Pareto fronts across hyper-reduction methods
        for the Taylor-Green vortex problem.}
    \label{fig:tg:pareto}
\end{figure}

Figure \ref{fig:tg:pareto}
focuses on the tradeoff between computational time and $L^2$-product norm of the error in the state $(\velocitySymbol, \energySymbol, \positionSymbol)$ across different hyper-reduction methods.
With the RK4 solver, the Pareto-optimal configurations progress from
DEIM to S-OPT and finally to EQP as the error level decreases.
The lowest total error is achieved by EQP, reaching approximately
$3\times10^{-3}$ with a relative online time slightly below
$4\times10^{-1}$.
With RK2Avg the overall Pareto front is dominated by the
interpolation methods.
They attain a total error of about $2\times10^{-3}$
at a relative online time of $4\times10^{-2}$,
and around $10^{-3}$ at $2\times10^{-1}$.
Among them, Q-DEIM generally forms the Pareto front.
To compare the performance of each method depending on the solver,
all interpolation methods perform better with RK2Avg than with RK4.
For EQP, the minimum obtained total error is slightly smaller
with the RK4 solver, while the corresponding relative online times
are of a similar magnitude.
Overall, the interpolation methods exhibit shorter relative online
times across both solvers.
However, when using RK4 the lowest error is obtained by EQP
at around $3\times10^{-3}$,
making it the preferred option for accuracy below $10^{-2}$.
The lowest total error among all configurations is attained by
Q-DEIM with the RK2Avg solver, reaching the $10^{-3}$ level.

\subsubsection{Triple point expansion}\label{sec:ex:tp}
The triple point expansion problem is a three dimensional hydrodynamic shock test
that features two materials in three different states.
More specifically, we consider a three state, two material, 2D Riemann
problem which generates vorticity, analogous to the two dimensional test
in \cite{Dobrev2012}.
The computational domain is
$\initialDomain=[0,7]\times [0,3]\times [0,1.5]$
with wall boundary conditions.
The initial velocity is given by $\velocitySymbol=0$.
Denoting
$\initialPosition=(\initialPosition_1,\initialPosition_2,\initialPosition_3)
\in\initialDomain$,
the initial density is given by
\begin{equation*}
\densitySymbol(\initialPosition,0)=\begin{cases}
    1& \text{ if }\initialPosition_1\leq 1\text{ or }
    \initialPosition_2\leq 1.5\\
    1/8& \text{ if }\initialPosition_1>1\text{ and }
    \initialPosition_2>1.5
    \end{cases}
\end{equation*}
and the initial thermodynamic pressure by
\begin{equation*}
\pressureSymbol(\initialPosition,0)=\begin{cases}
    1& \text{ if }\initialPosition_1\leq 1\\
    1/8& \text{ if }\initialPosition_1>1.
    \end{cases}
\end{equation*}
The adiabatic index is set to
\begin{equation*}
\adiabaticIndexSymbol(\initialPosition,0)=\begin{cases}
    1.4& \text{ if }\initialPosition_1>1\text{ and }
    \initialPosition_2\leq 1.5\\
    1.5& \text{ otherwise}
    \end{cases}
\end{equation*}
with the initial energy then determined by \eqref{eq:EOS}. 

\begin{figure}[t]
    \centering
    \includegraphics[width=0.85\textwidth]{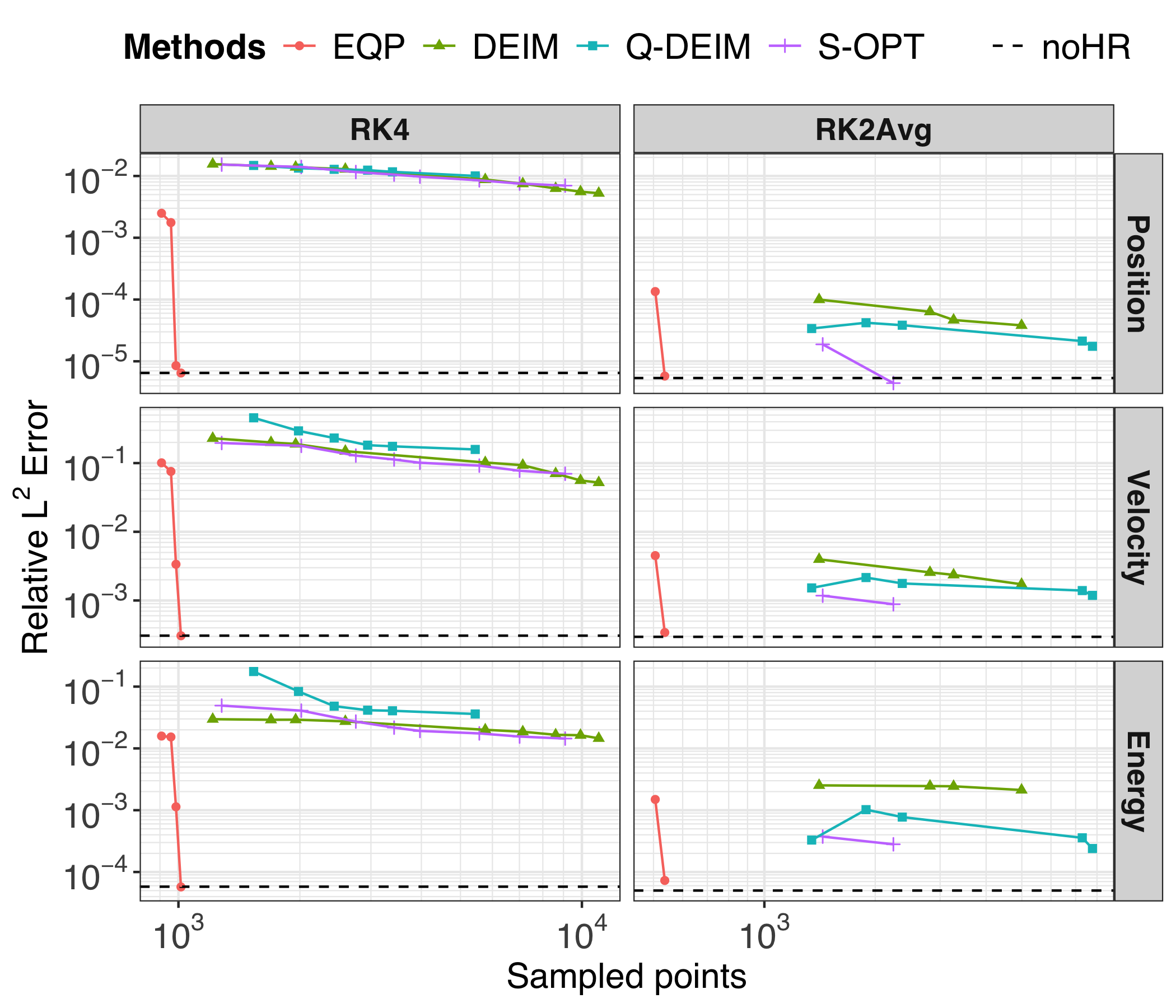}
    \caption{{ROM error with varying sampled points
    for the triple point expansion problem.}}
    \label{fig:tp:nqp}
\end{figure}

Beginning with Figure \ref{fig:tp:nqp}
we see that EQP achieves {the error level of ROM without hyper-reduction}
with both solvers.
With RK4 this error is obtained with $10^3$ quadrature points,
while with RK2Avg a comparable error level is reached with roughly half
as many quadrature points.
Contrary to EQP, the accuracy of the interpolation methods varies
significantly between the two solvers.
With RK4 these methods are less efficient, exhibiting minimum velocity
errors around $5\times 10^{-2}$, more than two orders of magnitude larger
than EQP.
With RK2Avg their performance improves considerably, reaching minimum
velocity errors of about $10^{-3}$, which represents a marked
improvement over RK4, albeit still higher than EQP.
The interpolation methods typically require between $10^3$ and $10^4$
sampled interpolation points, exceeding those required by EQP.
Overall, EQP attains low error levels using fewer quadrature points across both solvers.

\begin{figure}[t]
    \centering
    \includegraphics[width=0.85\textwidth]{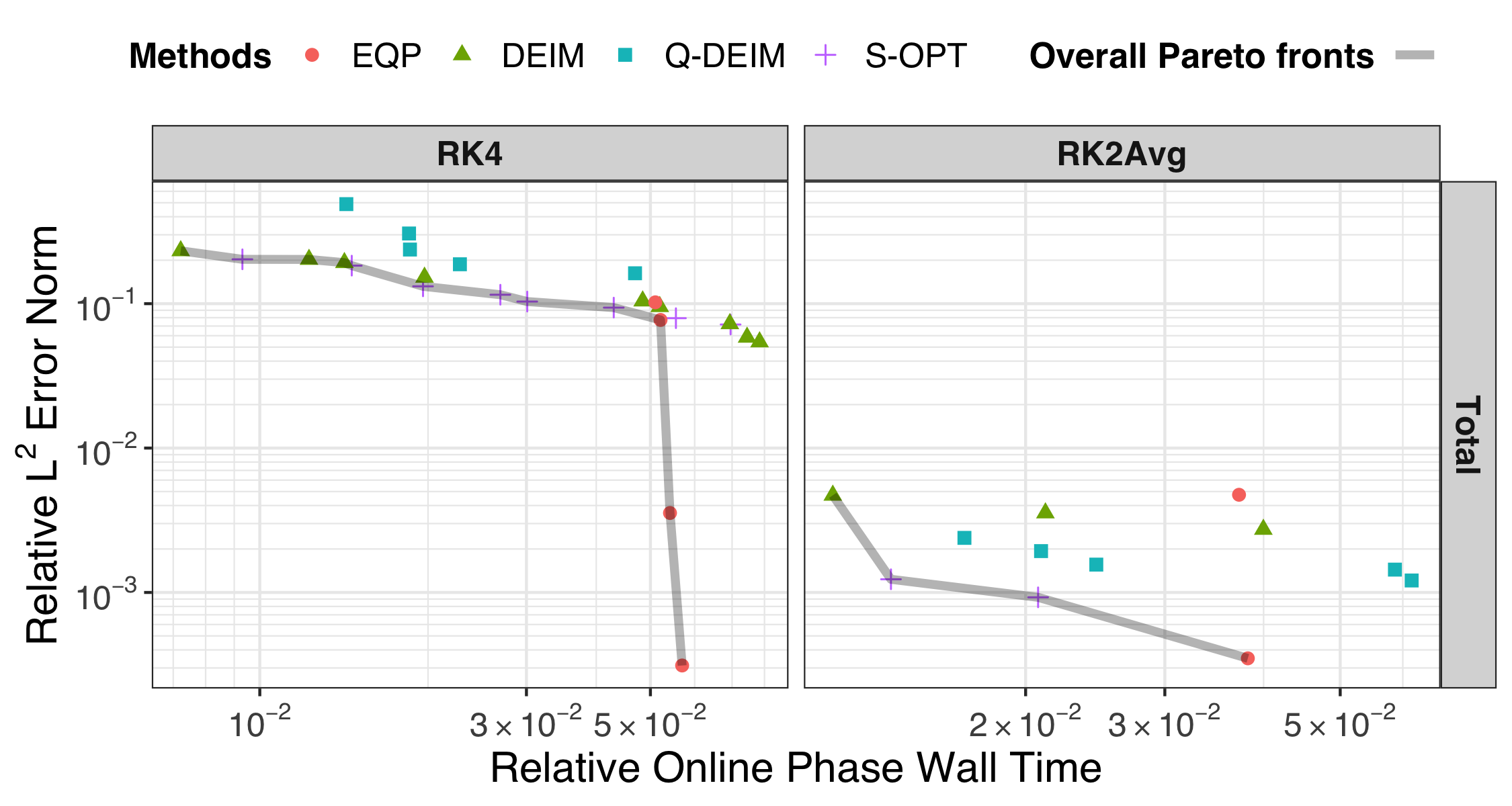} 
    \caption{Pareto fronts across hyper-reduction methods
        for the triple point expansion problem.}
    \label{fig:tp:pareto}
\end{figure}

Figure \ref{fig:tp:pareto} shows that,
although the interpolation methods use more sampled points than EQP,
their relative online times are comparable or even smaller for the
triple-point problem.
As discussed in \Cref{sec:ex:sedov},
this discrepancy between the number of quadrature points and relative online
wall time could be due to differences in sample meshes between EQP and
interpolation-based methods; see \Cref{fig:smesh}.
For both solvers, the Pareto optimal configurations are initially
composed of interpolation methods, with EQP taking over for lower
error levels.
At the lowest error levels, EQP achieves total errors of
approximately $3\times10^{-4}$ with relative online times of
about $4\times10^{-2}$ for RK2Avg and  $6\times10^{-2}$ for RK4.
On the contrary, the interpolation methods exhibit lower relative online
times and reduced accuracy.
With the RK4 solver they constitute the Pareto optimal set when the
relative online time is below $5\times10^{-2}$,
corresponding to total errors around $10^{-1}$.
With RK2Avg they remain Pareto optimal when the relative online time is
below $3.5\times10^{-2}$, with total errors of $9\times10^{-3}$ or higher.
Consistent with earlier observations, these methods perform
significantly better with RK2Avg than with RK4.
Overall, EQP remains the most accurate method across both solvers,
whereas interpolation methods offer a clear accuracy–efficiency tradeoff.

\section{Conclusion}
We introduced a variational framework for describing PDEs
solved by hyper-reduced models.
Five example problems were stated in terms of this framework and
solved using four hyper-reduction methods: three gappy POD-based
interpolation methods and the EQP quadrature method.
We compared the relative performance of these hyper-reduction methods
by computing the Pareto optimal front with respect to speed-up
and accuracy. 
{
We would like to remark that the scope of this comparison is limited to the selected methods and benchmark problems considered herein. While the presented results provide insight into the relative performance of the studied interpolation-based and quadrature-based methods, they should not be interpreted as a general comparison between these two classes of hyper-reduction approaches.
}

For the nonlinear diffusion and elasticity problems, 
we considered two types of simulations:
(1) reproductive, where the reduced model is tested for parameter values
that were used in its training;
(2) predictive, where the reduced model predicts solutions for
parameter values not used during its training.
In the Lagrangian hydrodynamics problems, we considered only
reproductive simulations.
Our numerical experiments demonstrate that the performance of the
tested hyper-reduction methods depends on the problem and time integration
method used, with no single method being universally superior.

In the diffusion problem, the EQP method was generally more efficient
in terms of relative online time in the reproductive scenario,
reaching similar error levels to the interpolation methods but with
higher speed-up.
In the predictive scenario, the differences among the methods were
less pronounced, with EQP and interpolation methods displaying
similar performance at lower error levels.

A similar trend appeared {for the nonlinear elasticity problem, both for the displacement and velocity fields}:
EQP was generally more efficient in the reproductive scenario and
less dominant in the predictive one.
The Q-DEIM method achieved the lowest errors across both scenarios {and for both displacement and velocity},
at the expense of a low relative speed-up of only 40\%.

For the Lagrangian hydrodynamics problems, the performance of the
interpolation methods was strongly influenced by the employed
time integration method:
interpolation methods performed consistently better with RK2Avg
than with RK4.
On the other hand, EQP did not demonstrate a clear dependency on
the selected time integration method.

For interpolation methods, each selected mesh element includes all
associated interpolation points, whereas EQP selects quadrature
points individually from the whole mesh.
This can explain why EQP achieved higher efficiency with respect to
the number of quadrature points in most of our numerical results.
However, due to computational overhead in the creation of the
sampled mesh elements, this advantage did not translate to the
expected efficiency in relative online time in our Lagrangian
hydrodynamics test cases. {The testing of these three hydrodynamics problems focuses on reproductive simulations to keep the comparisons as simple and clear as possible. As seen in the other problems, predictive simulations tend to have larger error than reproductive ones, with more error components. In particular, there is approximation of the solution using a fixed reduced basis, in addition to approximation of the hyper-reduced terms. The larger number of error components makes a straightforward comparison of hyper-reduction methods difficult for more complex problems sucah as the Lagrangian hydrodynamics ones. Testing on reproductive simulations enables comparison of the hyper-reduction approximation error. It also simplifies comparison of computational time, by avoiding the need to balance the choice of basis size and number of hyper-reduction samples for fast and accurate approximation of the predictive solution.}

Overall, our findings demonstrate that the EQP method can provide
a significant speed-up at high accuracy levels for a broad range
of problems.
Nevertheless, interpolation methods can achieve a higher speed-up
or better accuracy in certain cases.
The relative performance of the studied hyper-reduction methods
depends on the problem case, choice of time integration method,
and whether the tested simulation is reproductive or predictive.
These results highlight the need for problem-specific selection
of hyper-reduction strategies when choosing between interpolation
and quadrature methods. 

\section*{Acknowledgments}\label{sec:disc}
This work was supported in part by the U.S. Department of Energy,
Office of Science, Office of Advanced Scientific Computing Research,
Scientific Discovery through Advanced Computing (SciDAC) program through
the LEADS SciDAC Institute under Project Number SCW1933 at
Lawrence Livermore National Laboratory (LLNL). 
LLNL is operated by Lawrence Livermore National Security LLC, for the
U.S. Department of Energy, National Nuclear Security Administration
under Contract DE-AC52-07NA27344.
IM release number: LLNL-JRNL-2013738.
Axel Larsson was supported by the U.S. National Science Foundation
Grant award number 2122269.
Part of this work was performed while Axel Larsson, Minji Kim and
Chris Vales were research interns at LLNL supported by the
Mathematical Sciences Graduate Internship (MSGI)
program administered by the U.S. National Science Foundation.
Chris Vales also acknowledges support from the
U.S. Department of Energy under grant DE-SC0025101.
The authors thank Adrian Humphry and Masayuki Yano
from the University of Toronto for their valuable feedback on an
earlier version of this manuscript and for providing a piece of the
code used in our implementation of the empirical quadrature procedure.

\section*{Disclaimer}
This document was prepared as an account of work sponsored by an
agency of the United States government.
Neither the United States government nor Lawrence Livermore
National Security LLC nor any of their employees makes any warranty,
expressed or implied, or assumes any legal liability or responsibility
for the accuracy, completeness, or usefulness of any information,
apparatus, product, or process disclosed, or represents that its use
would not infringe privately owned rights.
Reference herein to any specific commercial product, process, or
service by trade name, trademark, manufacturer, or otherwise does not
necessarily constitute or imply its endorsement, recommendation,
or favoring by the United States government or
Lawrence Livermore National Security LLC.
The views and opinions of authors expressed herein do not necessarily
state or reflect those of the United States government or
Lawrence Livermore National Security LLC
and shall not be used for advertising or product endorsement purposes.

\appendix
\section{Command line options}\label{sec:appendix}
To facilitate the reproduction of our results, we present the
command line options used for each problem.
Examples are provided for the predictive scenarios in
Sections \ref{sec:ex:mnd} and \ref{sec:ex:nlel} and for the
reproductive scenario with the RK2Avg solver in
Section \ref{sec:ex:laghos}.
Each problem corresponds to a specific executable:
\texttt{mixed\_nonlinear\_diffusion} in \texttt{libROM},
\texttt{nonlinear\_elasticity\_global\_rom} in \texttt{libROM} and
\texttt{laghos} in \texttt{Laghos}.
Due to ongoing development in the repository,
the presented commands are subject to change.
The following commands are compatible with the recent commits of
the \texttt{master} branch of \texttt{MFEM}
\footnote{GitHub page, {\it  {https://github.com/mfem/mfem/tree/dc9128ef596e84daf1138aa3046b826bba9d259f}}},
the \texttt{master} branch of \texttt{libROM}
\footnote{GitHub page, {\it {https://github.com/LLNL/libROM/tree/d8cc32ce02f981fd5bf12ab32ef8d4c3445eab18}}},
and the \texttt{rom} branch of \texttt{Laghos}
\footnote{GitHub page, {\it  {https://github.com/CEED/Laghos/tree/ab24cfcdc6e7cd4747d9ddd7cf82fef9163f80e8}}}.

In order to use the ROM capability of \texttt{Laghos}, the user has to
navigate to the \texttt{rom} subdirectory.

The conventions used in the examples are as follows.
For Sections \ref{appx:cmd:mnd} and \ref{appx:cmd:nlel},
the following apply to the online phase:
\begin{itemize}
    \item EQP method: use the options \texttt{-eqp -ns};
        the number of quadrature points can be restricted with
        \texttt{-maxnnls}.
    \item Interpolation methods: specify the hyper-reduction method via
        \texttt{-hrtype} and the number of sampling points with
        \texttt{-nsr}.
\end{itemize}
For Section \ref{appx:cmd:laghos}:
\begin{itemize}
    \item \texttt{-s 7} corresponds to the RK2Avg solver and
        \texttt{-s 4} to RK4; RK2Avg is used in the examples.
    \item EQP method: requires the \texttt{-eqp} option in the merge phase;
        in the preprocessing and online phases use
        \texttt{-hrsamptype eqp} and \texttt{-lqnnls}.
    \item Interpolation methods: specify the hyper-reduction type in the
        preprocessing phase with \texttt{-hrsamptype};
        the number of sampling points is controlled by
        \texttt{-sfacv} and \texttt{-sface}.
\end{itemize}
\subsection{Nonlinear diffusion}\label{appx:cmd:mnd}
The following commands are used for the nonlinear diffusion problem
in Section \ref{sec:ex:mnd}.
The working directory is assumed to be \texttt{libROM/examples/prom}. 

\begin{enumerate}
\item Offline phase (predictive scenario):
\begin{lstlisting}
./mixed_nonlinear_diffusion -p 1 -offline -id 0 -sh 0.25
./mixed_nonlinear_diffusion -p 1 -offline -id 1 -sh 0.15
./mixed_nonlinear_diffusion -p 1 -offline -id 2 -sh 0.35
\end{lstlisting}

\item Merge phase:
\begin{lstlisting}
./mixed_nonlinear_diffusion -p 1 -merge -ns 3
\end{lstlisting}

\item Create the FOM comparison data:
\begin{lstlisting}
./mixed_nonlinear_diffusion -p 1 -offline -id 3 -sh 0.3
\end{lstlisting}

\item Online phase for EQP:
\begin{lstlisting}
./mixed_nonlinear_diffusion -p 1 -online -rrdim 6 -rwdim 3 -id 3 -sh 0.3 -eqp -ns 3
\end{lstlisting}

\noindent Online phase for S-OPT:
\begin{lstlisting}
./mixed_nonlinear_diffusion -p 1 -online -rrdim 6 -rwdim 3 -id 3 -sh 0.3 -nldim 6 -nsr 100 -hrtype sopt
\end{lstlisting}

\end{enumerate}

\subsection{Nonlinear elasticity}\label{appx:cmd:nlel}
The following commands are used for the nonlinear diffusion problem
in Section \ref{sec:ex:nlel}.
The working directory is again assumed to be \texttt{libROM/examples/prom}.

\begin{enumerate}
\item Offline phase (predictive scenario):
\begin{lstlisting}
./nonlinear_elasticity_global_rom -offline -dt 0.01 -tf 5.0 -s 14 -vs 100 -sc 0.9 -id 0
./nonlinear_elasticity_global_rom -offline -dt 0.01 -tf 5.0 -s 14 -vs 100 -sc 1.1 -id 1
\end{lstlisting}

\item Merge phase:
\begin{lstlisting}
./nonlinear_elasticity_global_rom -merge -ns 2 -dt 0.01 -tf 5.0
\end{lstlisting}

\item Create the FOM comparison data:
\begin{lstlisting}
./nonlinear_elasticity_global_rom -offline -dt 0.01 -tf 5.0 -s 14 -vs 100 -sc 1.0 -id 2
\end{lstlisting}

\item Online phase for EQP:
\begin{lstlisting}
./nonlinear_elasticity_global_rom -online -dt 0.01 -tf 5.0 -s 14 -vs 100 -eqp -ns 2 -rvdim 3 -rxdim 2 -hdim 1 -sc 1.0
\end{lstlisting}

\noindent Online phase for S-OPT:
\begin{lstlisting}
./nonlinear_elasticity_global_rom -online -dt 0.01 -tf 5.0 -s 14 -vs 100 -hyp -hrtype sopt -rvdim 3 -rxdim 2 -hdim 5 -nsr 100 -sc 1.0
\end{lstlisting}
\end{enumerate}

\subsection{Lagrangian hydrodynamics}\label{appx:cmd:laghos}
The following commands are used for the Lagrangian hydrodynamics
simulations in Section \ref{sec:ex:laghos}.
The working directory is assumed to be \texttt{Laghos/rom}

\textbf{Problem specification}:
First, we present the commands used to set up each problem.

\begin{itemize}
\item Sedov blast wave problem:
\begin{lstlisting}
./laghos -m data/cube01_hex.mesh  -p 1 -pt 211 -s 7 -tf 0.2 
\end{lstlisting}

\item Taylor-Green vortex problem:
\begin{lstlisting}
./laghos -m data/cube01_hex.mesh -p 0 -cfl 0.1 -s 7 -tf 0.1
\end{lstlisting}

\item Triple-point expansion problem:
\begin{lstlisting}
./laghos -m data/box01_hex.mesh -p 3  -cfl 0.5 -s 7 -tf 0.5
\end{lstlisting}
\end{itemize}

\textbf{ROM simulation}:
Next, we present the command line options used in each stage of the
ROM simulation for the Sedov blast wave problem.
These options should be appended to the problem setup command for Sedov,
except for the merge phase, which is executed independently.
The same procedure applies to the other problems by replacing the
problem specification accordingly.
Below, we provide example commands for EQP and S-OPT.

\begin{enumerate}
\item Offline phase:
\begin{lstlisting}
-offline -writesol -romsns -rpar 0 -rostype interpolate -sdim 10000 -sample-stages
\end{lstlisting}

\item Merge phase for EQP (executed independently):
\begin{lstlisting}
./merge -nset 1 -romsns -romos -rostype interpolate -nwinsamp 50 -ef 0.999999 -eqp
\end{lstlisting}

\noindent Merge phase for S-OPT (executed independently):
\begin{lstlisting}
./merge -nset 1 -romsns -romos -rostype interpolate -nwinsamp 50 -ef 0.999999
\end{lstlisting}

\item Hyper-reduction preprocessing phase for EQP:
\begin{lstlisting}
-online -romhrprep -romsns -rostype interpolate -nwin 15 -hrsamptype eqp -lqnnls
\end{lstlisting}

\noindent Hyper-reduction preprocessing phase for S-OPT:
\begin{lstlisting}
-online -romhrprep -romsns -rostype interpolate -nwin 15 -hrsamptype sopt -sfacv 2 -sface 2
\end{lstlisting}

\item ROM online phase for EQP:
\begin{lstlisting}
-online -romhr -romsns -rostype interpolate -nwin 15 -hrsamptype eqp -lqnnls
\end{lstlisting}

\noindent ROM online phase for S-OPT:
\begin{lstlisting}
-online -romhr -romsns -rostype interpolate -nwin 15 -sfacv 2 -sface 2
\end{lstlisting}

\item Postprocessing phase (solution reconstruction and error evaluation):
\begin{lstlisting}
-restore -romsns -rostype interpolate -nwin 15 -soldiff
\end{lstlisting}

\end{enumerate}

\bibliography{references}

@article{lauzon2024s,
  title={{S-OPT: a points selection algorithm for hyper-reduction in
      reduced order models}},
  author={Lauzon, Jessica T and Cheung, Siu Wun and Shin, Yeonjong
      and Choi, Youngsoo and Copeland, Dylan M and Huynh, Kevin},
  journal={SIAM J. Sci. Comput.},
  volume={46},
  number={4},
  pages={B474--B501},
  year={2024}
}

@article{peherstorfer2020stability,
  title={Stability of discrete empirical interpolation and gappy proper orthogonal decomposition with randomized and deterministic sampling points},
  author={Peherstorfer, Benjamin and Drmac, Zlatko and Gugercin, Serkan},
  journal={SIAM J. Sci. Comput.},
  volume={42},
  number={5},
  pages={A2837--A2864},
  year={2020}
}

@article{shin2016near,
  title={On a near optimal sampling strategy for least squares polynomial regression},
  author={Y. Shin and D. Xiu},
  journal = {J. Comput. Phys.},
  volume={326},
  pages={931--946},
  year={2016}
}

@misc{librom,
title = {lib{ROM}},
author = {Choi, Youngsoo and Arrighi, William J. and Copeland, Dylan M. and Anderson, Robert W. and Oxberry, Geoffrey M. and {USDOE NNSA}},
url = {https://www.osti.gov//servlets/purl/1505575},
year = {2019}
}

@misc{mfem-web,
  key = {mfem},
  title = {{MFEM}: Modular Finite Element Methods},
  howpublished = {\url{mfem.org}},
  doi = {10.11578/dc.20171025.1248}
}

@article{Copeland2022,
    title={Reduced order models for {Lagrangian} hydrodynamics},
    author={Copeland, Dylan Matthew and Cheung, Siu Wun
        and Huynh, Kevin and Choi, Youngsoo},
    journal={Comput. Methods Appl. Mech. Eng.},
    volume={388},
    pages={114259},
    year={2022},
    doi = {10.1016/j.cma.2021.114259}}

@article{Vales2025ceqp,
    title = {{Machine-precision energy conservative
        reduced models for Lagrangian hydrodynamics
        by quadrature methods}},
    author = {Vales, Chris and Cheung, Siu Wun
        and Copeland, Dylan M and Choi, Youngsoo},
    journal = {arXiv:2508.21279},
    year = {2025},
    doi = {10.48550/arXiv.2508.21279}
}

@article{everson1995,
author = {R. Everson and L. Sirovich},
journal = {J. Opt. Soc. Am. A},
number = {8},
pages = {1657--1664},
title = {{Karhunen--Lo\`{e}ve procedure for gappy data}},
volume = {12},
year = {1995}
}

@book{tarantola2005inverse,
  title={Inverse problem theory and methods for model parameter estimation},
  author={Tarantola, Albert},
  year={2005},
  publisher={SIAM},
  address = {Philadelphia}
}

@incollection{biegler2003large,
  title={Large-scale PDE-constrained optimization: an introduction},
  author={Biegler, Lorenz T and Ghattas, Omar and Heinkenschloss, Matthias and van Bloemen Waanders, Bart},
  booktitle={Large-scale PDE-constrained optimization},
  pages={3--13},
  year={2003},
  publisher={Springer},
  address = {Berlin}
}

@book{biegler2007real,
  title={Real-time PDE-constrained Optimization},
  author={Biegler, Lorenz T and Ghattas, Omar and Heinkenschloss, Matthias and Keyes, David and van Bloemen Waanders, Bart},
  year={2007},
  publisher={SIAM},
  address = {Philadelphia}
}

@book{gunzburger2002perspectives,
  title={Perspectives in flow control and optimization},
  author={Gunzburger, Max D},
  year={2002},
  publisher={SIAM},
  address = {Philadelphia}
}

@article{anirudh20232022,
  title={2022 review of data-driven plasma science},
  author={Anirudh, Rushil and Archibald, Rick and Asif, M Salman and Becker, Markus M and Benkadda, Sadruddin and Bremer, Peer-Timo and Bud{\'e}, Rick HS and Chang, Choong-Seock and Chen, Lei and Churchill, RM and others},
  journal={IEEE Trans. Plasma Sci.},
  year={2023}
}

@article{carlberg2013gnat,
  title={The {GNAT} method for nonlinear model reduction: Effective implementation and application to computational fluid dynamics and turbulent flows},
  author={Carlberg, Kevin and Farhat, Charbel and Cortial, Julien and Amsallem, David},
  journal={J. Comput. Phys.},
  volume={242},
  pages={623--647},
  year={2013}
}

@article{carlberg2011efficient,
  title={Efficient non-linear model reduction via a least--squares {P}etrov--{G}alerkin projection and compressive tensor approximations},
  author={Carlberg, Kevin and Bou‐Mosleh, Charbel and Farhat, Charbel},
  journal={Int. J. Numer. Methods Eng.},
  volume={86},
  pages={155--181},
  year={2011}
}

@article{Cheung2023,
    title = {{Local Lagrangian reduced-order modeling for the
        Rayleigh-Taylor instability by solution manifold decomposition}},
    author = {Cheung, Siu Wun and Choi, Youngsoo and
        Copeland, Dylan Matthew and Huynh, Kevin},
    journal = {J. Comput. Phys.},
    volume = {472},
    pages = {11655},
    year = {2023},
    doi = {10.1016/j.jcp.2022.111655}
}

@article{mojgani2017lagrangian,
  title={{L}agrangian basis method for dimensionality reduction of convection dominated nonlinear flows},
  author={Mojgani, Rambod and Balajewicz, Maciej},
  journal={arXiv:1701.04343},
  year={2017}
}

@article{burkardt2006pod,
  title={{POD} and {CVT}-based reduced-order modeling of {N}avier--{S}tokes flows},
  author={Burkardt, John and Gunzburger, Max and Lee, Hyung-Chun},
  journal={Comput. Methods Appl. Mech. Eng.},
  volume={196},
  number={1-3},
  pages={337--355},
  year={2006}
}

@article{mcbane2021component,
  title={Component-wise reduced order model lattice-type structure design},
  author={McBane, Sean and Choi, Youngsoo},
  journal={Comput. Methods Appl. Mech. Eng.},
  volume={381},
  pages={113813},
  year={2021}
}

@article{hoang2020domain,
  title={Domain-decomposition least-squares {P}etrov-{G}alerkin ({DD-LSPG}) nonlinear model reduction},
  author={Hoang, Chi and Choi, Youngsoo and Carlberg, Kevin},
  journal={Comput. Methods Appl. Mech. Eng.},
  volume={384},
  pages={113997},
  year={2021}
}

@article{choi2021space,
  title={Space-time reduced order model for large-scale linear dynamical systems with application to {B}oltzmann transport problems},
  author={Choi, Youngsoo and Brown, Peter and Arrighi, Bill and Anderson,
    Roberti and Huynh, Kevin},
  journal={J. Comput. Phys.},
  volume={424},
  pages={109845},
  year={2021}
}

@article{carlberg2018conservative,
  title={Conservative model reduction for finite-volume models},
  author={Carlberg, Kevin and Choi, Youngsoo and Sargsyan, Syuzanna},
  journal={J. Comput. Phys.},
  volume={371},
  pages={280--314},
  year={2018},
}

@article{Dobrev2012,
  title={High-order curvilinear finite element methods
      for {L}agrangian hydrodynamics},
  author={Dobrev, Veselin A and Kolev, Tzanio V and Rieben, Robert N},
  journal={SIAM J. Sci. Comput.},
  volume={34},
  number={5},
  pages={B606--B641},
  year={2012}
}

@article{amsallem2015design,
  title={Design optimization using hyper-reduced-order models},
  author={Amsallem, David and Zahr, Matthew and Choi, Youngsoo and Farhat, Charbel},
  journal={Struct. Multidiscip. Optim.},
  volume={51},
  number={4},
  pages={919--940},
  year={2015},
}

@article{choi2020gradient,
  title={Gradient-based constrained optimization using a database of linear reduced-order models},
  author={Choi, Youngsoo and Boncoraglio, Gabriele and Anderson, Spenser and Amsallem, David and Farhat, Charbel},
  journal={J. Comput. Phys.},
  volume={423},
  pages={109787},
  year={2020},
}

@article{choi2019accelerating,
  title={Accelerating design optimization using reduced order models},
  author={Choi, Youngsoo and Oxberry, Geoffrey and White, Daniel and Kirchdoerfer, Trenton},
  journal={arXiv:1909.11320},
  year={2019}
}

@article{glas2020reduced,
  title={A reduced basis method for the wave equation},
  author={Glas, Silke and Patera, Anthony T and Urban, Karsten},
  journal={Int. J. Comput. Fluid Dyn.},
  volume={34},
  number={2},
  pages={139--146},
  year={2020},
}

@article{cheung2021explicit,
    title={{Explicit and energy-conserving constraint energy minimizing
        generalized multiscale discontinuous Galerkin method for wave
        propagation in heterogeneous media}},
    author={Cheung, Siu Wun and Chung, Eric T and Efendiev, Yalchin
        and Leung, Wing Tat},
    journal={Multiscale Model. Simul.},
    volume={19},
    number={4},
    pages={1736--1759},
    year={2021}
}

@article{jiang2019implementation,
  title={Implementation and detailed assessment of a {GNAT} reduced-order model for subsurface flow simulation},
  author={Jiang, Rui and Durlofsky, Louis J},
  journal={J. Comput. Phys.},
  volume={379},
  pages={192--213},
  year={2019},
}

@article{yang2016fast,
  title={{Fast multiscale reservoir simulations with POD-DEIM model reduction}},
  author={Yang, Yanfang and Ghasemi, Mohammadreza and Gildin, Eduardo and Efendiev, Yalchin and Calo, Victor and others},
  journal={SPE Journal},
  volume={21},
  number={06},
  pages={2--141},
  year={2016},
}

@article{wang2020generalized,
  title ={Generalized multiscale multicontinuum model for fractured vuggy carbonate reservoirs},
  author={Wang, Min and Cheung, Siu Wun and Chung, Eric T. and Vasilyeva, Maria and Wang, Yuhe},
  journal={J. Comput. Appl. Math.},
  volume={366},
  pages={112370},
  year={2020}
}

@article{benner2015survey,
  title={A survey of projection-based model reduction methods for parametric dynamical systems},
  author={Benner, Peter and Gugercin, Serkan and Willcox, Karen},
  journal={SIAM Rev.},
  volume={57},
  number={4},
  pages={483--531},
  year={2015},
}

@article{gugercin2004survey,
  title={A survey of model reduction by balanced truncation and some new results},
  author={Gugercin, Serkan and Antoulas, Athanasios C},
  journal={Int. J. Control},
  volume={77},
  number={8},
  pages={748--766},
  year={2004},
}

@article{cheng2016reduced,
  title={A reduced-order representation of the {S}chr{\"o}dinger equation},
  author={Cheng, Ming-C},
  journal={AIP Adv.},
  volume={6},
  number={9},
  pages={095121},
  year={2016},
}

@article{choi2019space,
  title={Space--Time Least-Squares {P}etrov--{G}alerkin Projection for Nonlinear Model Reduction},
  author={Choi, Youngsoo and Carlberg, Kevin},
  journal={SIAM J. Sci. Comput.},
  volume={41},
  number={1},
  pages={A26--A58},
  year={2019},
}

@article{xiao2014non,
  title={Non-linear model reduction for the {N}avier--{S}tokes equations using residual {DEIM} method},
  author={Xiao, Dunhui and Fang, Fangxin and Buchan, Andrew G and Pain, Christopher C and Navon, Ionel Michael and Du, Juan and Hu, G},
  journal={J. Comput. Phys.},
  volume={263},
  pages={1--18},
  year={2014},
}

@article{choi2020sns,
  title={{SNS}: a solution-based nonlinear subspace method for time-dependent model order reduction},
  author={Choi, Youngsoo and Coombs, Deshawn and Anderson, Robert},
  journal={SIAM J. Sci. Comput.},
  volume={42},
  number={2},
  pages={A1116--A1146},
  year={2020},
}

@article{chaturantabut2010nonlinear,
  title={Nonlinear model reduction via discrete empirical interpolation},
  author={Chaturantabut, Saifon and Sorensen, Danny C},
  journal={SIAM J. Sci. Comput.},
  volume={32},
  number={5},
  pages={2737--2764},
  year={2010},
}

@article{drmac2016new,
  title={A new selection operator for the discrete empirical interpolation
  method---Improved a priori error bound and extensions},
  author={Drmac, Zlatko and Gugercin, Serkan},
  journal={SIAM J. Sci. Comput.},
  volume={38},
  number={2},
  pages={A631--A648},
  year={2016},
}

@article{drmac2018discrete,
  title={The discrete empirical interpolation method: canonical
      structure and formulation in weighted inner product spaces},
  author={Drmac, Zlatko and Saibaba, Arvind Krishna},
  journal={SIAM J. Matrix Anal. Appl.},
  volume={39},
  number={3},
  pages={1152--1180},
  year={2018},
}

@book{sedov1993similarity,
  title={Similarity and dimensional methods in mechanics},
  author={Sedov, Leonid Ivanovich and Volkovets, AG},
  year={2018},
  publisher={CRC Press},
  address = {Boca Raton}
}

@book{lawson1995solving,
  title={Solving least squares problems},
  author={Lawson, Charles L and Hanson, Richard J},
  year={1995},
  publisher={SIAM},
  address={Philadelphia}
}

@article{schmid2010dynamic,
  title={Dynamic mode decomposition of numerical and experimental data},
  author={Schmid, Peter J},
  journal={J. Fluid Mech.},
  volume={656},
  pages={5--28},
  year={2010}
}

@article{brunton2016discovering,
  title={Discovering governing equations from data by sparse identification of nonlinear dynamical systems},
  author={Brunton, Steven L and Proctor, Joshua L and Kutz, J Nathan},
  journal={Proc. Natl. Acad. Sci. USA},
  volume={113},
  number={15},
  pages={3932--3937},
  year={2016},
}

@article{van_der_merwe_fast_2007,
	title = {Fast neural network surrogates for very high dimensional physics-based models in computational oceanography},
	volume = {20},
	doi = {10.1016/j.neunet.2007.04.023},
	language = {en},
	number = {4},
	journal = {Neural Networks},
	author = {van der Merwe, Rudolph and Leen, Todd K. and Lu, Zhengdong and Frolov, Sergey and Baptista, Antonio M.},
	year = {2007},
	pages = {462--478},
}

@article{lagaris_artificial_1998,
	title = {Artificial neural networks for solving ordinary and partial differential equations},
	volume = {9},
	doi = {10.1109/72.712178},
	number = {5},
	journal = {IEEE Trans. Neural Networks},
	author = {Lagaris, I.E. and Likas, A. and Fotiadis, D.I.},
	year = {1998},
	pages = {987--1000}
}

@article{karniadakis2021physics,
  title={Physics-informed machine learning},
  author={Karniadakis, George Em and Kevrekidis, Ioannis G and Lu, Lu and Perdikaris, Paris and Wang, Sifan and Yang, Liu},
  journal={Nat. Rev. Phys.},
  volume={3},
  number={6},
  pages={422--440},
  year={2021}
}

@article{raissi2019physics,
  title={Physics-informed neural networks: a deep learning framework for solving forward and inverse problems involving nonlinear partial differential equations},
  author={Raissi, Maziar and Perdikaris, Paris and Karniadakis, George E},
  journal={J. Comput. Phys.},
  volume={378},
  pages={686--707},
  year={2019},
}

@article{kovachki2023neural,
  title={{Neural operator: learning maps between function spaces with applications to PDEs}},
  author={Kovachki, Nikola and Li, Zongyi and Liu, Burigede and Azizzadenesheli, Kamyar and Bhattacharya, Kaushik and Stuart, Andrew and Anandkumar, Anima},
  journal={J. Mach. Learn. Res.},
  volume={24},
  number={89},
  pages={1--97},
  year={2023}
}

@article{Yano2019,
  title={An {LP} empirical quadrature procedure for reduced basis
      treatment of parametrized nonlinear {PDEs}},
  author={Yano, Masayuki and Patera, Anthony T},
  journal={Comput. Methods Appl. Mech. Eng.},
  volume={344},
  pages={1104--1123},
  year={2019},
  doi = {10.1016/j.cma.2018.02.028}
}

@article{du2022,
	title = {Efficient hyperreduction of high-order discontinuous {Galerkin}
        methods: element-wise and point-wise reduced quadrature formulations},
	author = {Du, Eugene and Yano, Masayuki},
	journal = {J. Comput. Phys.},
	volume = {466},
	pages = {111399},
	year = {2022},
	doi = {10.1016/j.jcp.2022.111399}
}

@article{sleeman2022,
	title = {Goal-oriented model reduction for parametrized time-dependent nonlinear partial differential equations},
	volume = {388},
	doi = {10.1016/j.cma.2021.114206},
	journal = {Comput. Methods Appl. Mech. Eng.},
	author = {Sleeman, Michael K. and Yano, Masayuki},
	year = {2022},
	pages = {114206}
}

@article{grimberg2021mesh,
  title={{Mesh sampling and weighting for the hyperreduction of nonlinear
      Petrov-Galerkin reduced-order models with local reduced-order bases}},
  author={Grimberg, Sebastian and Farhat, Charbel and Tezaur, Radek
      and Bou-Mosleh, Charbel},
  journal={Int. J. Numer. Methods Eng.},
  volume={122},
  number={7},
  pages={1846--1874},
  year={2021}
}

@article{hernandez2017dimensional,
    title={Dimensional hyper-reduction of nonlinear finite element models via empirical cubature},
    author={Hernandez, Joaquin Alberto and Caicedo, Manuel Alejandro and Ferrer, Alex},
    journal = {Comput. Methods Appl. Mech. Eng.},
    volume={313},
    pages={687--722},
    year={2017}
}

@article{farhat2015structure,
  title={Structure-preserving, stability, and accuracy properties of the energy-conserving sampling and weighting method for the hyper reduction of nonlinear finite element dynamic models},
  author={Farhat, Charbel and Chapman, Todd and Avery, Philip},
  journal={Int. J. Numer. Methods Eng.},
  volume={102},
  number={5},
  pages={1077--1110},
  year={2015}
}

@article{sandu2021conservative,
  title={{Conservative high-order time integration for Lagrangian hydrodynamics}},
  author={Sandu, Adrian and Tomov, Vladimir and Cervena, Lenka and Kolev, Tzanio},
  journal={SIAM J. Sci. Comput.},
  volume={43},
  number={1},
  pages={A221--A241},
  year={2021}
}

@article{fritzen2018algorithmic,
  title={An algorithmic comparison of the hyper-reduction and the discrete empirical interpolation method for a nonlinear thermal problem},
  author={Fritzen, Felix and Haasdonk, Bernard and Ryckelynck, David and Sch{\"o}ps, Sebastian},
  journal={Math. Comput. Appl.},
  volume={23},
  number={1},
  pages={8},
  year={2018}
}

@article{brands2019reduced,
  title={Reduced-order modelling and homogenisation in magneto-mechanics:
      a numerical comparison of established hyper-reduction methods},
  author={Brands, Benjamin and Davydov, Denis and Mergheim, Julia
      and Steinmann, Paul},
  journal={Math. Comput. Appl.},
  volume={24},
  number={1},
  pages={20},
  year={2019},
}

@article{delagnes2024comparison,
  title={Comparison of Hyper-Reduction Methods Combined With POD: Model Order Reduction of a Squirrel Cage Induction Machine in Nonlinear Case},
  author={Delagnes, T and Henneron, Thomas and Clenet, Stephane and Fratila, M and Ducreux, JP},
  journal={IEEE Trans. Magn.},
  volume={60},
  number={6},
  pages={1--10},
  year={2024}
}

@article{bhattacharyya2025hyper,
  title={Hyper-Reduction Techniques for Efficient Simulation of Large-Scale Engineering Systems},
  author={Bhattacharyya, Suparno and Tao, Jian and Gildin, Eduardo and Ragusa, Jean C},
  journal={Arch. Comput. Methods Eng.},
  pages={1--43},
  year={2025}
}

@article{romor2025explicable,
  title={Explicable hyper-reduced order models on nonlinearly approximated solution manifolds of compressible and incompressible Navier-Stokes equations},
  author={Romor, Francesco and Stabile, Giovanni and Rozza, Gianluigi},
  journal={J. Comput. Phys.},
  volume={524},
  pages={113729},
  year={2025}
}

@book{Aris1989,
	title = {Vectors, tensors, and the basic equations of fluid mechanics},
	author = {Aris, Rutherford},
	publisher = {Dover},
    address = {New York},
	year = {1989}
}

@article{zhao2014pod,
  title={{POD-DEIM} based model order reduction for the spherical shallow water equations with {T}urkel-{Z}was finite difference discretization},
  author={Zhao, Pengfei and Liu, Cai and Feng, Xuan},
  journal={J. Appl. Math.},
  year={2014},
}

@article{cstefuanescu2013pod,
  title={{POD/DEIM} nonlinear model order reduction of an {ADI} implicit shallow water equations model},
  author={Stefanescu, R. and Navon, Ionel Michael},
  journal={J. Comput. Phys.},
  volume={237},
  pages={95--114},
  year={2013},
}

@book{pukelsheim2006optimal,
  title={Optimal design of experiments},
  author={F. Pukelsheim},
  year={2006},
  publisher={SIAM},
  address = {Philadelphia}
}

@article{Humphry2025,
	title = {Efficient hyperreduction for large‐scale problems:
        exploiting reducible constraint manifolds in empirical
        quadrature procedure},
	author = {Humphry, Adrian and Yano, Masayuki},
	journal = {Int. J. Numer. Meth. Eng.},
	volume = {126},
	number = {23},
	pages = {e70204},
	year = {2025},
	doi = {10.1002/nme.70204}
}

@article{Veroy2003,
  title = {Reduced-basis approximation of the viscous Burgers equation:
      rigorous a posteriori error bounds},
  author = {Veroy, Karen and Prud’homme, Christophe and Patera, Anthony T.},
  journal = {C.R. Math.},
  volume = {337},
  number = {9},
  pages = {619–624},
  year = {2003},
  doi = {10.1016/j.crma.2003.09.023}
}

@article{Barrault2004,
  title = {An ‘empirical interpolation’ method: application to efficient
      reduced-basis discretization of partial differential equations},
  author = {Barrault, Maxime and Maday, Yvon and Nguyen, Ngoc Cuong
      and Patera, Anthony T.},
  journal = {C.R. Math.},
  volume = {339},
  number = {9},
  pages = {667–672},
  year = {2004},
  doi = {10.1016/j.crma.2004.08.006}
}

@article{An2008,
    title={Optimizing cubature for efficient integration of
        subspace deformations},
    author={An, Steven S. and Kim, Theodore and James, Doug L.},
    journal={ACM Trans. Graph.},
    volume={27},
    number={5},
    pages={1--10},
    year={2008},
    doi = {10.1145/1409060.1409118}
}

@article{Patera2017,
	title = {An {LP} empirical quadrature procedure for parametrized
        functions},
	author = {Patera, Anthony T. and Yano, Masayuki},
	journal = {C.R. Math.},
	volume = {355},
	number = {11},
	pages = {1161--1167},
	year = {2017},
    doi = {10.1016/j.crma.2017.10.020}
}

\end{document}